\documentclass[11pt,a4paper]{article}
\pdfoutput=1

\usepackage{jheppub}
\usepackage{latexsym}
\usepackage{revsymb}
\usepackage{multirow}
\usepackage{color}
\usepackage[usenames,dvipsnames,svgnames,table]{xcolor}

\usepackage{graphicx}
\usepackage{epsfig}  
\usepackage{epsf}    
\usepackage{dcolumn}
\usepackage{bm}
\usepackage{dcolumn}
\usepackage{textcomp}
\usepackage{float}
\usepackage{hypcap}
\usepackage[]{hyperref}
\usepackage{adjustbox}
\usepackage{multirow}

\usepackage{subfigure}
\usepackage{alphalph}
\usepackage{morefloats}
\usepackage{textcomp}

\newcommand{\be}{\begin{equation}}
\newcommand{\ee}{\end{equation}}
\newcommand{\ba}{\begin{eqnarray}}
\newcommand{\ea}{\end{eqnarray}}
\def\nue{{\nu_e}}

\def\numu{{\nu_{\mu}}}

\newcommand{\stch}{\sin^2 2\theta_{13}}

\newcommand{\tmt}{\theta_{23}}
\newcommand{\tet}{\theta_{13}}

\def\nue{{\nu_e}}

\def\numu{{\nu_{\mu}}}

\preprint{IP/BBSR/2016-1}

\title{Discovery Potential of T2K and NO$\nu$A in the Presence of a Light Sterile Neutrino} 

\author[a]{Sanjib Kumar Agarwalla,}
\author[a]{Sabya Sachi Chatterjee,}
\author[a]{Arnab Dasgupta,}
\author[b]{Antonio Palazzo}

\affiliation[a]{Institute of Physics, Sachivalaya Marg, Sainik School Post, Bhubaneswar 751005, India}
\affiliation[b]{Dipartimento Interateneo di Fisica ``Michelangelo Merlin'',
Via Amendola 173, 70126 Bari, Italy}

\emailAdd{sanjib@iopb.res.in}
\emailAdd{sabya@iopb.res.in}
\emailAdd{arnab.d@iopb.res.in}
\emailAdd{palazzo@mpp.mpg.de}

\abstract{
We study the impact of one light sterile neutrino on the prospective data 
expected to come from the two presently running long-baseline experiments 
T2K and NO$\nu$A when they will accumulate their full planned exposure. 
Introducing for the first time, the bi-probability representation in the 4-flavor framework, 
commonly used in the 3-flavor scenario, we present a detailed discussion of 
the behavior of the $\nu_\mu \to \nu_e$ and $\bar\nu_\mu \to \bar\nu_e$ transition 
probabilities in the 3+1 scheme. We also perform a detailed sensitivity study of these
two experiments (both in the stand-alone and combined modes) to assess  
their discovery reach in the presence of a light sterile neutrino. For realistic 
benchmark values of the mass-mixing parameters (as inferred from the existing 
global short-baseline fits), we find that the performance of both these experiments
in claiming the discovery of the CP-violation induced by the standard CP-phase
$\delta_{13} \equiv \delta$, and the neutrino mass hierarchy 
get substantially deteriorated. The exact loss of sensitivity depends on the value 
of the unknown CP-phase $\delta_{14}$. Finally, we estimate the discovery potential 
of total CP-violation (i.e., induced simultaneously by the two CP-phases $\delta_{13}$ 
and $\delta_{14}$), and the capability of the two experiments of reconstructing the true 
values of such CP-phases. The typical (1$\sigma$ level) uncertainties on the 
reconstructed phases are approximately $40^0$ for $\delta_{13}$ and $50^0$ 
for $\delta_{14}$.}

\arxivnumber{1601.05995}

\begin{document}
\maketitle
\flushbottom

\section{Introduction and Motivation}
\label{introduction}

More than fifteen years ago, pioneering observations of neutrinos originating from natural sources (the sun core
and the earth atmosphere) led to the first evidence of neutrino oscillations 
establishing the massive nature of these fundamental particles. Such a discovery, recently awarded
with the Nobel Prize~\cite{Nobel}, has been the first of a long series of milestones in our 
understanding of neutrinos, whose properties have been gradually clarified by the subsequent 
findings of several other experiments performed with man-made neutrino sources (reactors and accelerators).

The two apparently disjoint effective 2-flavor descriptions initially introduced to explain separately the solar and the
atmospheric neutrino anomalies have been gradually recognized as two pieces of a single and 
more complex mosaic, which is currently accepted as the standard picture of neutrino oscillations. 
Such a 3-flavor framework involves two distinct mass-squared splittings ($\Delta m^2_{31}$, $\Delta m^2_{21}$), 
three mixing angles ($\theta_{12}$, $\theta_{23}$, $\theta_{13}$), and one CP-phase $\delta$.

The latest fundamental step in the establishment of the 3-flavor scheme has been accomplished
very recently (during 2012), with the determination of the long-sought third mixing angle $\theta_{13}$ by 
means of dedicated reactor experiments~\cite{An:2015rpe,RENO:2015ksa,Abe:2014bwa}.
This discovery  has opened 
the way to the measurement of the last unknown 3-flavor parameters: the CP-phase $\delta$ 
and the neutrino mass hierarchy (MH), i.e. the sign of $\Delta m^2_{31}$.  Both properties are at the center 
of an intense world-wide research program, which will be carried out with new accelerator, reactor 
and atmospheric neutrino experiments (for a recent review see~\cite{Stanco:2015ejj}). 

In spite of its tremendous success and of its beautiful structure the standard 3-flavor framework
may not constitute the ultimate description of neutrinos, which may reserve surprises. In fact, a few anomalies
have been recorded at the short baseline experiments, which cannot be accommodated in the 3-flavor scheme
(see~\cite{Abazajian:2012ys,Palazzo:2013me,Gariazzo:2015rra} for a review of the topic).
The two standard mass-squared splittings $\Delta m^2_{21} \equiv m^2_2 - m^2_1$ and $\Delta m^2_{31} \equiv m^2_3 - m^2_1$
are too small to produce observable effects in such setups and (at least) one new much larger mass-squared 
difference $O(\mathrm{eV}^2)$ must be introduced. The hypothetical fourth mass eigenstate must be essentially
sterile. A rich and diverse program of new more sensitive short-baseline experiments is underway in order to test
such an intriguing hypothesis (see the review in~\cite{Lasserre:2014ita}), which, if confirmed would represent a revolution in our understanding
of neutrinos, as important as the discovery of neutrino oscillations.

At a phenomenological level, the existence of sterile neutrinos would make necessary to extend
the standard 3-flavor framework. The enlarged scheme must be realized in such a way to preserve
the very good description of all the other (non short-baseline) data. In the minimal extension, involving only one
sterile species, the so-called  $3+ 1$ scheme, the new mass eigenstate $\nu_4$ is assumed to
be weakly mixed with the active neutrino flavors ($\nu_e, \nu_\mu, \nu_\tau$) and it is separated from the standard 
mass eigenstates $(\nu_1, \nu_2, \nu_3)$ by a large $O(\mathrm{eV}^2)$ splitting, giving rise to the hierarchical
pattern $|\Delta m^2_{21}| \ll |\Delta m^2_{31}| \ll |\Delta m^2_{41}|$. The 3+1 scheme involves six
mixing angles and three (Dirac) CP-violating phases. Hence, in case of discovery of a sterile neutrino, 
we would face the formidable challenge of identifying six more properties (3 mixing angles, 2 CP-phases and the sign of
$\Delta m^2_{41}$) in addition to those involved in the standard 3-flavor framework.

The $3+1$ scheme naturally predicts sizable effects at the short baselines, where the oscillating factor
$\Delta_{41} \equiv \Delta m^2_{41}L /4E$ ($L$ being the baseline and $E$ the neutrino energy) is of order one, 
and one expects an oscillating behavior with the characteristic $L/E$ dependency. However,
it must be emphasized that sterile neutrinos are observable in other (non-short-baseline) types of experiments
where they may manifest in a more subtle way. In the solar sector, for example, a non zero-value
of the electron neutrino mixing with $\nu_4$ (parametrized by the matrix element $U_{e4}$) can be felt as
a small deviation of the unitarity of the $(\nu_1, \nu_2,\nu_3)$ sub-system~\cite{Palazzo:2011rj,Palazzo:2012yf} 
(see also~\cite{Giunti:2009xz}). In the atmospheric sector, as first evidenced in~\cite{Nunokawa:2003ep}, at very 
high [$O$(TeV)] energies one expects a novel MSW resonance, which may reveal as a distortion of the zenith
angle distribution. To this regard we mention the dedicated analysis underway in the IceCube collaboration whose
preliminary results have been shown very recently in~\cite{IceCube-Sterile-CPAN-2015}. Complementary
information on sterile neutrino mixing using atmospheric neutrinos has been also extracted from  
Super-Kamiokande~\cite{Abe:2014gda}.

Sterile neutrino oscillations can influence also the long-baseline (LBL) accelerator experiments%
\footnote{In this paper we focus on the $\nu_\mu \to \nu_e$ channel. However,
information on sterile neutrinos can be obtained also from the LBL disappearance 
$\nu_\mu \to \nu_\mu$ searches and from the neutral current data. See the analyses performed
by MINOS~\cite{Timmons:2015lga,Adamson:2011ku}. Also the appearance $\nu_\mu \to \nu_\tau$ channel
can provide information, albeit currently the low statistics limits the sensitivity. See the
analysis performed by OPERA~\cite{Agafonova:2015neo}.}.
We recall that these setups, when working in the $\nu_\mu \to \nu_e$  (and $\bar \nu_\mu \to \bar \nu_e$)
appearance channel,
can probe the 3-flavor CP-violating phenomena encoded by the CP-phase $\delta$.
Their sensitivity is related to the fact that at long distances the $\nu_\mu \to \nu_e$ transition probability develops a 
small interference term (which is completely negligible at SBL) between the oscillations induced by the 
atmospheric splitting and those driven by the (much smaller) solar splitting. As first evidenced in~\cite{Klop:2014ima},
in the presence of sterile neutrinos a new interference term appears in the transition probability, which depends
on one additional CP-phase. Notably, the size of the new non-standard 4-flavor interference term is expected to be
similar to that of the standard 3-flavor interference term. 
Therefore, the sensitivity to the new interference term can disclose the possibility
to explore the new enlarged CP-violation (CPV) sector, which involves two additional CP-phases.  
It should  be stressed that both in the 3-flavor and 4-flavor schemes the CP-phases cannot be observed in SBL
experiments, since in such setups the two standard oscillating frequencies have negligible values.
Therefore, the LBL and SBL experiments are complementary in the exploration of the 3+1 scheme 
(and of any scheme involving more than one sterile neutrino).   
  
This basic observation provides the motivation for the study performed in the present paper, in which we
explore the physics potential of the two currently running LBL experiments T2K and NO$\nu$A in the presence of a 
hypothetical light sterile neutrino.%
\footnote{Old works on sterile neutrinos at LBL can be found in~\cite{Donini:2001xy,Donini:2001xp,Donini:2007yf,Dighe:2007uf,Donini:2008wz,Yasuda:2010rj,Meloni:2010zr,Bhattacharya:2011ee,Donini:2012tt}. More recent 
studies focusing on the future LBNE/DUNE experiment~\cite{Acciarri:2015uup} have been recently performed
in~\cite{Hollander:2014iha,Berryman:2015nua,Gandhi:2015xza}. In principle, the CERN-Pyh\"asalmi baseline of 2290 km 
actively studied under the umbrella of the LBNO collaboration~\cite{Agarwalla:2011hh,Agarwalla:2013kaa} can also be very sensitive to these issues. The same is also true for the future T2HK experiment~\cite{Abe:2014oxa,Abe:2015zbg}, 
which is a bigger version of T2K.}
The analyses performed in~\cite{Klop:2014ima,Palazzo:2015gja} with the existing data from T2K and NO$\nu$A have
already shown that these two experiments%
\footnote{In principle (see the 4-flavor analysis performed in~\cite{Palazzo:2015wea}), the CP-phases can impact 
also the $\nu_\mu \to \nu_e$ searches of ICARUS~\cite{Antonello:2012pq,Antonello:2015jxa} and OPERA~\cite{Agafonova:2013xsk}.
However, the very low statistics prevents to extract any information on the phases.}
can probe one of the new CP-phases for realistic values of the
mixing angles indicated by the global 3+1 fits~\cite{Giunti:2013aea, Kopp:2013vaa}. In addition, 
in~\cite{Palazzo:2015gja} it has been pointed out that the statistical significance of the current
indications concerning the standard CP-phase $\delta$ and the neutrino mass hierarchy is modified (reduced)
in the presence of sterile neutrinos. Therefore, it is timely and interesting to investigate if such features are expected
to persist even when the full exposure will be reached in both experiments. With this aim, we perform a prospective
study addressing in a quantitative way these questions.

The paper is organized as follows. In section~\ref{sec:probability}, we present a detailed discussion of the behavior of the 
4-flavor $\nu_\mu \to \nu_e$ and  $\bar\nu_\mu \to \bar\nu_e$ transition probabilities. 
For the first time we extend the bi-probability representation, commonly 
used in the 3-flavor framework, to the more general 3+1 scheme. 
Section~\ref{experimental-details} deals with the experimental details and also discusses the bi-events plots.
In section~\ref{simulation-details}, we describe the details of the statistical method that we use for the analysis.
Section~\ref{results} is devoted to the presentation of the results of the sensitivity study of T2K and NO$\nu$A. 
We draw our conclusions in section~\ref{Conclusions}.

\section{Conversion probability in the 3+1 scheme}
\label{sec:probability}
\subsection{Theoretical framework}

In the presence of a sterile neutrino $\nu_s$, the mixing among the flavor and the 
mass eigenstates is described by a $4\times4$ matrix. A convenient parameterization of the mixing matrix is
\begin{equation}
\label{eq:U}
U =   \tilde R_{34}  R_{24} \tilde R_{14} R_{23} \tilde R_{13} R_{12}\,, 
\end{equation} 
where $R_{ij}$ ($\tilde R_{ij}$) are real (complex) $4\times4$ rotations in the ($i,j$) plane
containing the $2\times2$ submatrix 
\begin{eqnarray}
\label{eq:R_ij_2dim}
     R^{2\times2}_{ij} =
    \begin{pmatrix}
         c_{ij} &  s_{ij}  \\
         - s_{ij}  &  c_{ij}
    \end{pmatrix}
\,\,\,\,\,\,\,   
     \tilde R^{2\times2}_{ij} =
    \begin{pmatrix}
         c_{ij} &  \tilde s_{ij}  \\
         - \tilde s_{ij}^*  &  c_{ij}
    \end{pmatrix}
\,,    
\end{eqnarray}
in the  $(i,j)$ sub-block, with
\begin{eqnarray}
 c_{ij} \equiv \cos \theta_{ij} \qquad s_{ij} \equiv \sin \theta_{ij}\qquad  \tilde s_{ij} \equiv s_{ij} e^{-i\delta_{ij}}.
\end{eqnarray}
The  parameterization in Eq.~(\ref{eq:U}) has the following properties: i) When the mixing
invoving the fourth state is zero $(\theta_{14} = \theta_{24} = \theta_{34} =0)$ 
it returns the 3-flavor matrix in its common parameterization.
ii) With the leftmost positioning of the matrix $\tilde R_{34}$ the vacuum $\nu_{\mu} \to \nu_{e}$    
conversion probability is independent of $\theta_{34}$ and of the related CP-phase
 $\delta_{34}$ (see~\cite{Klop:2014ima}). iii)  For small values of $\theta_{13}$ and of the mixing angles 
involving $\nu_4$, one has $|U_{e3}|^2 \simeq s^2_{13}$, $|U_{e4}|^2 = s^2_{14}$, 
$|U_{\mu4}|^2  \simeq s^2_{24}$ and $|U_{\tau4}|^2 \simeq s^2_{34}$, 
with an immediate physical interpretation of the new mixing angles.

\subsection{Analytical Expressions in Vacuum and Matter}
\label{analytical-expressions}

Let us now consider the transition probability relevant for T2K and NO$\nu$A.
As shown in~\cite{Klop:2014ima}, the $\nu_{\mu} \to \nu_{e}$ conversion probability 
can be written as the sum of three contributions
\begin{eqnarray}
\label{eq:Pme_4nu_3_terms}
P^{4\nu}_{\mu e}  \simeq  P^{\rm{ATM}} + P^{\rm {INT}}_{\rm I}+   P^{\rm {INT}}_{\rm II}\,.
\end{eqnarray}
The first (positive-definite) term is driven by the atmospheric frequency and it 
gives the leading contribution to the probability. The second and third terms 
are related to the interference of two distinct frequencies  and can assume both
positive and negative values. The first of the two interference terms is connected to the standard 
solar-atmospheric interference, while the second one is driven by the atmospheric-sterile interference.
The conversion probability depends on the three small mixing angles $\theta_{13}$, $\theta_{14}$, $\theta_{24}$,
whose best estimates, derived from the global 3-flavor (for $\theta_{13}$) analyses~\cite{Capozzi:2013csa,Forero:2014bxa,Gonzalez-Garcia:2014bfa} and from the 3+1 
fits~\cite{Giunti:2013aea, Kopp:2013vaa} (for $\theta_{14}$ and $\theta_{24}$), turn out to be very similar 
and we have approximately $s_{13} \sim s_{14} \sim s_{24} \sim 0.15$ (see table~\ref{tab:benchmark-parameters}).
Therefore, it is meaningful to treat all such three mixing angles as small quantities of the same order $\epsilon$. An other 
small quantity involved in the transition probability is the ratio of the solar and the atmospheric
mass-squared splitting  $\alpha \equiv \Delta m^2_{21}/ \Delta m^2_{31} \simeq \pm 0.03$,  which
can be assumed to be of order $\epsilon^2$. Keeping terms up to the third order, in vacuum, one finds 
\begin{eqnarray}
\label{eq:Pme_atm}
 &\!\! \!\! \!\! \!\! \!\! \!\! \!\!  P^{\rm {ATM}} &\!\! \simeq\,  4 s_{23}^2 s^2_{13}  \sin^2{\Delta}\,,\\
 \label{eq:Pme_int_1}
 &\!\! \!\! \!\! \!\! \!\! \!\! \!\! \!\! P^{\rm {INT}}_{\rm I} &\!\!  \simeq\,   8 s_{13} s_{12} c_{12} s_{23} c_{23} (\alpha \Delta)\sin \Delta \cos({\Delta + \delta_{13}})\,,\\
 \label{eq:Pme_int_2}
 &\!\! \!\! \!\! \!\! \!\! \!\! \!\! \!\! P^{\rm {INT}}_{\rm II} &\!\!  \simeq\,   4 s_{14} s_{24} s_{13} s_{23} \sin\Delta \sin (\Delta + \delta_{13} - \delta_{14})\,,
\end{eqnarray}
where $\Delta \equiv  \Delta m^2_{31}L/4E$ is the atmospheric oscillating factor, which depends
on the baseline $L$ and the neutrino energy $E$. The two LBL experiments under consideration,
T2K and NO$\nu$A, make use of an off-axis configuration, which leads to a narrow-band
sharply-peaked energy spectrum of the emitted neutrinos. In theory, the off-axis angle should 
be tuned exactly to match (at the peak energy) the condition $\Delta \sim \pi/2$, corresponding
to the first oscillation maximum. In practice this condition holds only approximately.  In T2K, the neutrino
flux is peaked at $E = 0.6$ GeV and the condition $\Delta = \pi/2$ is exactly 
matched. In NO$\nu$A, the peak of flux is located at  $E = 2$ GeV, while 
the oscillations maximum is at $E = 1.5$ GeV. 
In the following, when discussing the behavior of the conversion probability we will chose the 
peak value for both experiments, i.e. we will use $E =0.6$\, GeV for T2K and $E=2$\, GeV for NO$\nu$A. 
This will allow a better understanding of the subsequent discussion at the events level presented 
in section~\ref{experimental-details}. In fact, in both experiments the total rate keeps the leading contribution
from the energies close to the peak. Also, it should be stressed that the sensitivity to the spectral distortions is quite limited 
(due to systematic errors and the limited statistics) and the total rate suffices to understand the
basic feature of the numerical results, albeit our analysis includes a full treatment of the 
spectrum (see section~\ref{experimental-details}).

The presence of matter slightly modifies the transition probability through the MSW effect, which introduces 
a dependency on the ratio
\begin{equation}
\label{eq:v}\,
v = \frac{V}{k} \equiv \frac{2VE}{\Delta m^2_{31}}\,,
\end{equation}
where 
\begin{equation}
\label{eq:Pme_atm}
 V = \sqrt 2 G_F N_e\,
\end{equation}
is the constant matter potential along the neutrino trajectory in the earth crust.  
Both in T2K and in NO$\nu$A the value of $v$ is relatively small, being $v\sim 0.05$ in T2K,  
and $v\sim 0.17$ in NO$\nu$A, where we have taken as a benchmark  value the peak energy 
($E = 0.6$ GeV in T2K, $E = 2$ GeV in NO$\nu$A). Therefore, 
$v$ can be treated as a small parameter of order $\epsilon$. 
The $\nu_\mu \to \nu_e$ conversion probability in matter can be obtained (see the appendix in~\cite{Klop:2014ima} and the 
works~\cite{Cervera:2000kp,Asano:2011nj,Agarwalla:2013tza}) by performing, in the leading
term of the the vacuum probability, the following substitution
\begin{equation}
\label{eq:Pme_atm_matt}
P^{\rm {ATM}}_m \simeq  (1+ 2 v) P^{\rm {ATM}}\,,
\end{equation}
which incorporates the (third order) corrections due to matter effects. It can be
shown that the two interference terms acquire corrections which are of the 
fourth order. In this work, we will limit the expansion at the third order in $\epsilon$.
Therefore, the interference terms will have the vacuum expression.  

\begin{table}[t]
\begin{center}
{
\newcommand{\mc}[3]{\multicolumn{#1}{#2}{#3}}
\newcommand{\mr}[3]{\multirow{#1}{#2}{#3}}
\begin{tabular}{|c|c|c|}
\hline\hline
\mr{2}{*}{\bf Parameter} & \mr{2}{*}{\bf True value} & \mr{2}{*}{\bf Marginalization Range} \\
  & &  \\
\hline\hline
\mr{2}{*}{$\sin^2{\theta_{12}}$} & \mr{2}{*}{0.304} & \mr{2}{*}{Not marginalized} \\
  & &  \\
\hline
\mr{2}{*}{$\sin^22\theta_{13}$} & \mr{2}{*}{$0.085$} & \mr{2}{*}{Not marginalized} \\ 
  & &  \\
\hline
\mr{2}{*}{$\sin^2{\theta_{23}}$} & \mr{2}{*}{0.50} & \mr{2}{*}{[0.34, 0.68]} \\
  & &  \\
\hline
\mr{2}{*}{$\sin^2{\theta_{14}}$} & \mr{2}{*}{0.025} & \mr{2}{*}{Not marginalized} \\
  & &  \\  
\hline
\mr{2}{*}{$\sin^2{\theta_{24}}$} & \mr{2}{*}{0.025} & \mr{2}{*}{Not marginalized} \\
  & &  \\ 
\hline
\mr{2}{*}{$\sin^2{\theta_{34}}$} & \mr{2}{*}{0.0} & \mr{2}{*}{Not marginalized} \\
  & &  \\  
\hline
\mr{2}{*}{$\delta_{13}/^{\circ}$} & \mr{2}{*}{[- 180, 180]} & \mr{2}{*}{[- 180, 180]} \\
  & &  \\
\hline
\mr{2}{*}{$\delta_{14}/^{\circ}$} & \mr{2}{*}{[- 180, 180]} & \mr{2}{*}{[- 180, 180]} \\
  & &  \\
\hline
\mr{2}{*}{$\delta_{34}/^{\circ}$} & \mr{2}{*}{0} & \mr{2}{*}{Not marginalized} \\
  & &  \\  
\hline
\mr{2}{*}{$\frac{\Delta{m^2_{21}}}{10^{-5} \, \rm{eV}^2}$} & \mr{2}{*}{7.50} & \mr{2}{*}{Not marginalized} \\
  & &  \\
\hline
\mr{2}{*}{$\frac{|\Delta{m^2_{32}}|}{10^{-3} \, \rm{eV}^2}$} & \mr{2}{*}{2.4} & \mr{2}{*}{Not marginalized} \\
  & &  \\
\hline
\mr{2}{*}{$\frac{\Delta{m^2_{31}}}{10^{-3} \, \rm{eV}^2}$ (NH)} & \mr{2}{*}{(2.4 + 0.075)} &\mr{2}{*}{Not marginalized} \\
  & & \\
\hline
\mr{2}{*}{$\frac{\Delta{m^2_{31}}}{10^{-3} \, \rm{eV}^2}$ (IH)} & \mr{2}{*}{- 2.4} &\mr{2}{*}{Not marginalized} \\
  & & \\
\hline
\mr{2}{*}{$\frac{\Delta{m^2_{41}}}{\rm{eV}^2}$} & \mr{2}{*}{1.0} & \mr{2}{*}{Not marginalized} \\
  & &  \\
\hline\hline
\end{tabular}
}
\caption{Parameter values/ranges used in the numerical calculations. The second column reports 
the true values of the oscillation parameters used to simulate the 
``observed'' data set. The third column depicts the range 
over which $\sin^2\theta_{23}$, $\delta_{13}$, and $\delta_{14}$ are varied 
while minimizing the $\chi^{2}$ to obtain the final results.}
\label{tab:benchmark-parameters}
\end{center}
\end{table}

Before closing this section we recall that a swap in the neutrino mass hierarchy 
is parametrized by the replacements 
\begin{eqnarray} 
\label{eq:phase_symm}
\Delta   \to   -\Delta, \qquad \alpha \to -\alpha, \qquad v \to -v. 
\end{eqnarray} 
Due to the change of sign of $v$ in Eq.~(\ref{eq:Pme_atm_matt}), the transition probability 
(which acquires the dominant contribution from the atmospheric term) tend to increase (decrease) 
with respect to the vacuum case  in the NH (IH) case. NO$\nu$A is expected to be more sensitive than T2K to
the MH because of the larger value of the ratio $v$.  
\begin{figure}[t]
\centerline{
\includegraphics[width=0.49\textwidth]{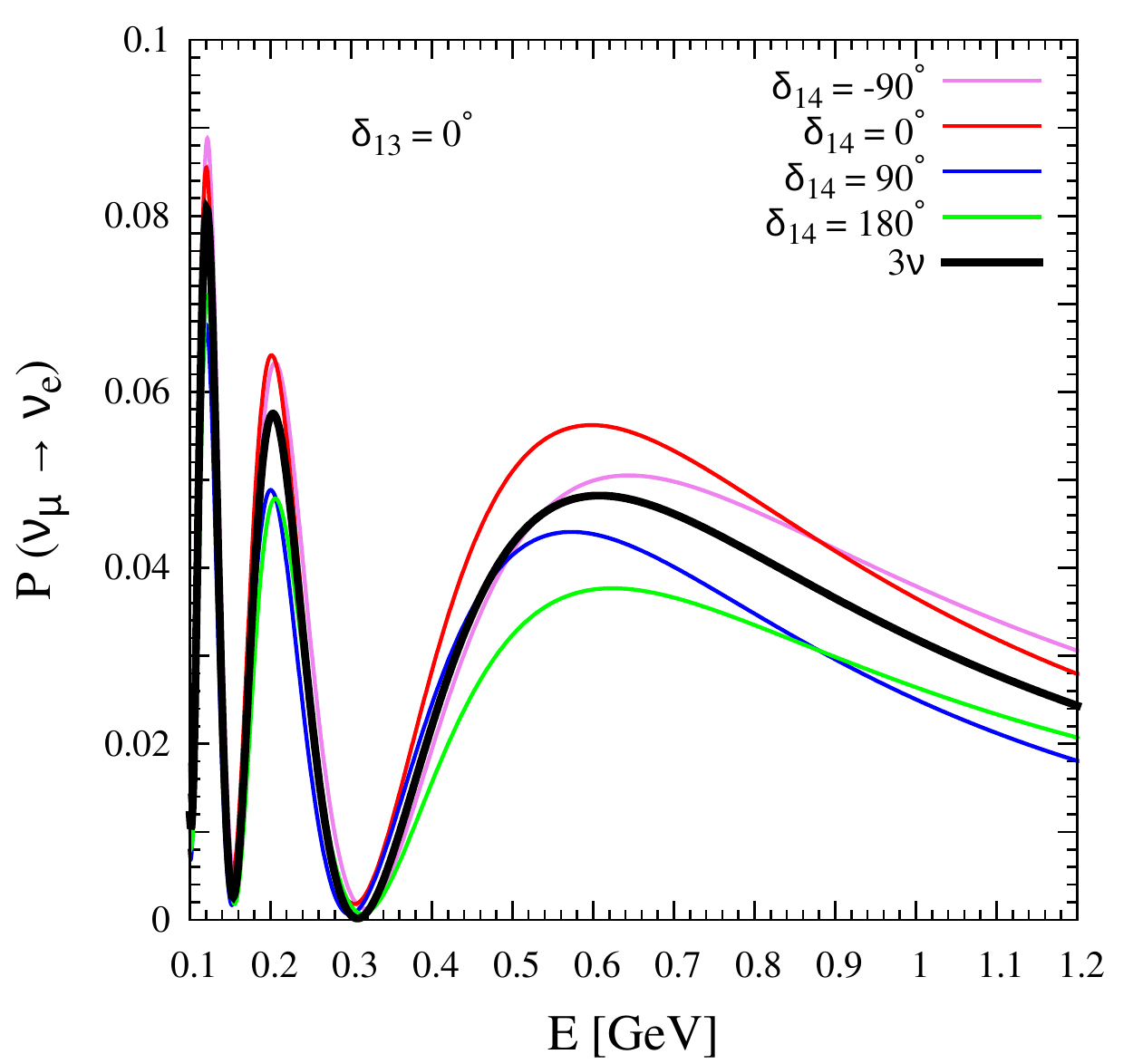}
\includegraphics[width=0.49\textwidth]{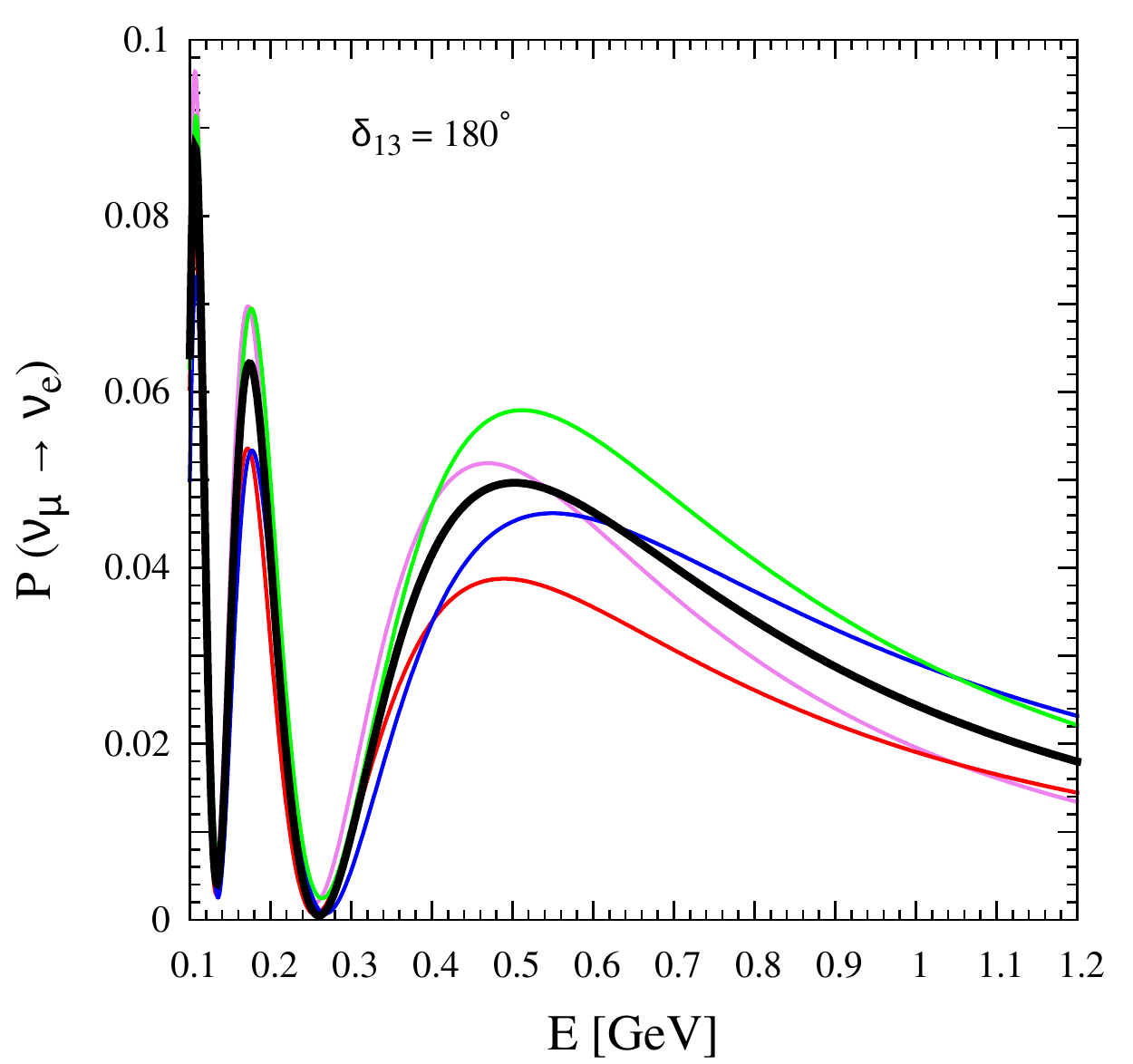}
}
\centerline{
\includegraphics[width=0.49\textwidth]{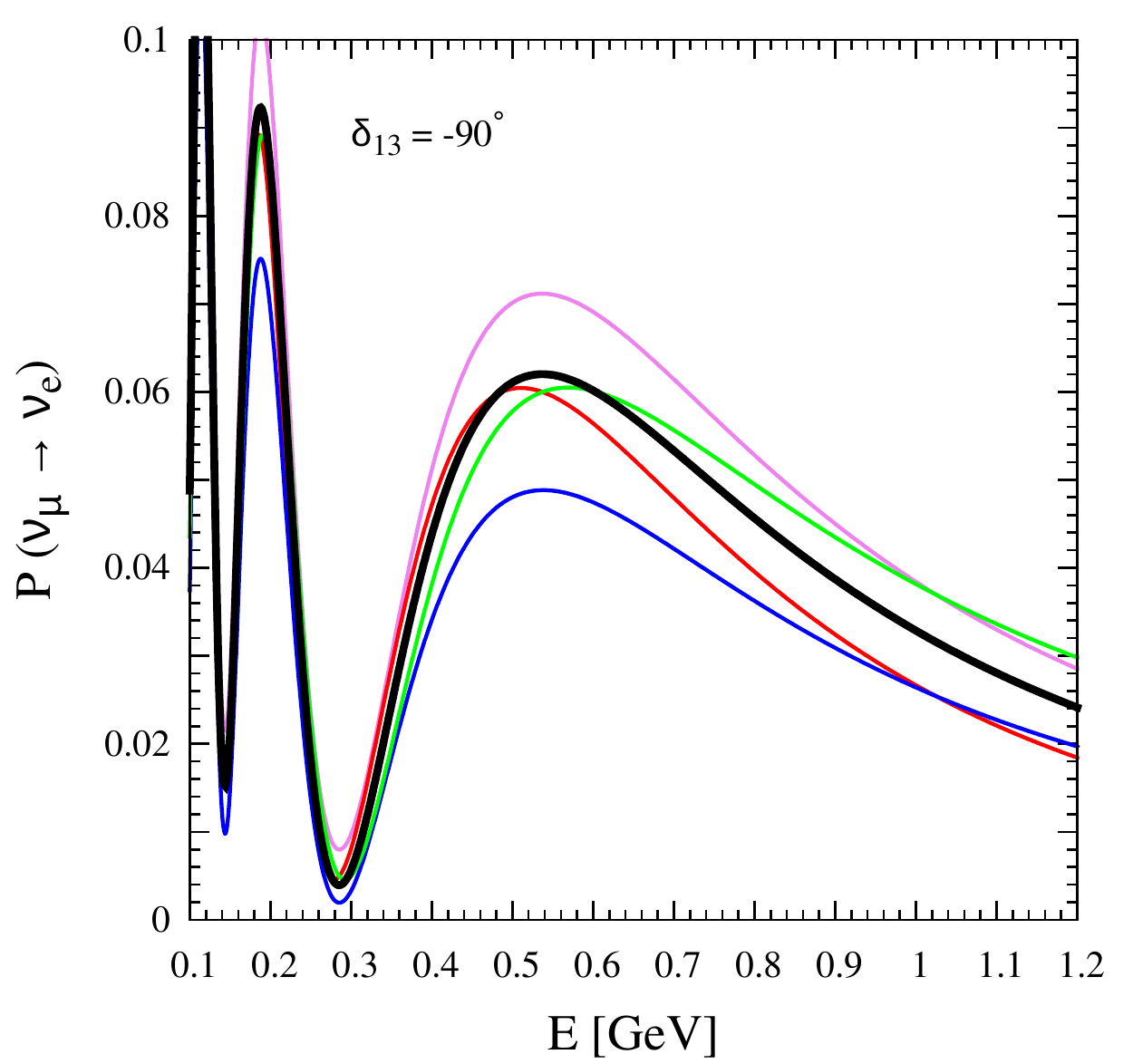}
\includegraphics[width=0.49\textwidth]{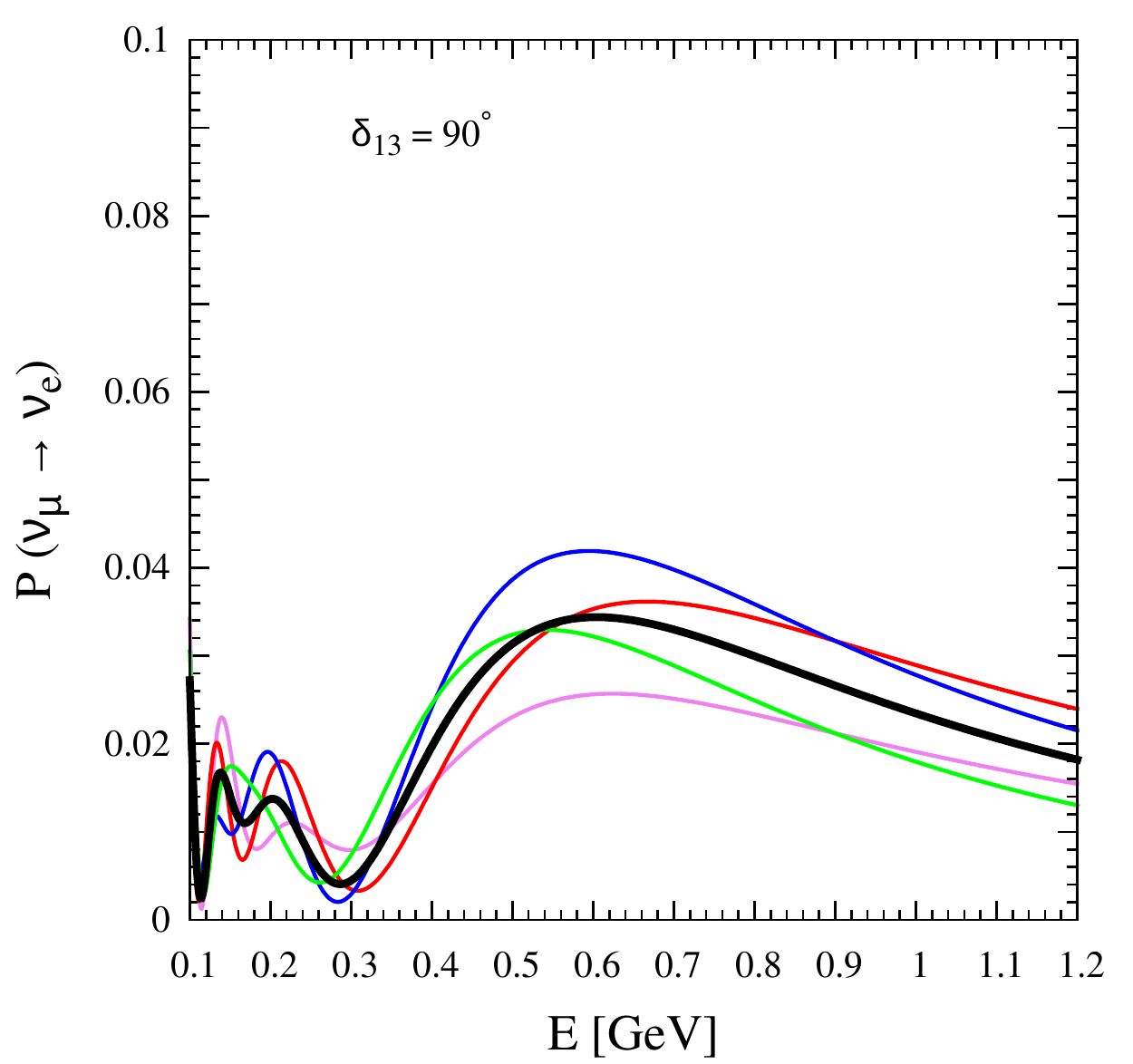}
}
\caption{$\numu \rightarrow \nue$ transition probability as a function 
of neutrino energy for T2K after performing the averaging over the fast oscillations.}
\label{fig:T2K-appearance-probability}
\end{figure}
Finally, we recall that the transition probability for antineutrinos is obtained from 
that of neutrinos with a change in the sign of the MSW potential $V$ and of all the CP-phases.
This, for a given choice of the MH, corresponds to the substitutions   
\begin{eqnarray} 
\label{eq:phase_symm}
\delta_{13}   \to   -\delta_{13}, \qquad \delta_{14} \to -\delta_{14}, \qquad v \to -v. 
\end{eqnarray} 
 In the NH case $v>0$ for neutrinos and $v<0$ for antineutrinos. 
 According to Eq.~(\ref{eq:Pme_atm_matt}), in the NH case
 the leading contribution to the transition probability will increase (decrease) 
 for neutrinos (antineutrinos). In the IH case the opposite conclusion holds. 
 
\begin{figure}[t!]
\centerline{
\includegraphics[width=0.49\textwidth]{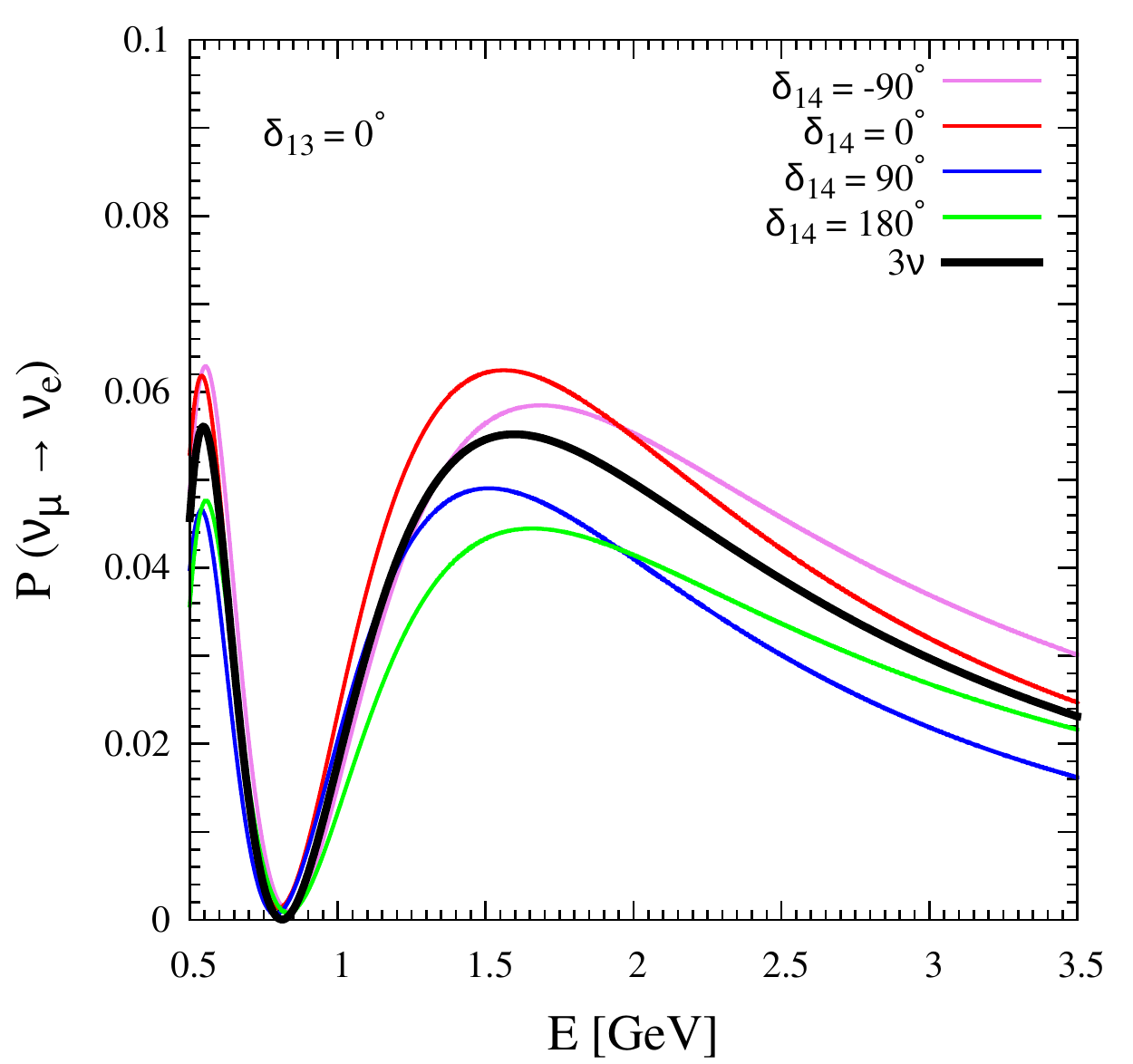}
\includegraphics[width=0.49\textwidth]{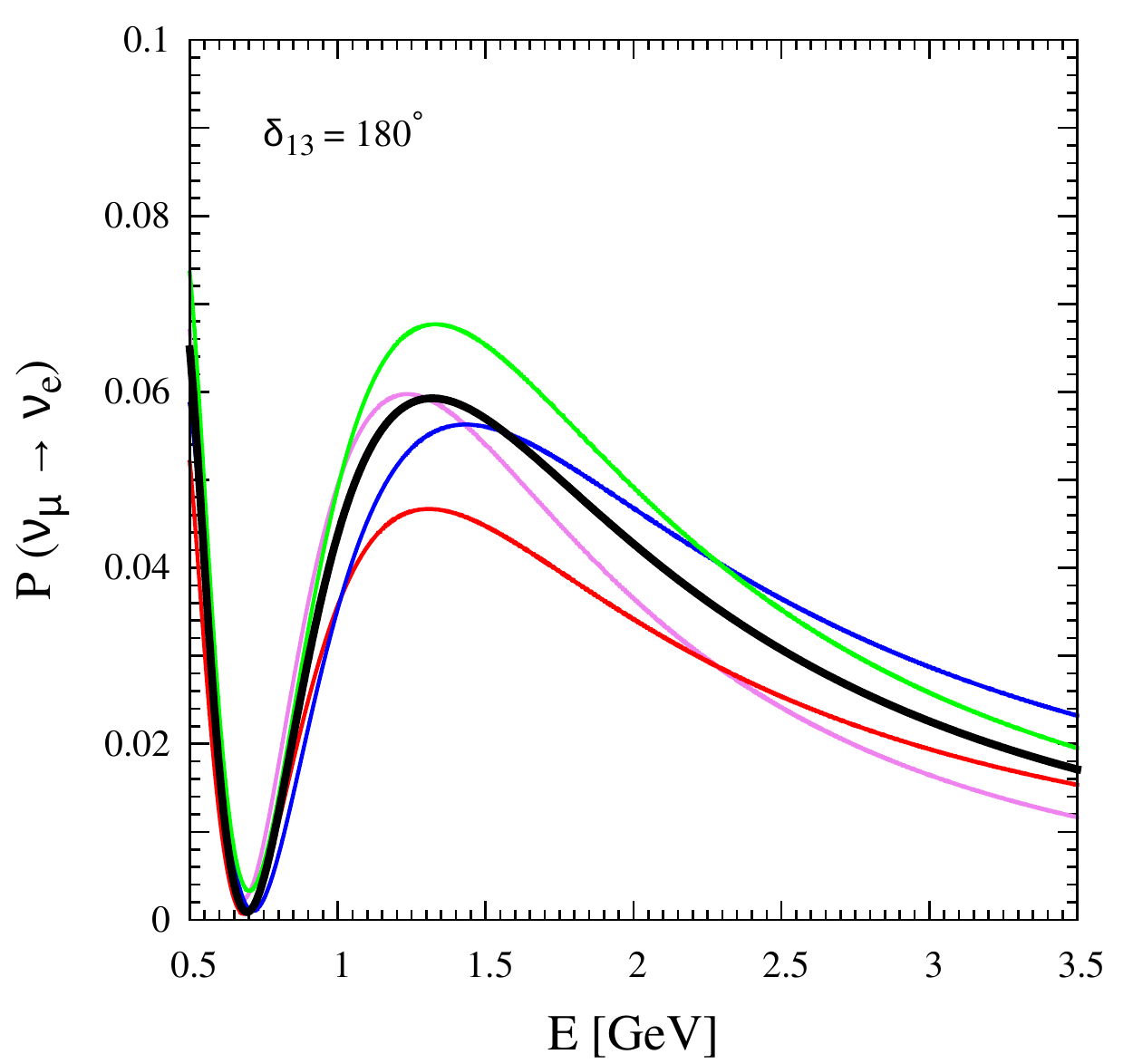}
}
\centerline{
\includegraphics[width=0.49\textwidth]{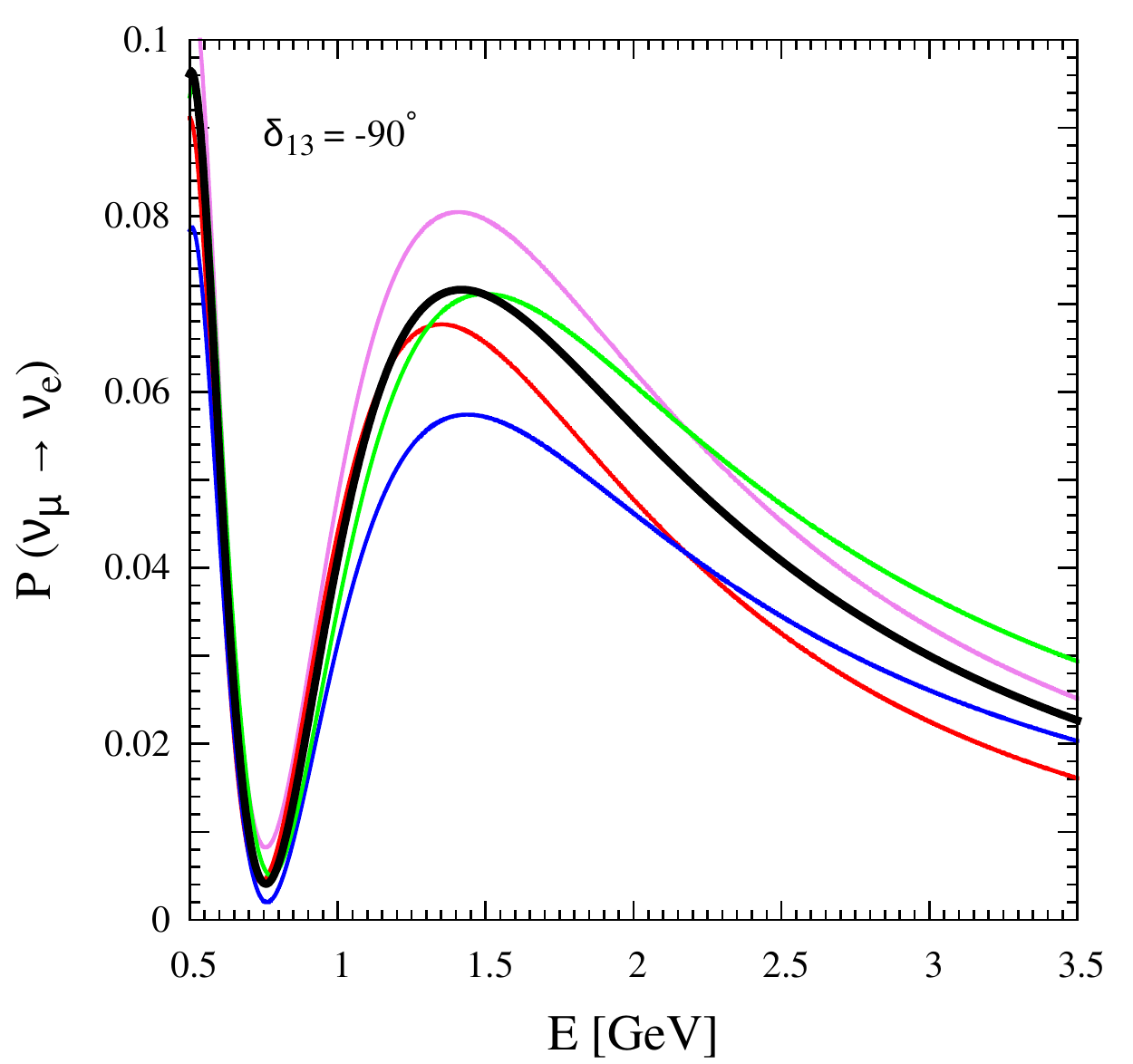}
\includegraphics[width=0.49\textwidth]{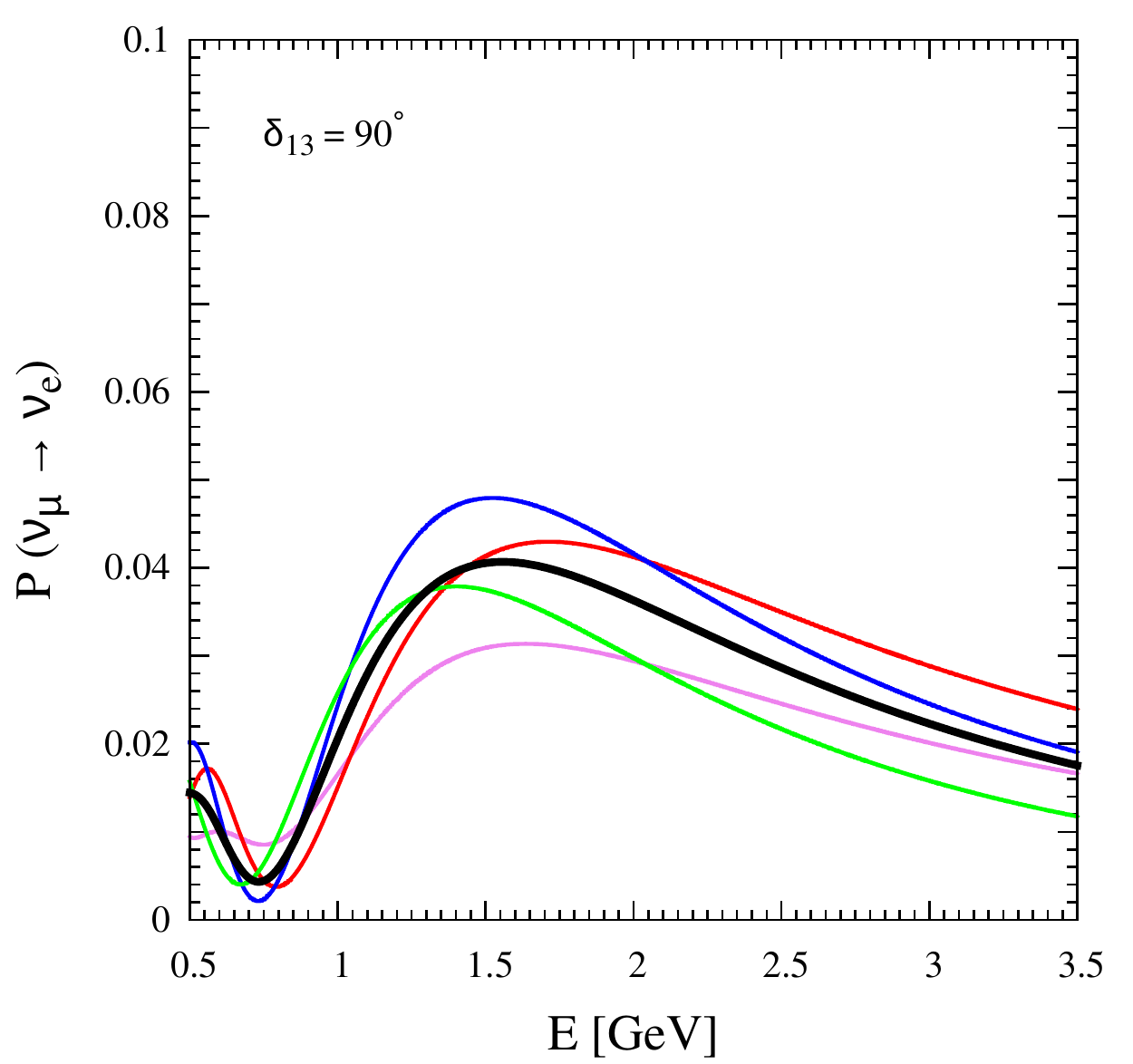}
}
\caption{$\numu \rightarrow \nue$ transition probability as a function 
of neutrino energy for NO$\nu$A after performing the averaging over the fast oscillations.}
\label{fig:NOvA-appearance-probability}
\end{figure}

Figures~\ref{fig:T2K-appearance-probability} and \ref{fig:NOvA-appearance-probability}
show the transition probability for neutrinos as a function of the energy 
for both experiments T2K and NO$\nu$A. In both figures we have assumed 
normal hierarchy and we have taken the mass-squared difference 
$\Delta m^2_{41} = 1$ eV$^2$ and fixed the mixing angle $\theta_{13}$ and the
two mixing angles $\theta_{14}$ and $\theta_{24}$ at the benchmark values
indicated in the second column of table~\ref{tab:benchmark-parameters}, where we also report all the other mass-mixing 
parameters involved in the calculations. For such high values of the 
mass-squared splitting, the oscillating factor $\Delta_{14}$ is very large and the 
sterile-induced oscillations are completely averaged out by the finite energy resolution of the detector.
Hence, we report the transition probability obtained after that such an averaging process
has been taken into account. In each plot of Figs.~\ref{fig:T2K-appearance-probability} and \ref{fig:NOvA-appearance-probability}
the value of the standard CP-phase $\delta_{13}$ is kept fixed  at the value displayed in the legend,
while the new phase $\delta_{14}$ assumes four representative values.
In each panel, the 3-flavor probability is represented by a thick black line,
while the four 3+1 cases are displayed by thin colored lines. The magenta
curve corresponds to $\delta_{14} = -\pi/2$, the blue one to  $\delta_{14} = \pi/2$,
the red one to $\delta_{14} = 0$ and the green one to $\delta_{14} = \pi$.
For clarity, we will adopt such color convention in all the figures presented in the paper.
From Figs.~\ref{fig:T2K-appearance-probability} and \ref{fig:NOvA-appearance-probability}
 it clearly emerges that the impact of the 4-flavor corrections
induced by a non-zero value of the mixing angles $\theta_{14}$ and $\theta_{24}$ 
is sizable. Their amplitude and shape depend on the particular value of the 
new CP-phase $\delta_{14}$.  The plots show that the most evident 
effect of the 4-flavor corrections is a change in the overall normalization
of the transition probability with respect to the standard 3-flavor case. In addition,  
an appreciably different energy dependence is also present, which reflects
the different dependency from the $L/E$ ratio of the standard interference
term [Eq.~(\ref{eq:Pme_int_1})] and the non-standard one [Eq.~(\ref{eq:Pme_int_2})].
The changes induced  in the overall
normalization are as big as the modifications induced by varying $\delta_{13}$
(compare the excursion of the black curves between different panels with the
excursion of the colored curve in a given panel). This confirms the analytical
estimates made in the previous section.

\subsection{Bi-Probability Plots}
\label{bi-probability}

In the 3-flavor framework, the behavior of the transition probability is often represented
with the CP-phase trajectory diagrams in bi-probability space, first introduced in~\cite{Minakata:2001qm}.
Such plots, commonly dubbed as bi-probability plots, represent the 
parametric curves of the two transition probabilities ($\nu_\mu \to \nu_e$ and $\bar\nu_\mu \to \bar\nu_e$)
where the varying parameter is the CP-phase $\delta_{13}$. Since the two
transition probabilities are cyclic functions of the phase $\delta_{13}$ the resulting contours form
a closed curve. This representation is particularly advantageous as it gives a bird-eye view
of the salient features of a given experimental setups, in particular its sensitivity to MH and CPV.
Here we attempt to generalize the bi-probability representation to the more general 4-flavor scheme.
This makes sense because also in the 4-flavor scheme the probability remains a cyclic function
of the (more numerous) CP-phases. Also in this case, as we will show, this representation is very useful for the interpretation of the numerical results. In the following we first recall the basic features of the bi-probability plots in the standard 3-flavor framework, then we generalize our study to the 4-flavor scheme.

\subsubsection{The 3-flavor case}
\label{bi-probability_3nu}

 In the 3-flavor case the neutrino and antineutrino transition probabilities can be written as
\begin{eqnarray}
\label{eq:Parametric_3nu_nu}
 P &=& P_0 + A (\cos \Delta \cos{\delta_{13}} - \sin \Delta \sin{\delta_{13}}) \, \\
 \label{eq:Parametric+3nu_nubar}
 \bar P &=& \bar P_0 + \bar A (\cos \Delta \cos{\delta_{13}} + \sin \Delta \sin{\delta_{13}})\,.
 \end{eqnarray}
In general, due to the presence of matter effects one has $P_0 \ne \bar P_0$ and $ A\ne \bar A$.
As discussed in the previous section the matter effects shift%
\footnote{$P_0$ and $\bar P_0$ can be identified with $P^{\rm {ATM}}_m$ of Eq.~(\ref{eq:Pme_atm_matt}), where
the sign of the adimensional quantity $v$ related to the matter
effects (for a fixed choice of the mass hierarchy) is opposite for neutrinos and antineutrinos.}
$P_0$ by an amount  proportional to 
the small parameter $v = 2VE/\Delta m^2_{31}$, which is of order $O(\epsilon)$. Therefore,
since $P_0$ in vacuum is $O(\epsilon^2)$, the absolute shift of $P_0$ is $O(\epsilon^3)$, which
is thus comparable with the amplitude ($A \simeq \bar A$) of the interference term.
The amplitude of the interference term is also modified with respect to the vacuum case,  its
relative change being proportional to $v$. Since the amplitude $A$ is of order $O(\epsilon^3)$, the corrections
are of order $O(\epsilon^4)$. Therefore, truncating the expansion of the probability at the third order
corresponds to consider $A = \bar A$, with
\begin{eqnarray}
\label{eq:Parametric_3nu_nu_A}
A \simeq 8 s_{13} s_{12} c_{12} s_{23} c_{23} (\alpha\Delta)\sin\Delta \,.
\end{eqnarray}
The relations (\ref{eq:Parametric_3nu_nu})-(\ref{eq:Parametric+3nu_nubar}) represent the parametric equations 
of an ellipse of center $(P_0,\bar P_0)$. 
Under the assumption $A = \bar A$, as already shown in~\cite{Minakata:2001qm}, the major (minor)
axis of the ellipse is proportional to $\sin \Delta$ ($\cos \Delta$) and has 
an inclination of $-\pi/4$ ($\pi/4$). To see this one can perform a counter-clockwise rotation $R$
of the parametric curve around its center $(P_0,\bar P_0)$ by the angle $\omega = \pi/4$
\begin{eqnarray}
\label{eq:R_ij_2dim}
     R =
    \begin{bmatrix}
         \cos \omega  &  -\sin \omega   \\
         \sin \omega  &  \cos \omega
    \end{bmatrix}
    =
     \frac{1}{\sqrt 2}\begin{bmatrix}
         1  &  -1   \\
         1  &  1
    \end{bmatrix}\,,    
\end{eqnarray}
obtaining for the rotated probabilities
\begin{eqnarray}
\label{eq:Parametric_3nu_nu_rotated}
 P' &=& P_0 - \sqrt{2} A \sin \Delta \sin{\delta_{13}} \, \\
 \label{eq:Parametric_3nu_nubar_rotated}
 \bar P' &=& \bar P_0 + \sqrt{2} A\cos \Delta \cos{\delta_{13}}\,.
 \end{eqnarray}
From these relations one arrives at the equation of an ellipse in the canonical form 
\begin{eqnarray}
 \frac{(P'-P_0)^2}{a^2} +\frac{(\bar P'-P_0)^2}{b^2} = 1 \,,
 \end{eqnarray}
with the two semi-axes having lengths   
\begin{eqnarray}
a &=& \sqrt{2}A\sin \Delta \,, \\ 
b &=& \sqrt{2}A\cos \Delta \,.
\end{eqnarray}
The combination of signs in the parametric equations (\ref{eq:Parametric_3nu_nu_rotated})-(\ref{eq:Parametric_3nu_nubar_rotated}) 
implies that for the NH case the ``chirality'' of the ellipse is positive, i.e. the 
trajectory winds in the counter-clockwise sense as the phase $\delta_{13}$ increases.
The chirality is opposite (negative) in the IH case since the coefficient 
A changes its sign under a swap of the mass hierarchy [see Eq.~(\ref{eq:Parametric_3nu_nu_A})].

The energy spectrum of the neutrino beams employed in typical LBL experiments is peaked
around the first oscillation maximum, where $\Delta \sim \pi/2$ and therefore one expects  $a\gg b$,
i.e. the major axis much bigger than the minor one. This feature is particularly pronounced in 
T2K since, as already remarked in the previous section, the peak energy ($E=0.6\,$GeV) almost 
exactly matches the condition $\Delta = \pi/2$. In this case the ellipse becomes almost degenerate with a line.
This behavior can be observed in Fig.~\ref{fig:T2K-appearance-bi-probability}, where in all
four panels the two black curves correspond to the 3-flavor limit for the two cases of NH (solid line) 
and IH (dashed line).  The colored curves correspond to four representative  4-flavor cases
that will be discussed later. In the NO$\nu$A experiment, at the peak energy ($E=2\,$GeV)
we have $\Delta = 0.4 \pi$ and the ratio of the major over the minor axis is given by $a/b = \tan \Delta \sim 3$, as one can 
appreciate from the  plots in Fig.~\ref{fig:NOvA-appearance-bi-probability}, where again
like for T2K, we display the two cases of NH (black solid line) and IH (black dashed line).
  
The bi-probability representation is particularly useful because it neatly shows that 
the presence of matter effects tend to split the two ellipses corresponding to the two mass
hierarchies, thus giving a qualitative bird-eye view of the sensitivity
of a given experiment to the MH. In addition, the ellipse curves show pictorially 
the effect of the genuine (or intrinsic) CPV due to $\sin \delta_{13}$,
disentangling it from the fake (or extrinsic) CPV induced by the matter effects.
In particular, for $\delta_{13} =  (-\pi/2, \pi/2 )$ the representative point in the bi-probability
space (respectively a circle and a square) lies on the intercepts of the ellipse with the
major axis and one has the maximal (intrinsic) CPV.
Conversely, the effect from the CP conserving $\cos \delta_{13}$ term is proportional
to the length of the minor axis. For $\delta_{13} = (0, \pi)$ the representative point on the ellipse
(respectively a triangle and an asterisk)  basically coincide.

From the comparison of  Fig.~\ref{fig:T2K-appearance-bi-probability} and 
Fig.~\ref{fig:NOvA-appearance-bi-probability}, it emerges that the splitting between 
the NH and IH curves is less pronounced in T2K than in NO$\nu$A. This 
is a consequence of the fact that, as already discussed in the previous section, 
the matter effects are larger in the second experiment. It is useful to recall [see Eq.~(\ref{eq:Pme_atm_matt})]
that the matter effects induce modifications proportional to the dimensionless quantity 
$v = 2 VE/\Delta m^2_{31}$, which is $v\simeq 0.05$ at the T2K peak energy and $v \simeq 0.17$ at
the NO$\nu$A peak energy.

\subsubsection{The 4-flavor case}
\label{bi-probability_4nu}

In the 3+1 scheme the transition probabilities have the general form
\begin{eqnarray}
\label{eq:Parametric_4nu_nu_v1}
P &=& P_0 + A \cos (\Delta + \delta_{13}) + B \sin (\Delta - \delta_{14} + \delta_{13})\,, \\
\label{eq:Parametric_4nu_nubar_v1}
\bar P &=& \bar P_0 + \bar A \cos(\Delta - \delta_{13}) + \bar B \sin (\Delta + \delta_{14} - \delta_{13})\,,
\end{eqnarray}
where, neglecting $O(\epsilon^4)$ corrections, we have
\begin{eqnarray}
\label{eq:Parametric_4nu_coeff_A}
A &\simeq& \bar A \simeq S_A \frac{\alpha}{|\alpha|}\Delta\sin\Delta\,, \\
\label{eq:Parametric_4nu_coeff_B}
B &\simeq& \bar B \simeq  S_B \sin \Delta\,.
\end{eqnarray}
We have introduced the two (positive definite) auxiliary quantities 
\begin{eqnarray}
\label{eq:Parametric_4nu_coeff_SA}
S_A &=& 8 s_{13} s_{12} c_{12} s_{23} c_{23}|\alpha| \,, \\
\label{eq:Parametric_4nu_coeff_SB}
S_B &=& 4 s_{14} s_{24} s_{13} s_{23}\,, 
\end{eqnarray}
which, for the specific values of the mixing angles under consideration (see the second 
table of~\ref{tab:benchmark-parameters}) yield
\begin{eqnarray}
\label{eq:Parametric_4nu_coeff_SAnum}
S_A  &\simeq& 0.8 \times 10^{-2}\,, \\
\label{eq:Parametric_4nu_coeff_SBnum}
S_B  &\simeq& 10^{-2}\,.
\end{eqnarray}
In the expression of the coefficients $A$ and $B$ in Eqs.~(\ref{eq:Parametric_4nu_coeff_A})-(\ref{eq:Parametric_4nu_coeff_B})
we have left evident the dependency from the oscillation factor $\Delta$ and from the sign of the ratio $\alpha$.
This will be useful when discussing the role of the neutrino mass hierarchy. 
We note that, under a swap of the MH (implying
$\Delta \to -\Delta$ and $\alpha \to -\alpha$), both coefficients $A$ and $B$ change sign
and therefore their product $AB$ remains unaltered and positive definite. 
The equations~(\ref{eq:Parametric_4nu_nu_v1})-(\ref{eq:Parametric_4nu_nubar_v1})  
can be re-expressed in the form
\begin{eqnarray}
\label{eq:Parametric_4nu_nu_v2}
P &=& P_0 + C \cos{\delta_{13}} + D \sin{\delta_{13}} \, \\
\label{eq:Parametric_4nu_nubar_v2}
\bar P &=& \bar P_0 + \bar C \cos{\delta_{13}} + \bar D \sin{\delta_{13}}\,,
\end{eqnarray}
where the new coefficients $C,D, \bar C, \bar D$ depend on $\Delta$ and 
on the CP-phase $\delta_{14}$ as follows
\begin{eqnarray}
\label{eq:Parametric_4nu_coeff_C}
C &=& + A \cos{\Delta} + B \sin{(\Delta - \delta_{14}}) \,, \\
\label{eq:Parametric_4nu_coeff_D}
D &=&  - A \sin{\Delta} + B \cos{(\Delta - \delta_{14}})\,,\\
\label{eq:Parametric_4nu_coeff_Cb}
\bar C &=& + \bar A \cos{\Delta} + \bar B \sin{(\Delta + \delta_{14}}) \,, \\
\label{eq:Parametric_4nu_coeff_Db}
\bar D &=&  + \bar A \sin{\Delta} - \bar B \cos{(\Delta + \delta_{14}})\,.
\end{eqnarray}
By eliminating the CP-phase $\delta_{13}$ from the expressions~(\ref{eq:Parametric_4nu_nu_v2})-(\ref{eq:Parametric_4nu_nubar_v2}), one easily arrives at the equation of an ellipse.%
\footnote{If one makes explicit the dependency on $\delta_{14}$ (instead of $\delta_{13}$)
and treats $\delta_{14}$ as the varying parameter one still obtains 
a (different) ellipse. In this case, the center of the ellipse depends on the value of $\delta_{13}$.}
However, the situation is more involved with respect to the 3-flavor case, because the geometrical 
properties of the 4-flavor ellipse (length and inclination of the two axes) depend not only from $\Delta$ 
but also from the new CP-phase $\delta_{14}$.

\begin{figure}[t]
\centerline{
\includegraphics[width=0.49\textwidth]{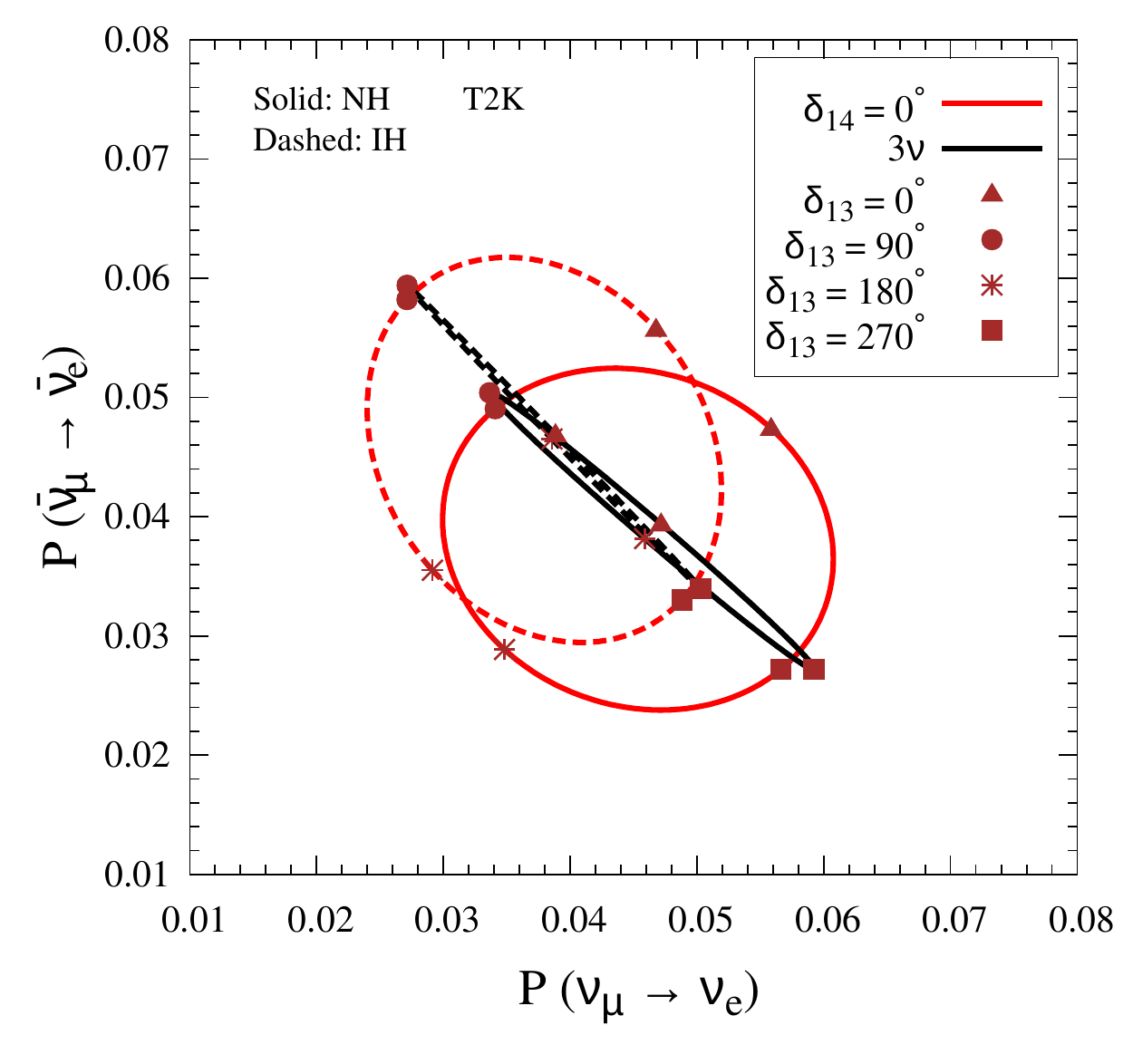}
\includegraphics[width=0.49\textwidth]{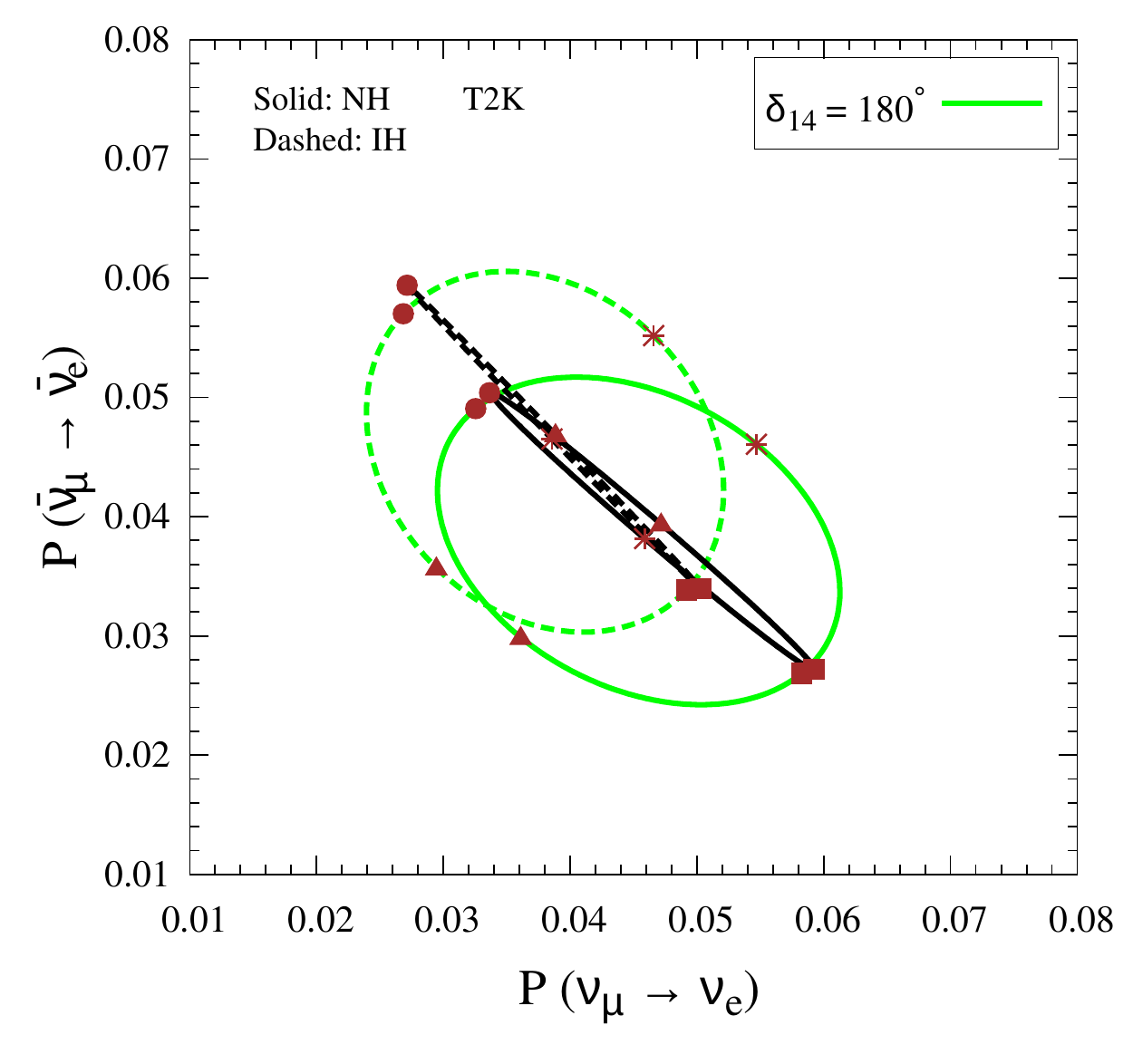}
}
\centerline{
\includegraphics[width=0.49\textwidth]{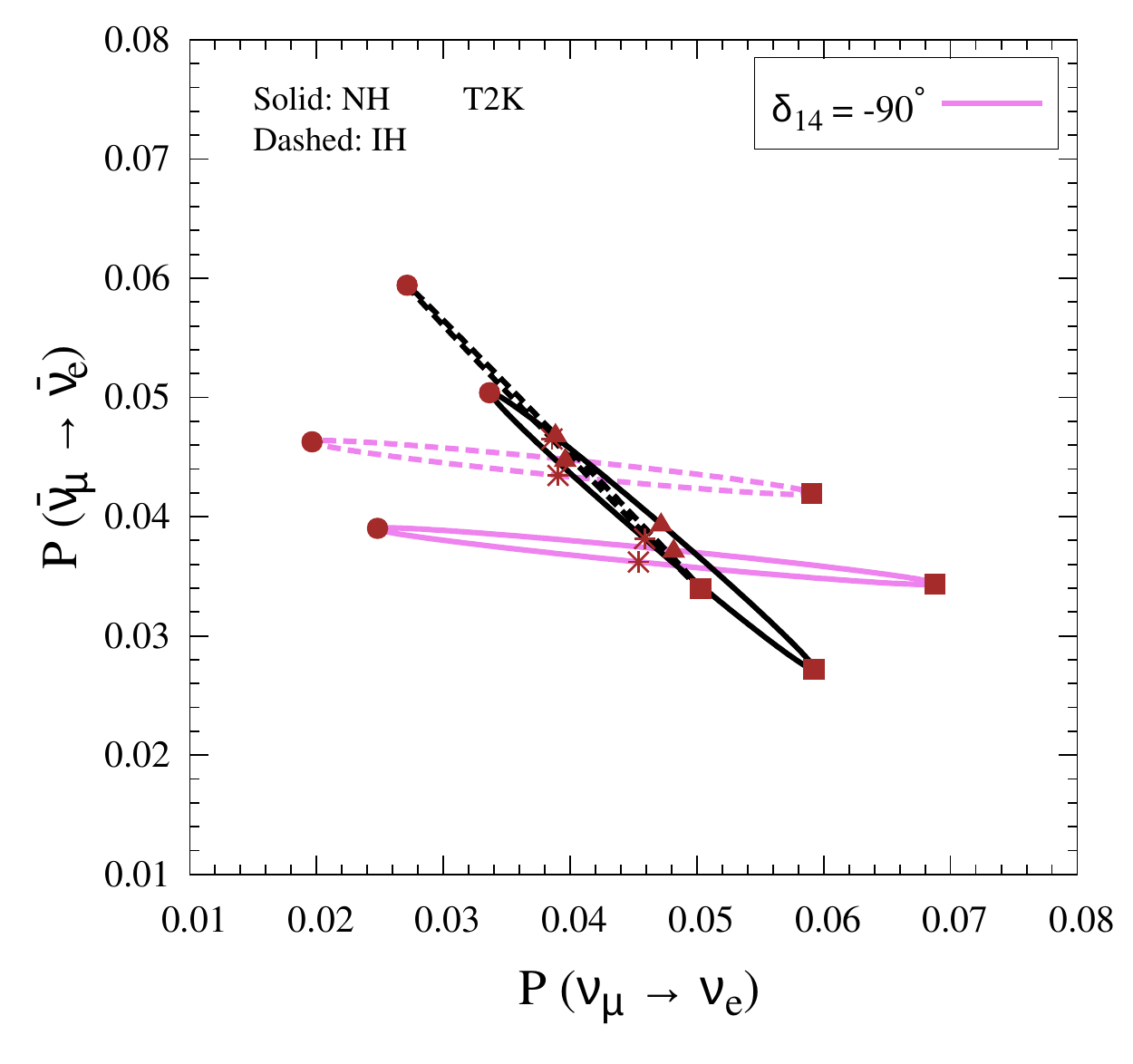}
\includegraphics[width=0.49\textwidth]{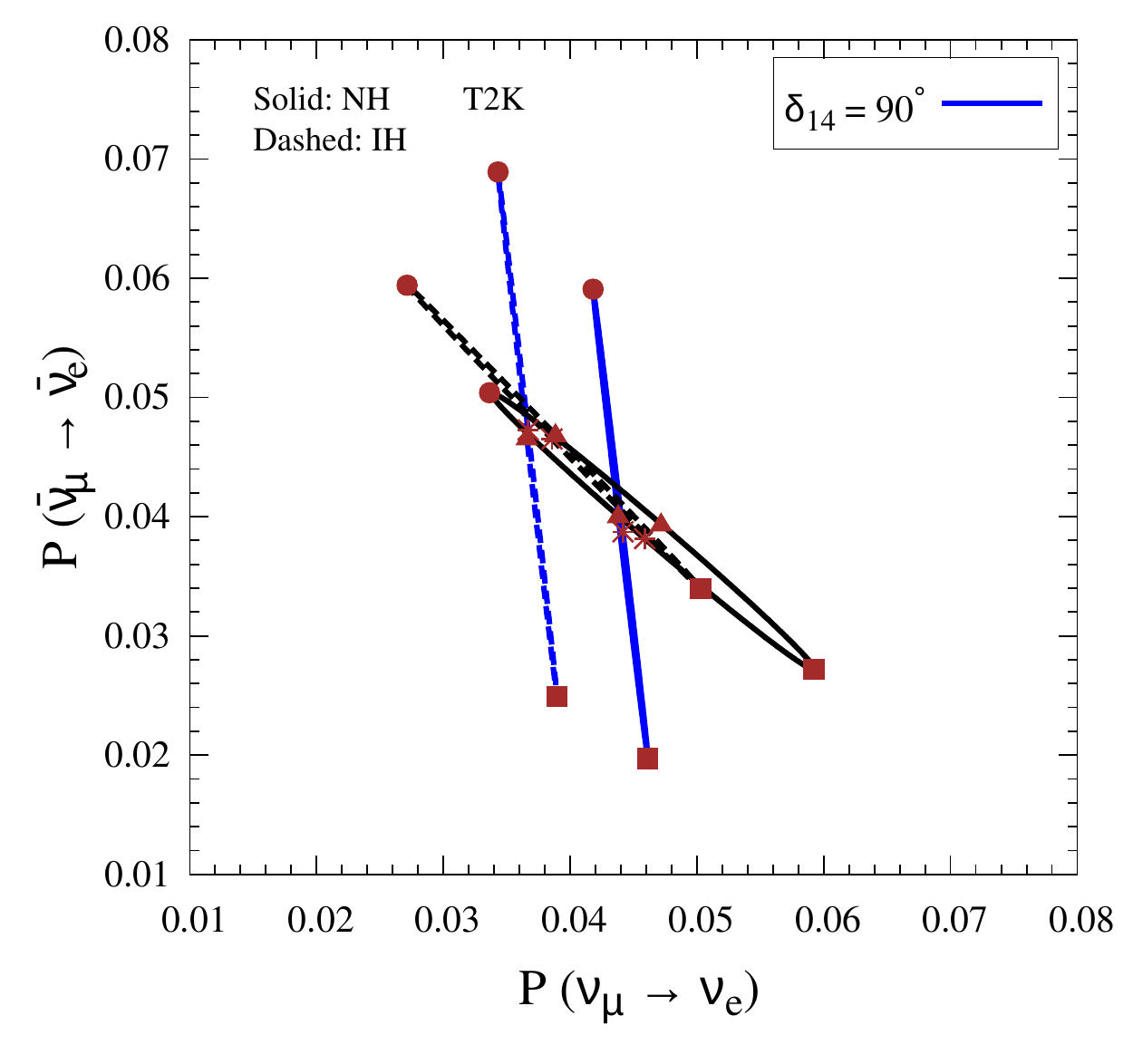}
}
\caption{Bi-probability plots for T2K for four fixed values of 
$\delta_{14}$ and neutrino energy 0.6 GeV. In each panel, we also show the 3-flavor ellipses for the sake
of comparison. In both the 3-flavor and 4-flavor ellipses, the running parameter is
the CP-phase $\delta_{13}$ varying in the range $[-\pi,\pi]$. The solid (dashed) curves refer to NH
(IH).}
\label{fig:T2K-appearance-bi-probability}
\end{figure}

\begin{figure}[t]
\centerline{
\includegraphics[width=0.49\textwidth]{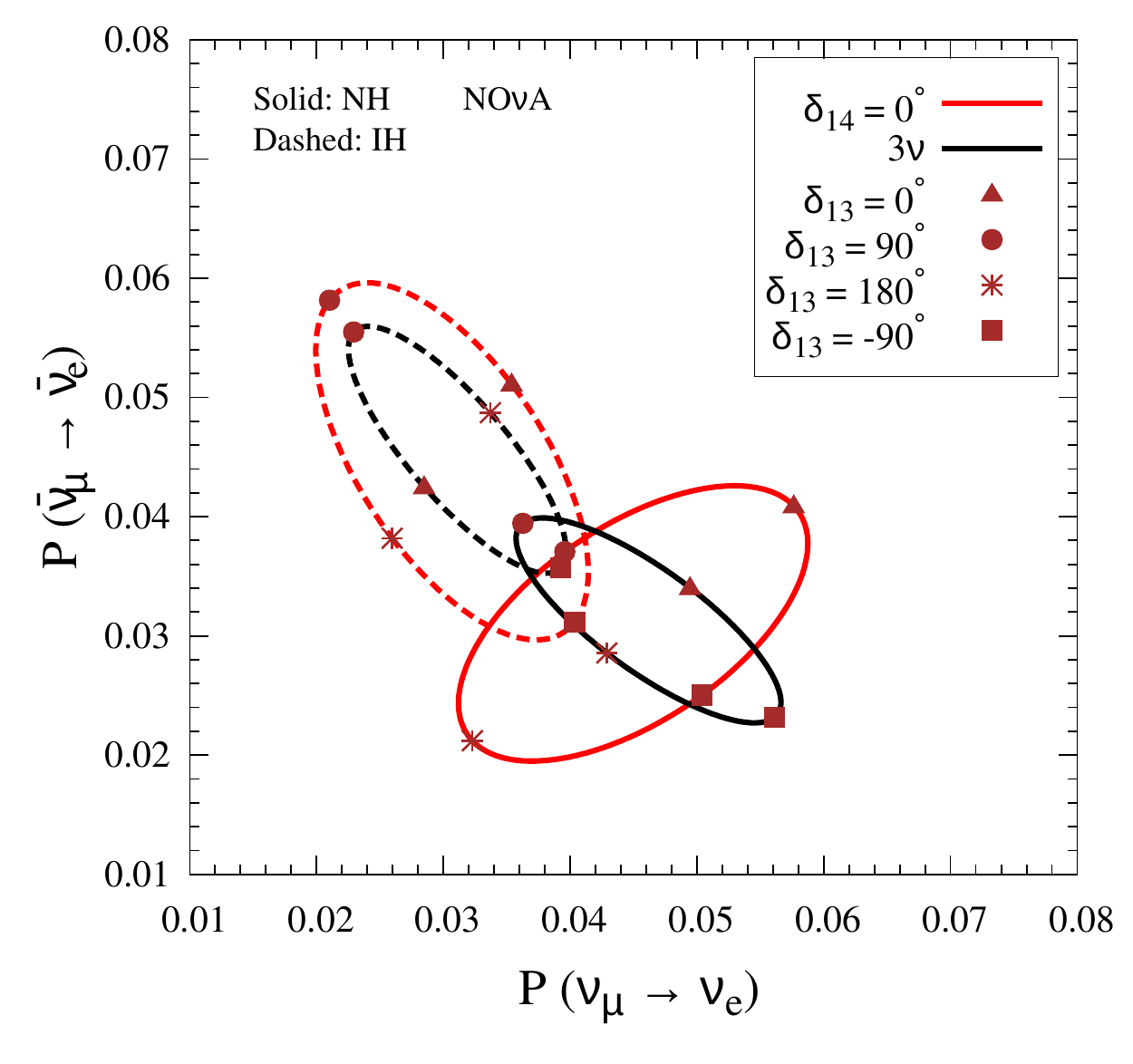}
\includegraphics[width=0.49\textwidth]{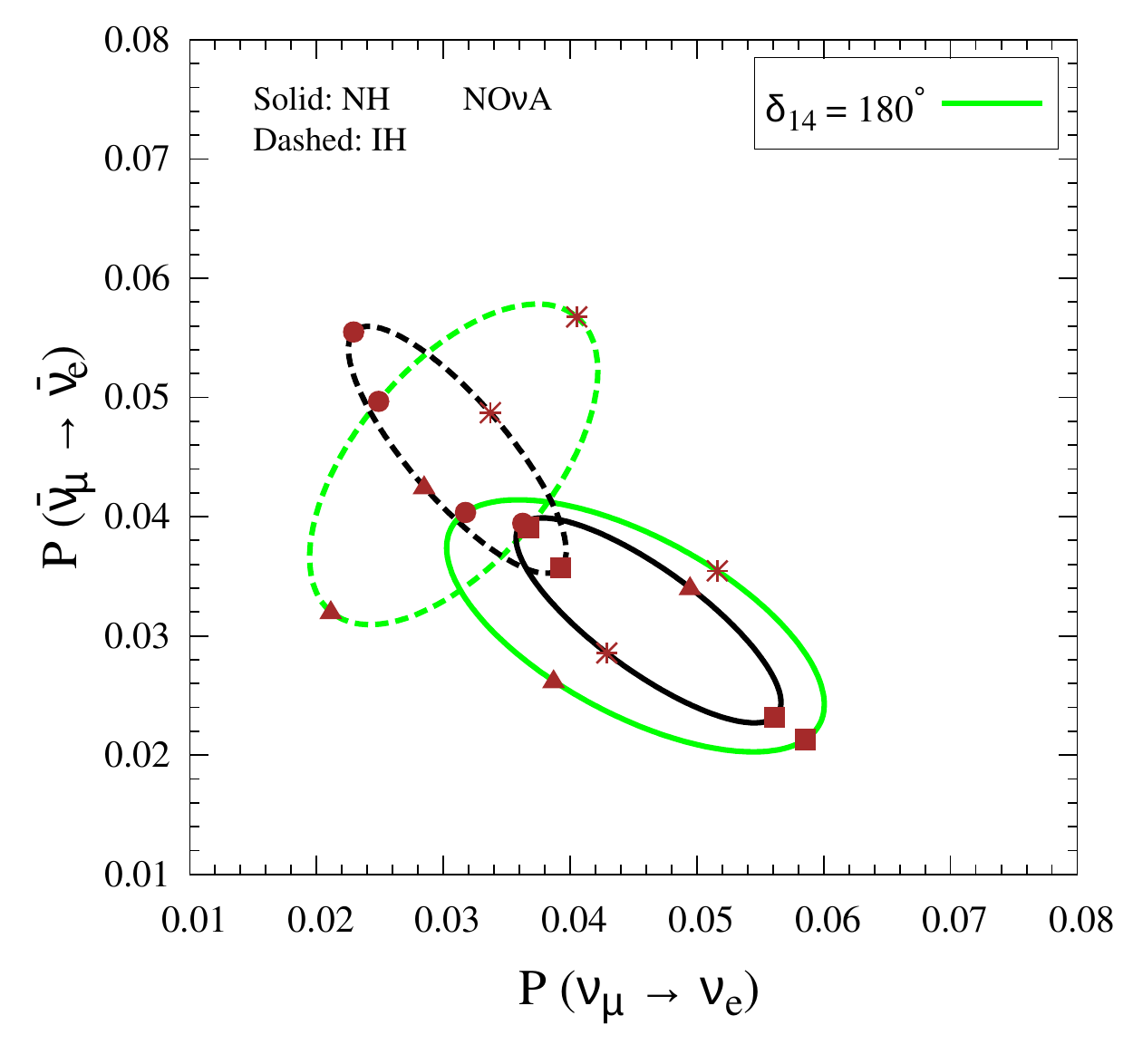}
}
\centerline{
\includegraphics[width=0.49\textwidth]{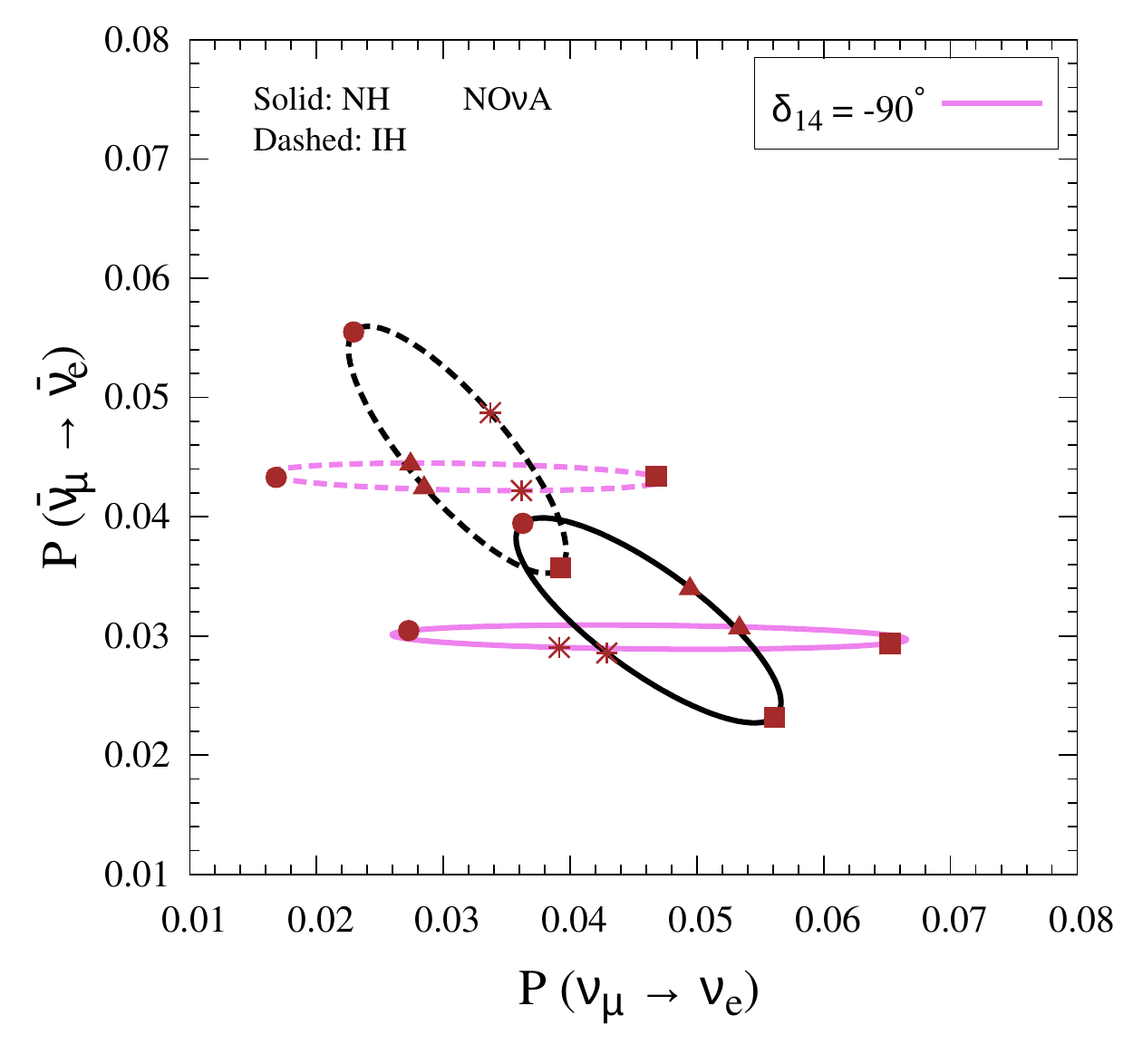}
\includegraphics[width=0.49\textwidth]{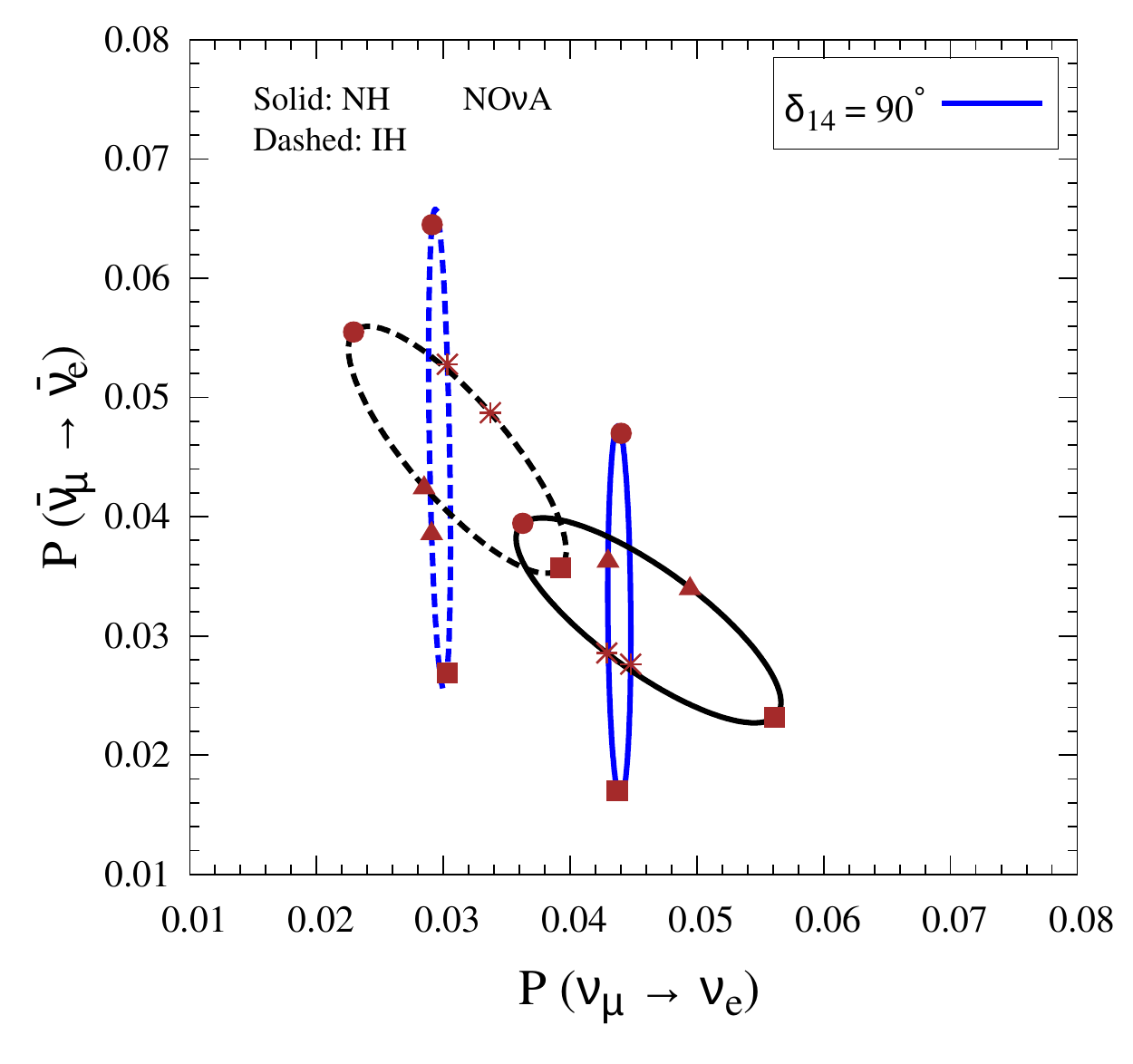}
}
\caption{Bi-probability plots for NO$\nu$A for four fixed values of 
$\delta_{14}$ and neutrino energy of 2 GeV. In each panel, we also show the 3-flavor ellipses 
for the sake of comparison. In both the 3-flavor and 4-flavor ellipses, the running parameter is
the CP-phase $\delta_{13}$ varying in the range $[-\pi,\pi]$. The solid (dashed) curves refer to NH
(IH).}
\label{fig:NOvA-appearance-bi-probability}
\end{figure}

Figures~\ref{fig:T2K-appearance-bi-probability} and \ref{fig:NOvA-appearance-bi-probability}
represent the 4-flavor ellipses obtained respectively for T2K (at the energy $E= 0.6$\,GeV)
and NO$\nu$A (at the energy $E= 2.0$\,GeV) for four fixed values of the phase $\delta_{14}$.
In each panel the solid (dashed) curve represents the NH (IH) case. From the figures
it is evident that the properties of the ellipses depend on: i) the particular experiment 
(due to the different value of the oscillation factor $\Delta$); ii) the value of the phase $\delta_{14}$;
iii) the neutrino mass hierarchy (only in NO$\nu$A).

We can understand the basic behavior of a 4-flavor ellipse using the following relation for its inclination,%
\footnote{For the derivation one has to write the equation of the ellipse by eliminating the
parameter $\delta_{13}$ and than use the general formulae available in textbooks.}
valid under the assumption that the perturbations induced by the matter effects on the 
interference terms are negligible (i.e. $A =\bar A$ and $B = \bar B$),
\begin{eqnarray}
\label{eq:Parametric_4nu_omega_v1}
\tan 2 \omega =  \frac{(B^2 - A^2)\cos 2\Delta - 2 AB \sin 2\Delta \cos \delta_{14}}{2 AB \sin \delta_{14}} \,,
\end{eqnarray}
which, making use of the definitions of $A$ and $B$ in Eqs.~(\ref{eq:Parametric_4nu_coeff_A})-(\ref{eq:Parametric_4nu_coeff_B}), and in particular of the invariance of the (positive) sign of the product AB, 
becomes
\begin{eqnarray}
\label{eq:Parametric_4nu_omega_v2}
\tan 2 \omega = \frac{(S_B^2 -S_A^2\Delta^2) \cos 2\Delta -2S_AS_B|\Delta| \sin 2 \Delta \cos \delta_{14}}{2S_AS_B |\Delta| \sin \delta_{14}}\,. 
\end{eqnarray}
The angle $\omega \in ]-\pi/4, \pi/4[$ represents the inclination (with respect to 
the axis of the abscissas) of the major (minor) axis depending on the negative (positive) sign
of the denominator in Eq.~(\ref{eq:Parametric_4nu_omega_v2}) (i.e. the sign of $\sin \delta_{14}$ 
since the product AB is positive definite). In the limit $\sin \delta_{14} \to 0$ the inclination of 
the major axis is $|\omega| = \pi/4$. In this case, the sign of $\omega$ can be determined by looking
at the sign of the numerator in Eq.~(\ref{eq:Parametric_4nu_omega_v2}). If the numerator is positive,
one has $\omega = -\pi/4$, if it is negative one has $\omega = \pi/4$. 

In T2K we have $\Delta \simeq \pi/2$ and Eq.~(\ref{eq:Parametric_4nu_omega_v2}) takes the simpler form 
\begin{eqnarray}
\label{eq:Parametric_4nu_omega_v3}
\tan 2 \omega  =  \frac{S_A^2 \Delta^2 -S_B^2}{2S_AS_B |\Delta| \sin \delta_{14}} \simeq \frac{0.22}{\sin \delta_{14}}\,.
\end{eqnarray}
This result is independent of the mass hierarchy and therefore the inclination of the ellipses
will be identical in the two cases of NH and IH. This is confirmed by Fig.~\ref{fig:T2K-appearance-bi-probability}.
Indeed, in each panel the solid ellipse has almost the same orientation of the dashed one. The very small difference
in the inclinations is due to the matter effects [O($\epsilon^4$)] that we are neglecting at the level of the interference terms.
The matter effects are instead retained at the level of the leading terms $P_0$ and $\bar P_0$ where
they induce O($\epsilon^3$) corrections. As a result the centers of the ellipses are shifted
in opposite directions (with respect to the vacuum case) for the two cases of NH and IH. 
We also observe that the centers of the 4-flavor ellipses almost coincide with those of the
3-flavor ones since, as discussed in section~\ref{sec:probability} [see Eq.~(\ref{eq:Pme_atm_matt})], 
the matter effects enter in a similar way in the two schemes. The very small differences in the location 
of the centers of the 3-flavor and 4-flavor ellipses is imputable to corrections of order O($\epsilon^4$), which 
are neglected  in our treatment.
For the two values $\delta_{14} = (0, \pi)$, the inclination of the major axis is $\omega = -\pi/4$
since the numerator in Eq.~(\ref{eq:Parametric_4nu_omega_v3}) is positive in both cases. 
This is confirmed by the first (red curves) and second (green curves) panel of Fig.~\ref{fig:T2K-appearance-bi-probability}. 
For $\delta_{14} = \pm \pi/2$, one has $\tan 2 \omega = \pm 0.22$, approximately corresponding to $\omega \simeq \pm 0.11$ (or $\pm 6^0$). In the case $\delta_{14} = -\pi/2$, the sign of the denominator in Eq.~(\ref{eq:Parametric_4nu_omega_v3}) 
 is negative and the inclination of $-6^0$ is that of the major axis. In the case $\delta_{14} = \pi/2$, the sign of the denominator in Eq.~(\ref{eq:Parametric_4nu_omega_v3}) is positive and the inclination of $+6^0$ is that of the minor axis. This behavior is corroborated by the third panel (magenta curves) and fourth panel (blue curves) of 
Fig.~\ref{fig:T2K-appearance-bi-probability}.

In NO$\nu$A we have $\Delta \simeq 0.4 \pi$ and Eq.~(\ref{eq:Parametric_4nu_omega_v2}) takes the form 
\begin{eqnarray}
\label{eq:Parametric_4nu_omega_v4}
\tan 2 \omega = k_1 \frac{1 \mp k_2 \cos \delta_{14}}{\sin \delta_{14}} \,.
\end{eqnarray}
where the two constants $k_1,k_2$ are given by
\begin{eqnarray}
\label{eq:contants_ab}
k_1 &\simeq&  3.85 \times 10^{-3}\,,\\ 
k_2 &\simeq& 1.53 \times  10^{2}\,.
\end{eqnarray}
The minus (plus) sign in Eq.~(\ref{eq:Parametric_4nu_omega_v4}) refers to the case of NH (IH).
So at the NO$\nu$A peak energy, which corresponds to a value of $\Delta$ different
from $\pi/2$, differently from T2K, we expect a dependency of the orientation of the ellipse from the mass hierarchy. For the two values $\delta_{14} = (0, \pi)$ the denominator goes to zero so the absolute inclination of the ellipses is $|\omega| = \pi/4$. The sign of $\omega$ is determined by the sign of the numerator, which in the normal hierarchy case is negative for $\delta_{14} = 0$ and
positive for $\delta_{14} = \pi$. Therefore, in the NH case the inclination of the major axis is
$\pi/4$ for $\delta_{14} = 0$ and $-\pi/4$ for $\delta_{14} = \pi$. In the IH case the situation is reversed, since the sign in the numerator in  
Eq.~(\ref{eq:Parametric_4nu_omega_v4}) is opposite.
This behavior is basically confirmed by the first two panels of Fig.~\ref{fig:NOvA-appearance-bi-probability}.
Coming now to the two cases $\delta_{14} = \pm \pi/2$,  one has $\tan 2\omega = k_1/\sin \delta_{14}$, which is a relation independent of the neutrino mass hierarchy. Due to the small value
of the coefficient $k_1$, the value of $\omega$ is approximately zero. The inclination refers to the 
major (minor) axis for $\delta_{14} = -\pi/2$ ($\delta_{14} = \pi/2$). This behavior is confirmed 
by the numerical results displayed in the third and fourth panel in Fig.~\ref{fig:NOvA-appearance-bi-probability}. 
 
Hence, one can see that for both experiments the relatively simple formulae illustrated above allow us to explain analytically 
all the properties of the ellipses displayed in Figs.~\ref{fig:T2K-appearance-bi-probability} and~\ref{fig:NOvA-appearance-bi-probability}, 
which are obtained by a full numerical calculation. In the case of T2K the formula for the inclination of the ellipse
is accurate at the level of less than one degree. In NO$\nu$A the accuracy, in some cases, is at the level of a few degrees, 
due to the larger impact of the fourth order corrections related to matter effects. For clarity, in table~\ref{table_ellipses}
we report the approximated properties of the ellipses for the 3-flavor and the 4-flavor cases.

\begin{table}[t]
{%
\newcommand{\mc}[3]{\multicolumn{#1}{#2}{#3}}
\newcommand{\mr}[3]{\multirow{#1}{#2}{#3}}
\begin{center}
\begin{tabular}{|c|c|c|c|c|c|}
\hline
& $\delta_{14} (\rm true) $ & MH & Chirality & Inclination (T2K) & Inclination (NO$\nu$A) \\
\hline
 \mr{2}{*}{3$\nu$}&  & NH & + & $-45^0$ & $-45^0$\\
 &  & IH & - & $-45^0$ & $-45^0$ \\
\hline
 \mr{8}{*}{4$\nu$}& \mr{2}{*}{$0^0$} & NH & + & $-45^0$ & $+45^0$ \\
 &  & IH & + & $-45^0$  & $-45^0$ \\
\cline{2-6}
 & \mr{2}{*}{$180^0$} & NH & - & $-45^0$ & $-45^0$ \\
 &  & IH  & - & $-45^0$ &$+45^0$ \\
 \cline{2-6}
  & \mr{2}{*}{$-90^0$} & NH & + & $-6^0$ & $0^0$ \\
 &  &IH  &+ & $-6^0$ & $0^0$\\
 \cline{2-6}
 & \mr{2}{*}{$90^0$} & NH & - & $-84^0$ & $-90^0$ \\
 &  & IH  & - & $-84^0$ & $-90^0$\\
 \hline
\end{tabular}
\end{center}
}%
\caption{Geometrical properties of the ellipses for the 3-flavor and 4-flavor schemes. The first column reports the value 
of the CP-phase $\delta_{14}$ (not defined in the 3-flavor case). The second column reports the neutrino
mass hierarchy. The third column reports the chirality of the ellipse (which is the same for T2K and NO$\nu$A).
The plus (minus) sign means that the trajectory winds in the counter-clockwise (clockwise) sense as the phase $\delta_{13}$ increases.
The fourth and fifth columns report the inclination of the major axis of the ellipse for T2K and NO$\nu$A, respectively.
The values of the inclinations are those found with the third order expansion of the transition probabilities.}
\label{table_ellipses}
\end{table}

\begin{figure}[t]
\centerline{
\includegraphics[width=0.49\textwidth]{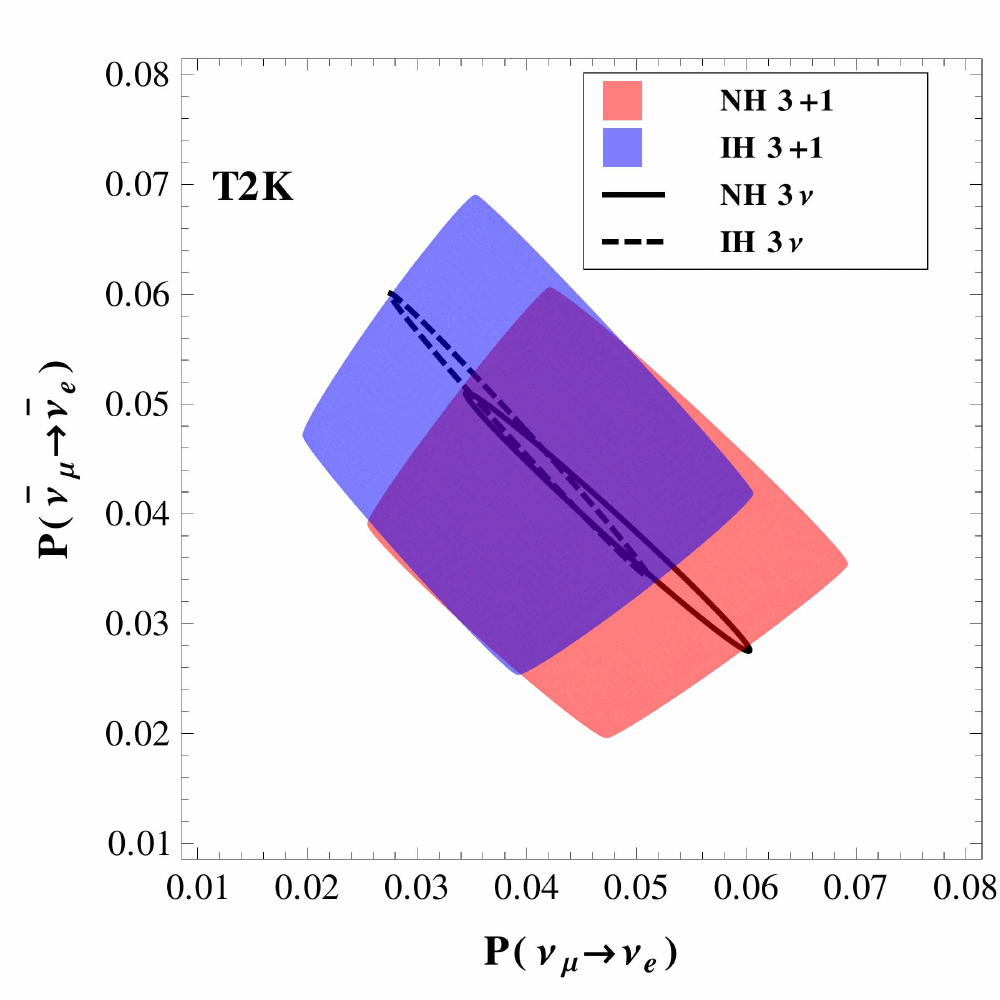}
\includegraphics[width=0.49\textwidth]{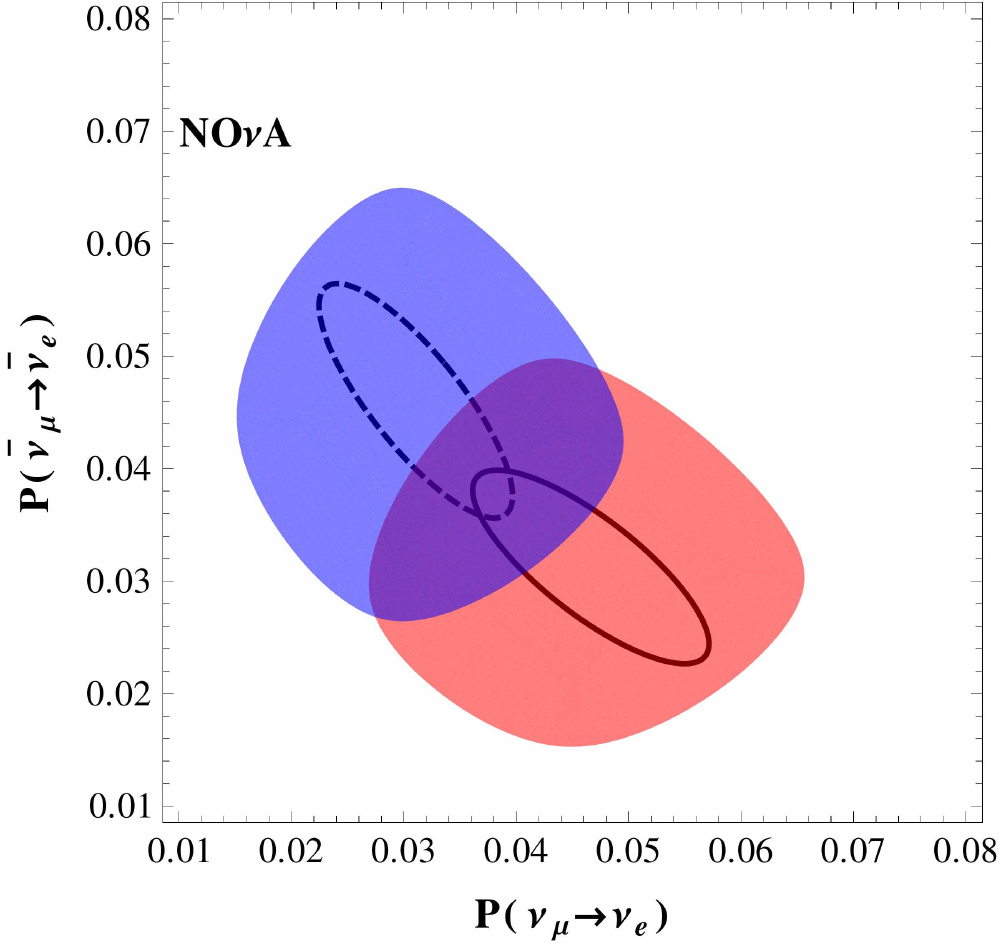}
}
\caption{The colored shaded blobs represent the convolution of the bi-probability graphs 
for T2K (left panel) and NO$\nu$A (right panel) in the 3+1 scheme. The blobs are
obtained by superimposing several ellipses, each corresponding to 
a different value of the CP-phase $\delta_{14}$ taken in its range of variability $[-\pi,\pi]$.
The black curves show the 3-flavor ellipses as 
a benchmark. In both the 3-flavor and 4-flavor cases, the running parameter 
on the ellipses is the CP-phase $\delta_{13}$.}
\label{fig:convoluted-bi-probability}
\end{figure}

The bi-probability plots shown in Figs.~\ref{fig:T2K-appearance-bi-probability} and~\ref{fig:NOvA-appearance-bi-probability}
are obtained for fixed values of the CP-phase $\delta_{14}$. Since the value of $\delta_{14}$ is unknown, 
it is interesting to ask what happens if one superimposes all the (theoretically infinite) ellipses corresponding to 
all the possible choices of $\delta_{14}$. The result of this exercise is shown in figure~\ref{fig:convoluted-bi-probability},
which has been produced by drawing the convolution of all the ellipses%
\footnote{A similar plot has been shown in~\cite{Friedland:2012tq} for the experiment NO$\nu$A to visualize 
the impact of new CP-phases potentially related to non-standard neutrino interactions.}
obtained with a dense grid for the parameter $\delta_{14}$ 
in its range of variability $[-\pi, \pi]$. Alternatively, Fig.~\ref{fig:convoluted-bi-probability} may be seen as a dense 
scatter plot obtained by varying simultaneously both CP-phases $\delta_{13}$ and $\delta_{14}$.
This plot provides a bird-eye view of the degree of separation of the two 
neutrino mass hierarchies in the 3+1 scheme. We see 
that a separation persists also in such an enlarged scheme. This means that there 
will exist some combinations of the two CP-phases $\delta_{13}$ and $\delta_{14}$ (corresponding
to those points which do not lie in the superposition area of the blue and orange regions) for which it will 
be possible to distinguish between the two hierarchies at some non-zero confidence level. 
The numerical analysis of the section~\ref{results} will allow us to determine such specific combinations
of the two CP-phases and the exact confidence level of the separation of the two hierarchies.

\section{Experimental features and discussion at the events level}
\label{experimental-details}

\subsection{The off-axis experiments: T2K and NO$\nu$A}
\label{T2K-NOvA}

In this section, we briefly mention the key experimental features 
of the currently running T2K~\cite{Itow:2001ee,Abe:2011ks} and 
NO$\nu$A~\cite{Ayres:2002ws,Ayres:2004js,Ayres:2007tu,Patterson:2012zs} 
experiments that go into carrying out the simulation. The T2K experiment in Japan
is collecting the data since 2010. Neutrinos are being produced at the 
J-PARC accelerator facility in Tokai, and are being observed in the
22.5 kton (fiducial) Super-Kamiokande water \v{C}erenkov detector 
at Kamioka, at a distance of 295 km from the source at an off-axis 
angle of $2.5^\circ$~\cite{Itow:2001ee}. Due to the off-axis nature 
of the beam~\cite{Para:2001cu}, it peaks sharply at the first oscillation 
maximum of 0.6 GeV. Another major benefit of using the off-axis 
technique is that it helps to reduce the intrinsic $\nue$ contamination 
in the beam and also the background coming from neutral current 
events, improving the signal-to-background ratio by great extent.
As a result, the T2K experiment has already been able to provide
an important breakthrough to establish the three-flavor paradigm by 
observing the $\theta_{13}$-driven appearance signal in 
$\nu_{\mu}\rightarrow\nu_{e}$ oscillation channel~\cite{Abe:2013hdq}. 
In May 2014, T2K started its operation in the antineutrino mode, and 
after collecting 10\% of their expected antineutrino data set, recently
they have announced the first appearance results in the antineutrino 
channel~\cite{T2K_antineutrino_EPS_HEP_2015,Salzgeber:2015gua}, 
clearly taking a first step towards probing the CP symmetry in a direct 
fashion. As mentioned earlier, in this paper,
we consider the full projected exposure of $7.8 \times 10^{21}$ 
protons on target (p.o.t.) which the T2K experiment plans to 
achieve during their entire run with a proton beam power of 
750 kW and with a proton energy of 30 GeV. We also assume 
that the T2K experiment would use half of its full exposure in 
the neutrino mode which is $3.9 \times 10^{21}$ p.o.t. and the
remaining half would be used during antineutrino run. We follow
the recent publication by the T2K collaboration~\cite{Abe:2014tzr} 
in great detail to simulate the signal and background event spectra 
and their total rates to obtain our final results. Following the same
reference~\cite{Abe:2014tzr}, we assume an uncorrelated 
5\% normalization error on signal and 10\% normalization error on 
background for both the appearance and disappearance channels
to analyze the prospective data from the T2K experiment. We use
the same set of systematics for both the neutrino and antineutrino 
channels which are also uncorrelated.

The US-based long-baseline experiment NO$\nu$A is currently taking
data. It uses a 14 kton liquid scintillator far detector at Ash River, Minnesota 
to detect the oscillated NuMI\footnote{Neutrinos at the Main Injector.} 
muon neutrino beam produced at Fermilab~\cite{Ayres:2007tu,Patterson:2012zs,Childress:2013npa}.
The NO$\nu$A far detector is placed 810 km away from the source at 
an off-axis angle of 14 mrad ($0.8^\circ$) with respect to the beam line, 
and sees a narrow-band beam which peaks around 2 GeV.
Based on the exposure of $2.74 \times 10^{20}$ p.o.t., recently, 
the NO$\nu$A experiment has released their first $\nue$ 
appearance data providing a solid evidence of $\numu \to \nue$
oscillation over a baseline of 810 km which is the longest baseline 
in operation now~\cite{NOvA_appearance_seminar_FNAL_2015,Bian:2015opa,Adamson:2016tbq,Adamson:2016xxw}.
In this work, we take the full projected exposure of $3.6 \times 10^{21}$ p.o.t. 
which the NO$\nu$A experiment aims to use during their full running time
with a NuMI beam power of 700 kW and 120 GeV proton energy~\cite{Ayres:2007tu}.
In our simulation, we assume that NO$\nu$A would also use 50\%
of its full exposure in the neutrino mode which is $1.8 \times 10^{21}$ p.o.t. 
and the remaining 50\% would be utilized to collect the data in the 
anti-neutrino mode. Following references~\cite{Agarwalla:2012bv,Patterson:2012zs,Agarwalla:2013ju},
we estimate the signal and background event spectra and their total
rates in our calculations. We use a simplified systematic treatment 
for NO$\nu$A: an uncorrelated 5\% normalization uncertainty on signal 
and 10\% normalization uncertainty on background for both the appearance 
and disappearance channels. This is true for both the neutrino and 
antineutrino modes which are also assumed to be uncorrelated.

\begin{figure}[t]
\centerline{
\includegraphics[width=0.49\textwidth, height=0.49\textwidth]{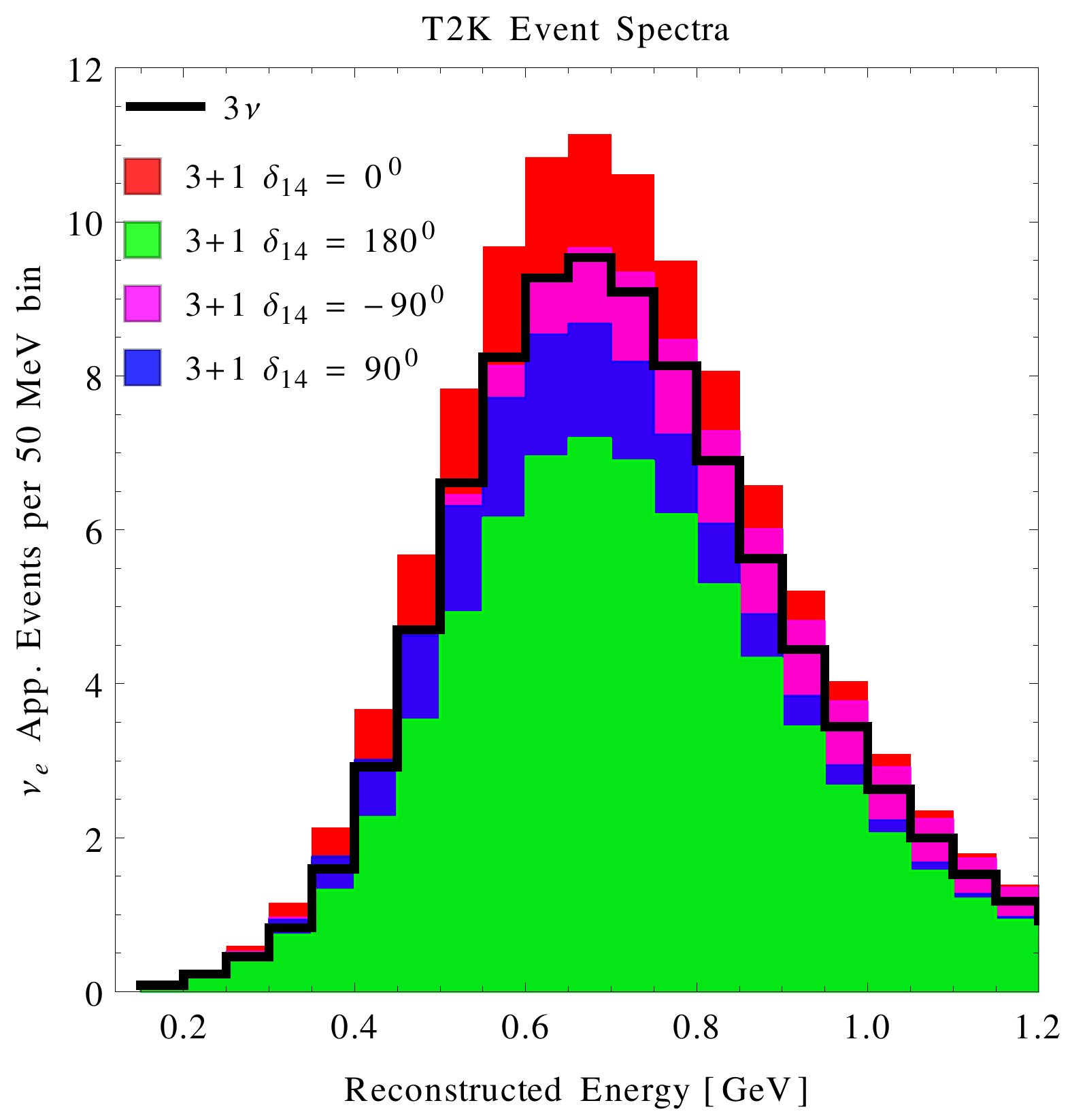}
\includegraphics[width=0.49\textwidth, height=0.49\textwidth]{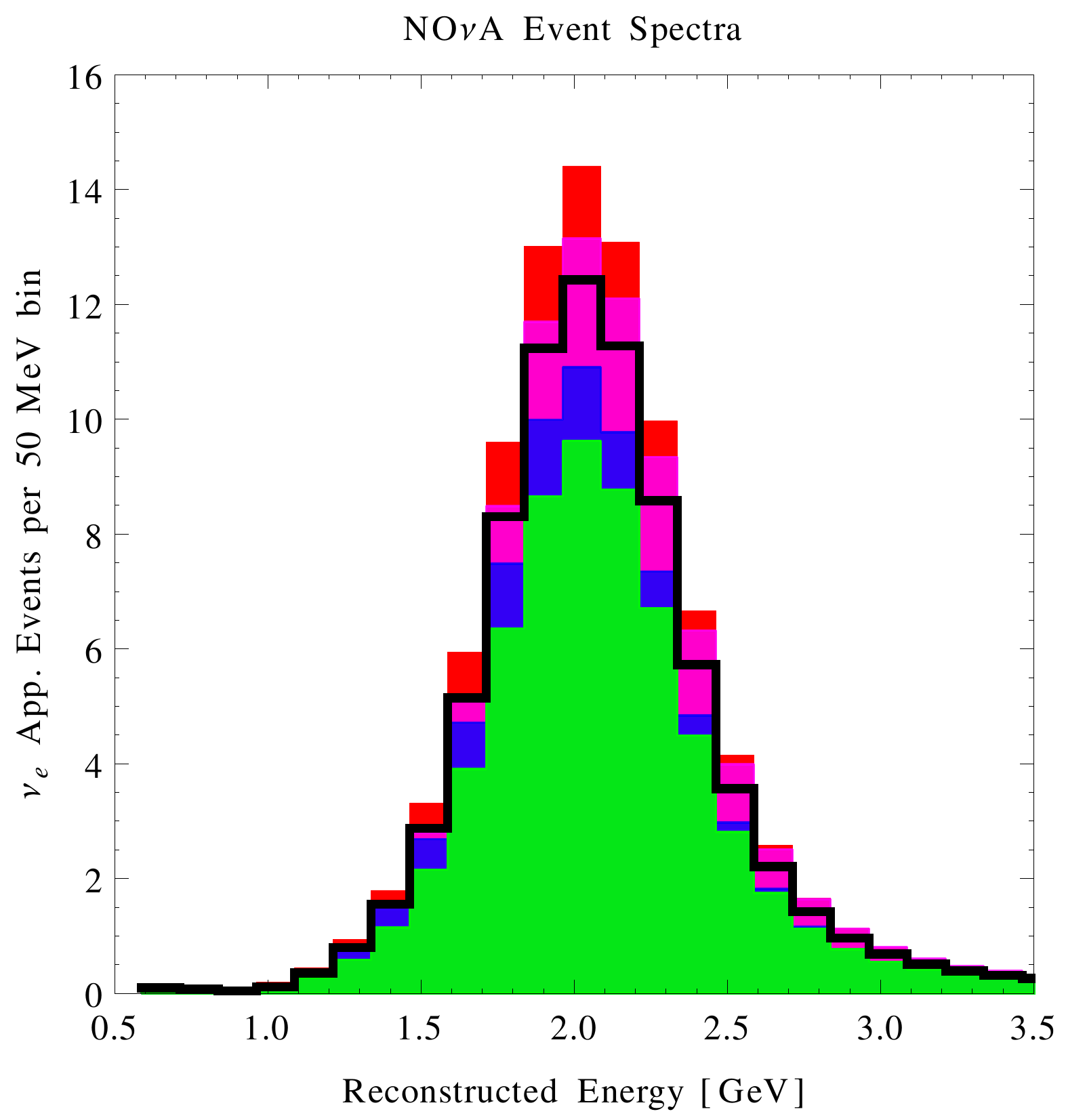}
}
\caption{Expected signal event spectra in the $\nue$ appearance channel 
as a function of the reconstructed neutrino energy. The left (right) panel refers 
for T2K (NO$\nu$A). The black line corresponds to the 3-flavor case. 
The colored histograms refer to the 3+1 scheme for 
the four different values of $\delta_{14}$ indicated in the legend.
In all cases (both 3-flavor and 4-flavor) we have set $\delta_{13} = 0$.} 
\label{fig:event-spectrum}
\end{figure}

\subsection{Event spectra}
\label{event-spectra}

We devote this section to discuss the expected event spectra in 3$\nu$ and 3+1
schemes for both the T2K and NO$\nu$A setups using their full projected 
exposures as mentioned in the previous section. The number of expected 
appearance electron events\footnote{We can calculate the number of positron 
events using Eq.~(\ref{eq:events}), by taking into account appropriate oscillation 
probability and cross-section. The same strategy can be applied to estimate 
$\mu^{\pm}$ events.} in the $i$-th energy bin in the detector is estimated 
using the following well known expression
\begin{equation}
N_{i} = \frac{T\, n_n\, \epsilon}{4\pi L^2}~ \int_0^{E_{\rm max}}
dE \int_{E_{A_i}^{\rm min}}^{E_{A_i}^{\rm max}} dE_A \,\phi(E)
\,\sigma_\nue(E) \,R(E,E_A)\, P_{\mu e}(E) \, ,
\label{eq:events}
\end{equation}
where $\phi(E)$ is the neutrino flux, $T$ is the total running time,
$n_n$ is the number of target
nucleons in the detector, $\epsilon$ is the detector efficiency,
$\sigma_\nue$ is the neutrino interaction cross-section, and 
$R(E,E_A)$ is the Gau\ss ian energy resolution function of the 
detector. The quantities $E$ and $E_A$ are the true and reconstructed 
(anti-)neutrino energies respectively, and $L$ is the baseline.
In Fig.~\ref{fig:event-spectrum}, we show the expected signal event
spectra for the $\nue$ appearance channel as a function of 
reconstructed neutrino energy for both the experiments under
consideration. As expected due to their off-axis nature, we see 
a narrow peak in the projected event spectrum around 0.6 GeV 
for the T2K experiment (see the left panel), and for the 
NO$\nu$A experiment (see the right panel), the events mainly 
occur around 2 GeV where the flux is maximum. In both panels, 
the thick black lines correspond to the 3$\nu$ case assuming 
$\delta_{13} = 0^{\circ}$. The other colored histograms 
(red, green, magenta, and blue) are drawn in the 3+1 scheme
assuming different values of $\delta_{14}$ which are mentioned
in the figure legends. Next, we discuss the bi-events plots 
to get more physics insight.

\begin{figure}[t]
\centerline{
\includegraphics[width=0.49\textwidth]{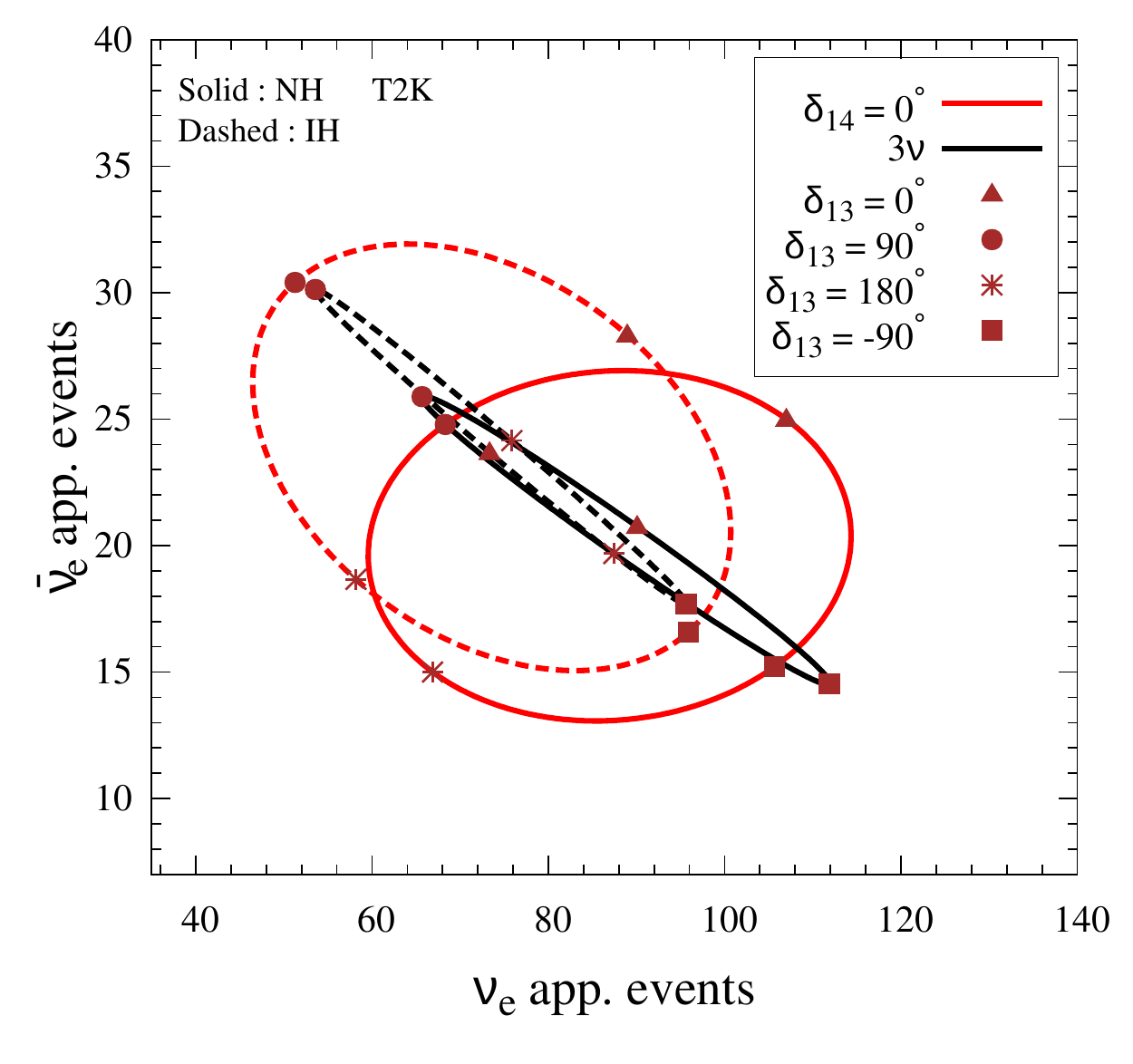}
\includegraphics[width=0.49\textwidth]{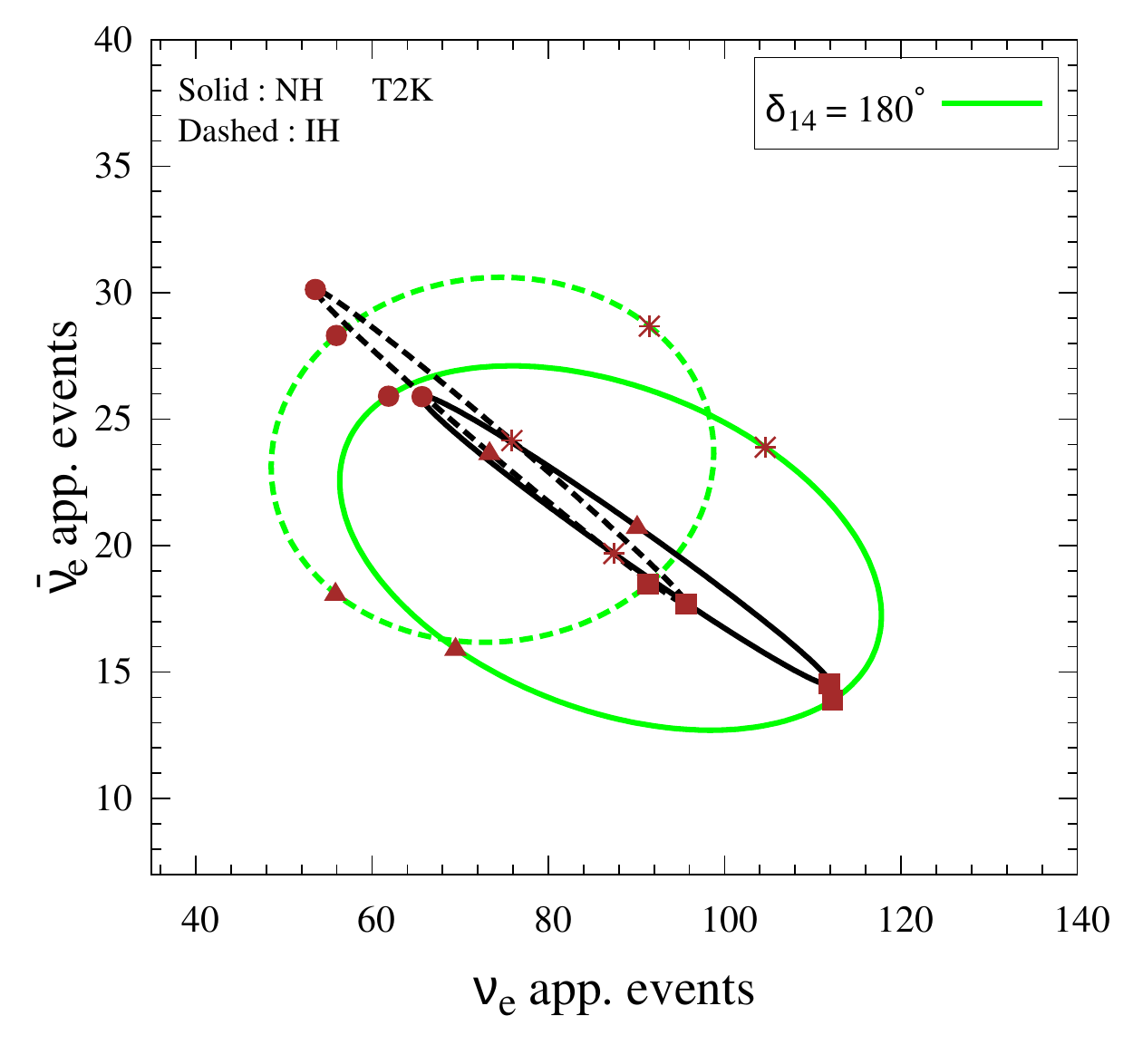}
}
\centerline{
\includegraphics[width=0.49\textwidth]{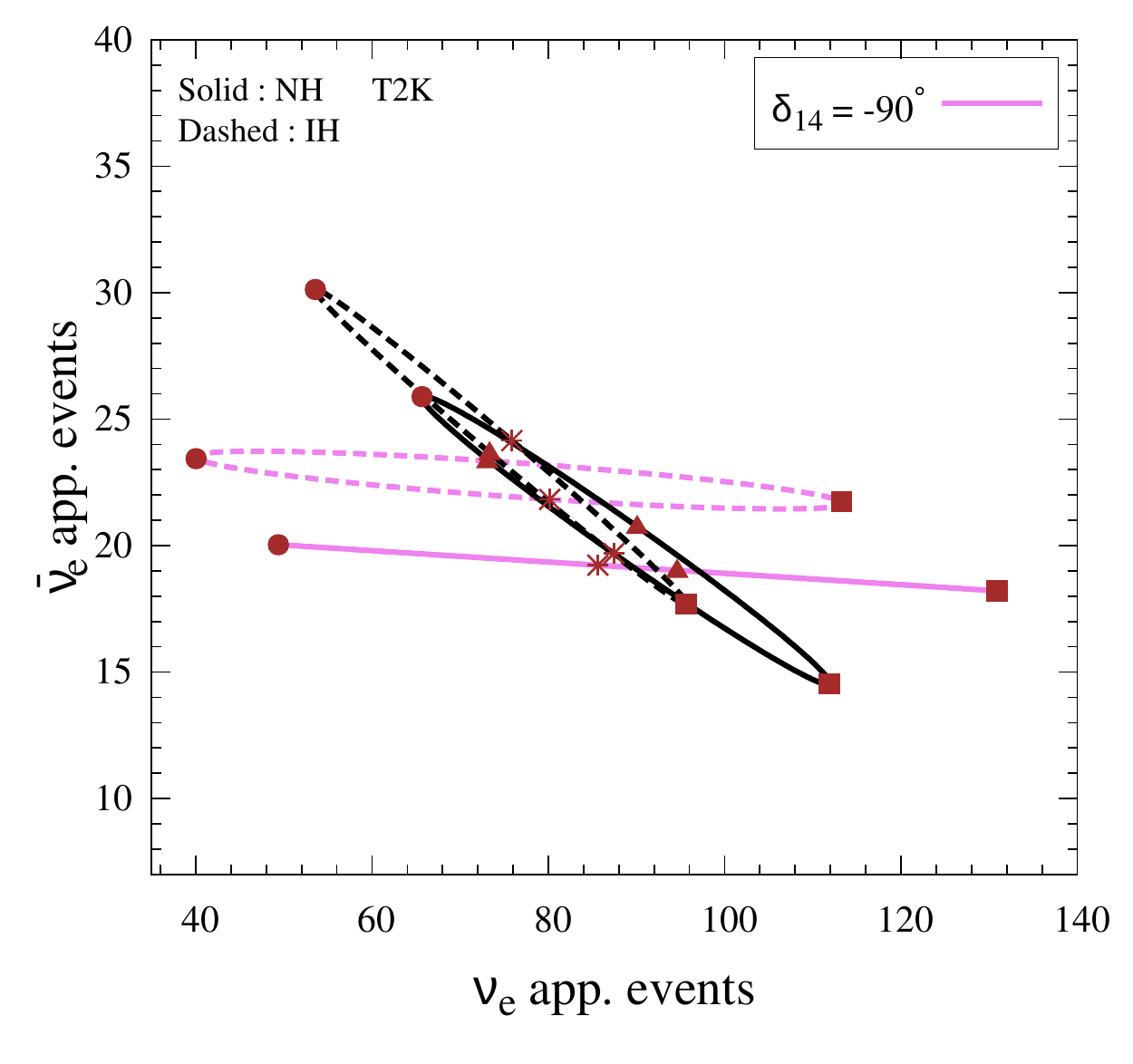}
\includegraphics[width=0.49\textwidth]{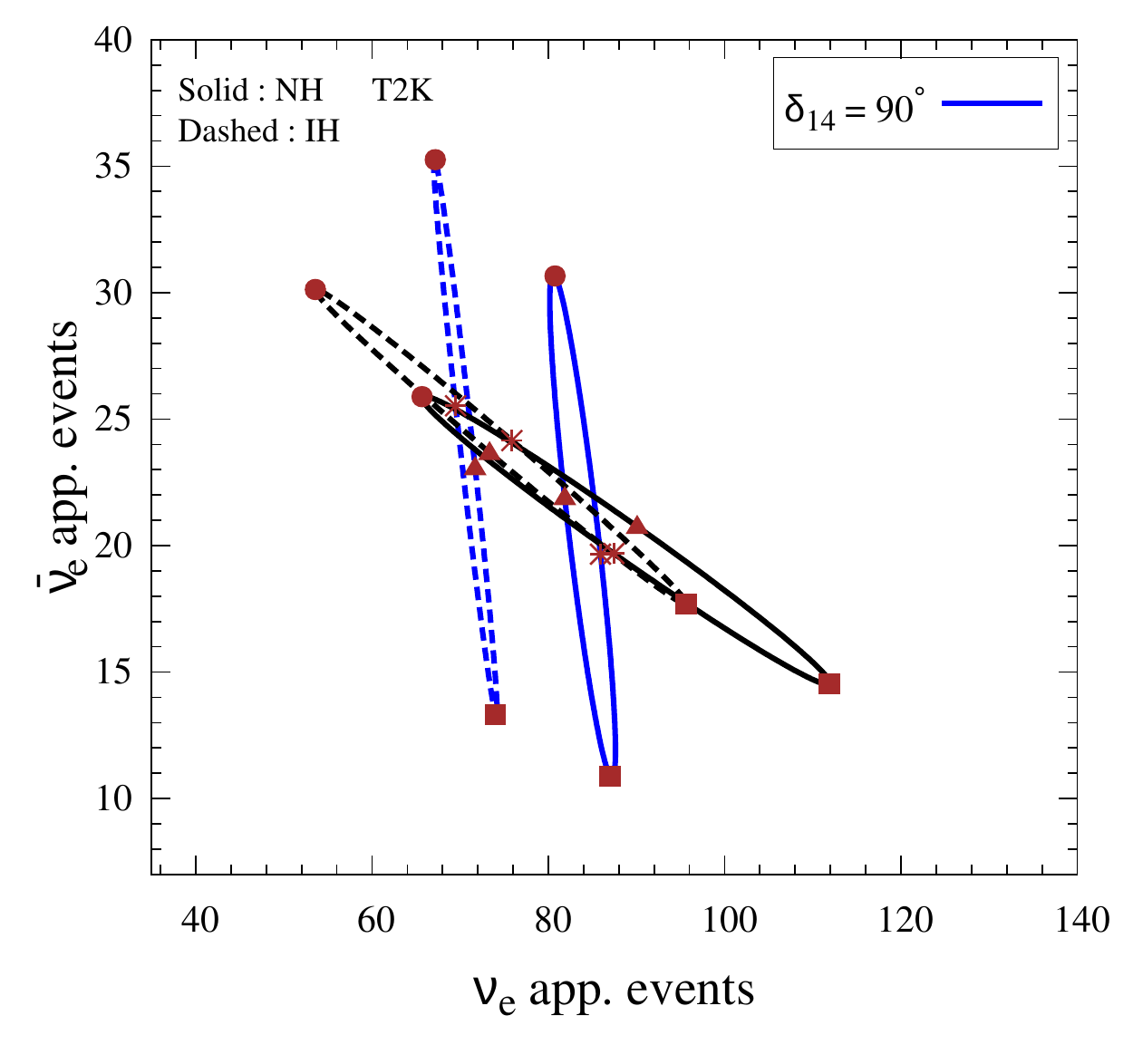}
}
\caption{Bi-events plots for T2K for four fixed values of the CP-phase $\delta_{14}$.
In each panel, we also show the 3-flavor ellipses for the sake
of comparison.
In both the 3-flavor and 4-flavor ellipses, the running parameter is
the CP-phase $\delta_{13}$ varying in the range $[-\pi,\pi]$. The solid (dashed) curves refer to NH (IH). 
We have assumed that half of the full T2K exposure will be used in the neutrino mode and
the other half in the antineutrino mode.}
\label{fig:T2K-appearance-bi-events}
\end{figure}

\subsection{Bi-events plots}
\label{bi-events}

The bi-probability plots presented in section~\ref{bi-probability_4nu} give a very clear idea of the behavior
of the transition probability at the specific value of the energy corresponding
to the peak of the spectrum and allow us to approximately predict the 
behavior of a given off-axis experiment, since the dominant contribution to the
total rate comes from the energies close to the peak. In this section we present,
for completeness, also the bi-events plots, where on the two axes it is represented
the theoretical value of number of events ($\nu_e$ on the x-axis, $\bar \nu_e$ on the y-axis)
expected in a given experiment. Such plots provide a more precise information on the behavior of a given experiment,
because the event rates take into account the complete energy spectrum 
and provide the information on the statistics involved in the experiment.  

\begin{figure}[t]
\centerline{
\includegraphics[width=0.49\textwidth]{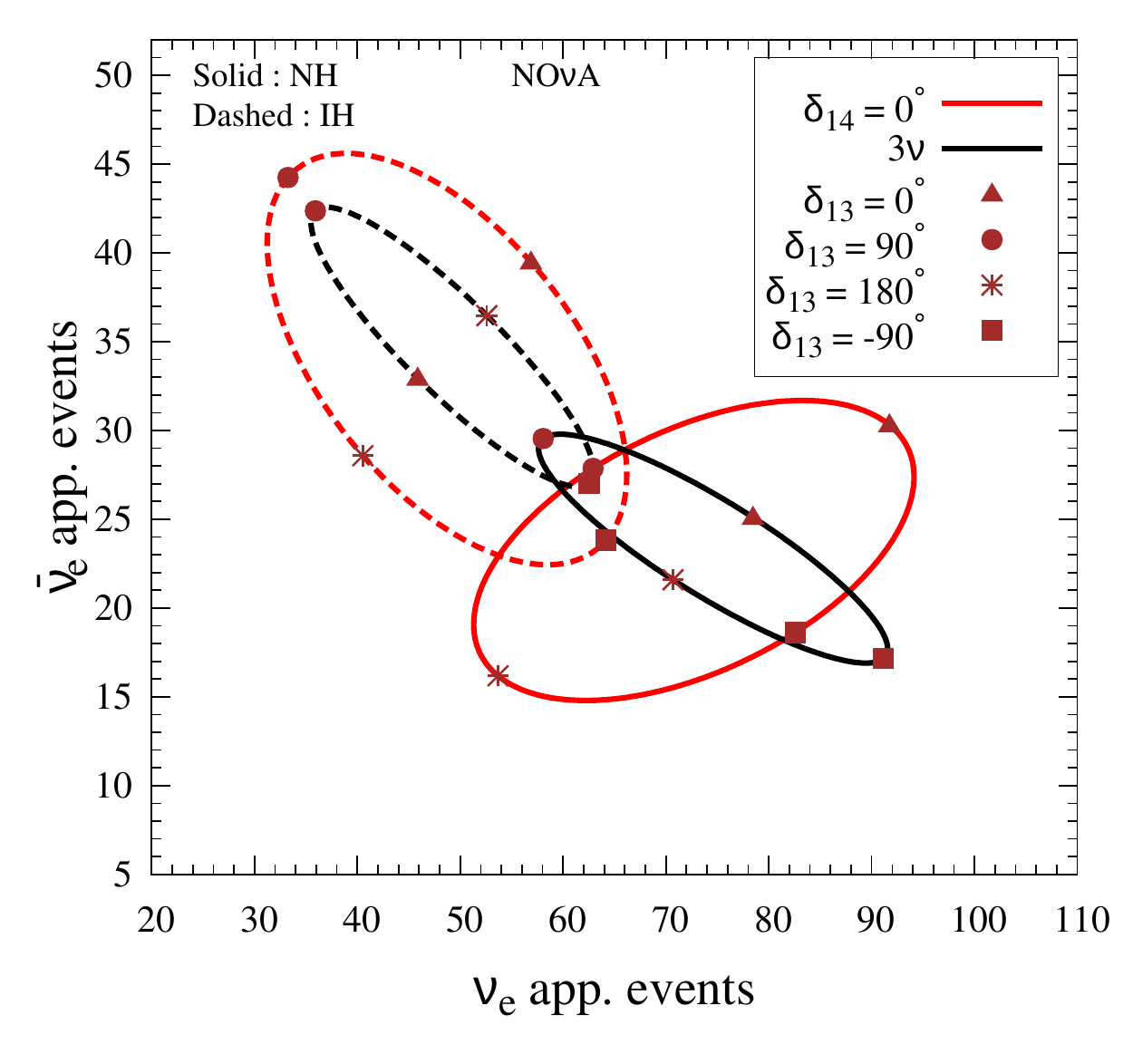}
\includegraphics[width=0.49\textwidth]{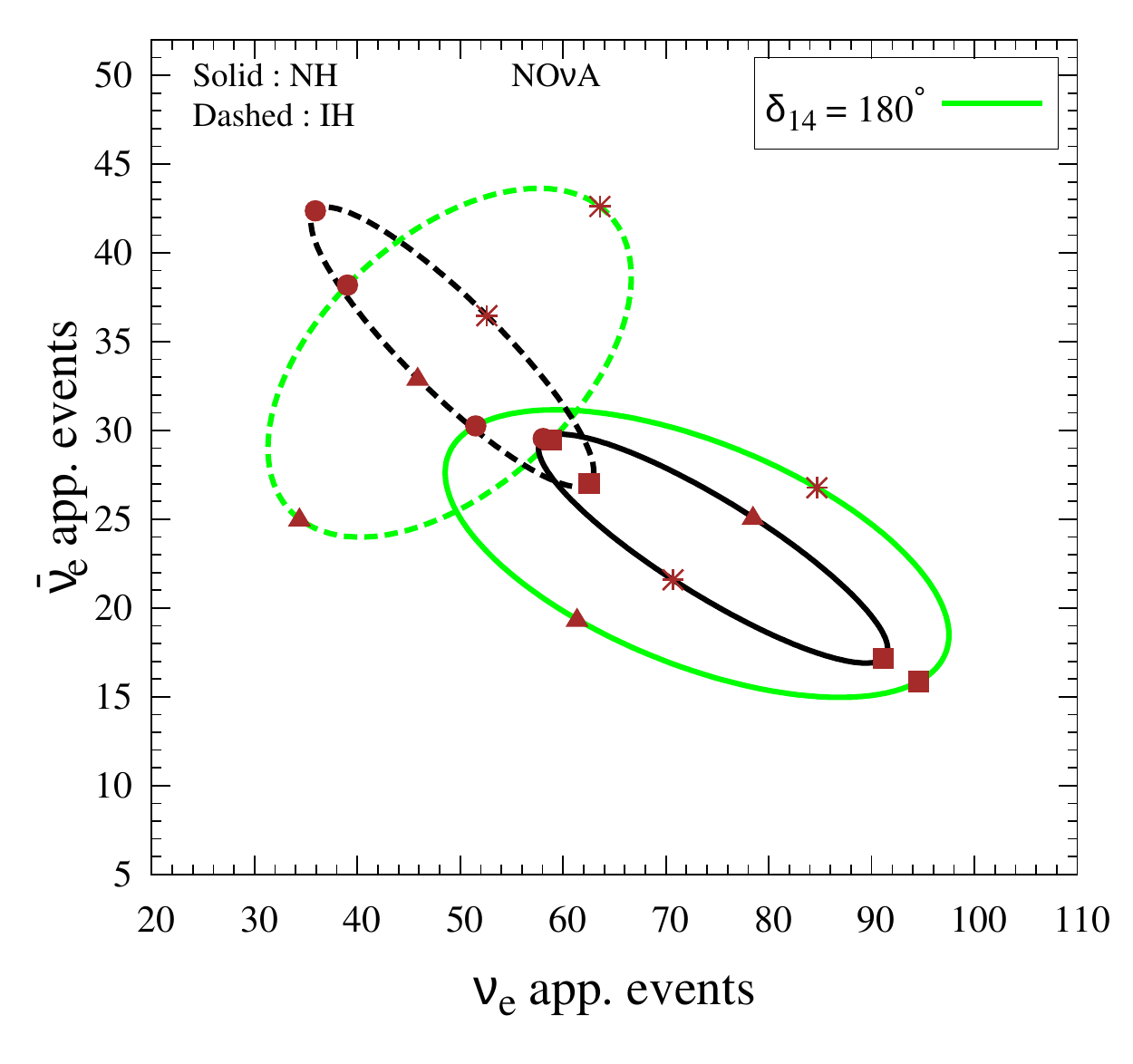}
}
\centerline{
\includegraphics[width=0.49\textwidth]{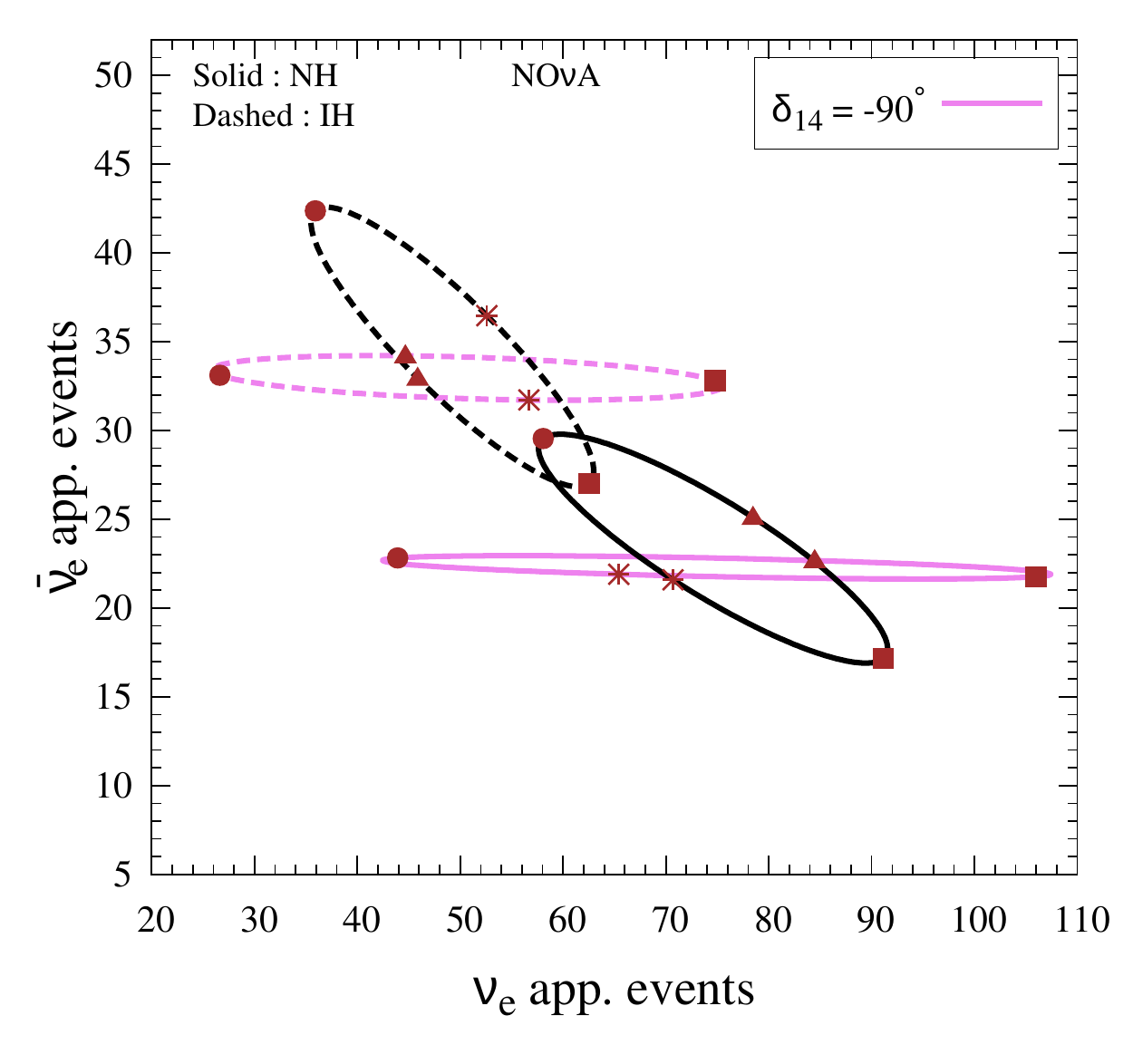}
\includegraphics[width=0.49\textwidth]{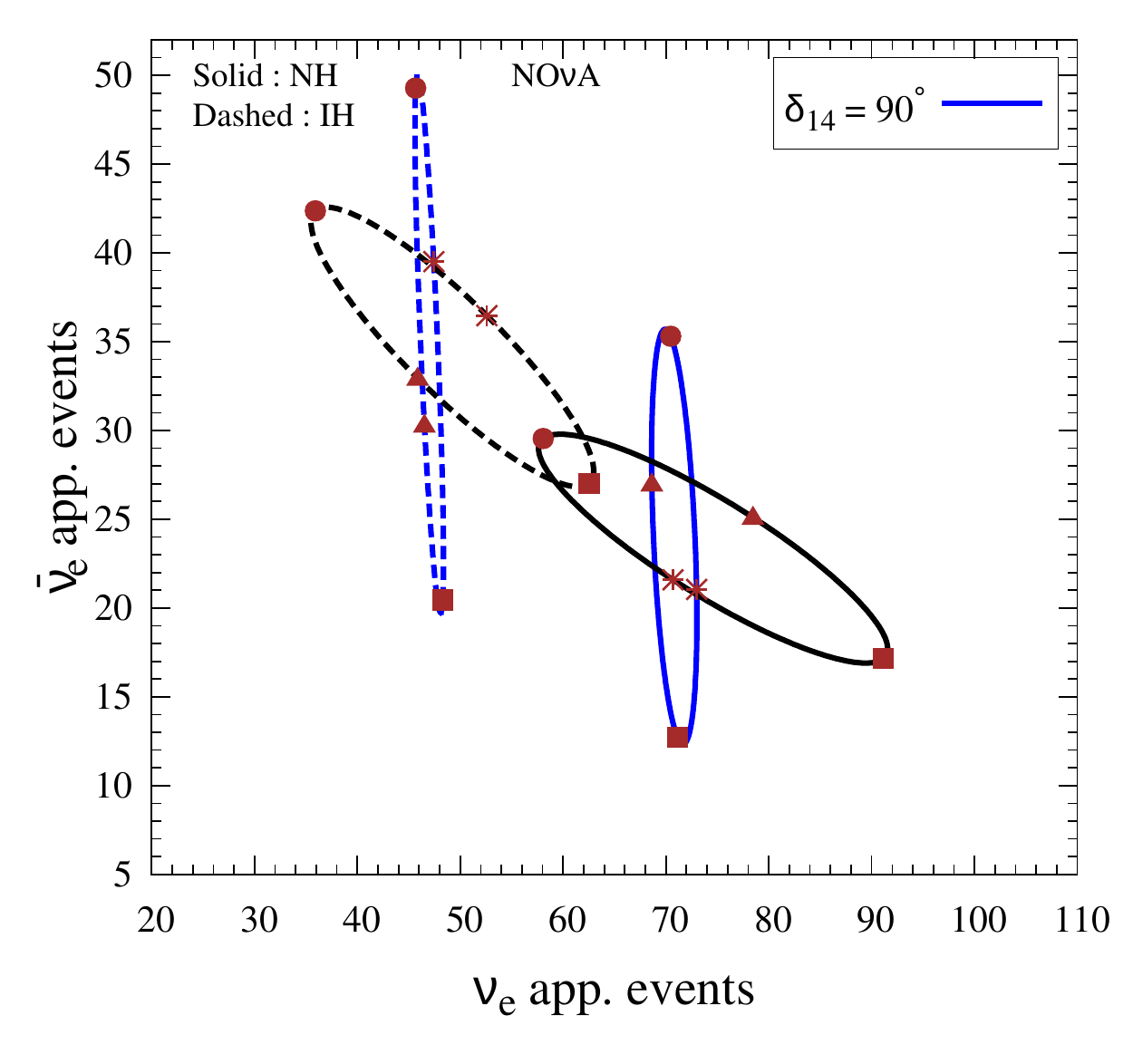}
}
\caption{Bi-events plots for NO$\nu$A for four fixed values of $\delta_{14}$.
In each panel, we also show the 3-flavor ellipses for the sake of comparison.
In both the 3-flavor and 4-flavor ellipses, the running parameter is
the CP-phase $\delta_{13}$ varying in the range $[-\pi,\pi]$. The solid (dashed) curves refer to NH (IH). 
We have assumed that half of the full NO$\nu$A exposure will be used in the neutrino mode
and the other half in the antineutrino mode.}
\label{fig:NOvA-appearance-bi-events}
\end{figure}
 
Figure~\ref{fig:T2K-appearance-bi-events} shows the bi-events plots for T2K,
where we have used the same contour style convention of the bi-probability plots.
We still have elliptical curves, since we are just replacing the coefficients 
in the parametric equations~(\ref{eq:Parametric_4nu_coeff_C})-(\ref{eq:Parametric_4nu_coeff_Db})
with appropriate weighted averages.
A quick comparison of the T2K bi-events plot in Fig.~\ref{fig:T2K-appearance-bi-events} with the corresponding bi-probability 
one (Fig.~\ref{fig:T2K-appearance-bi-probability}) shows that the geometrical
properties of the ellipses are slightly different from those obtained 
for the probabilities. Apart from an obvious deformation factor due 
to the different scale used for the events, we can appreciate other
differences, which are introduced by the contribution of several 
energies in the integration. In particular, appreciable differences
are now visible between the two cases of NH and IH. 
Most importantly, figure~\ref{fig:T2K-appearance-bi-events} gives a clear feeling
on the number of events expected in the experiment T2K.

\begin{figure}[t]
\centerline{
\includegraphics[width=0.49\textwidth]{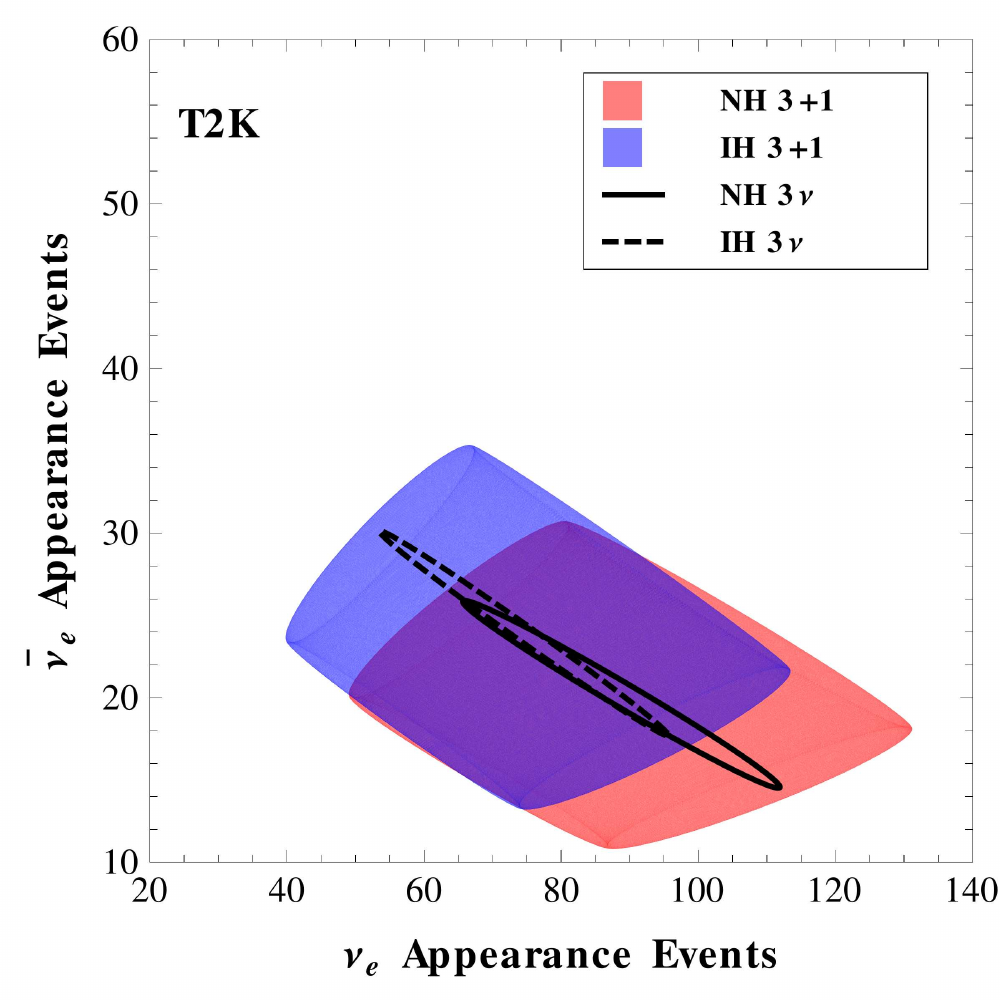}
\includegraphics[width=0.49\textwidth]{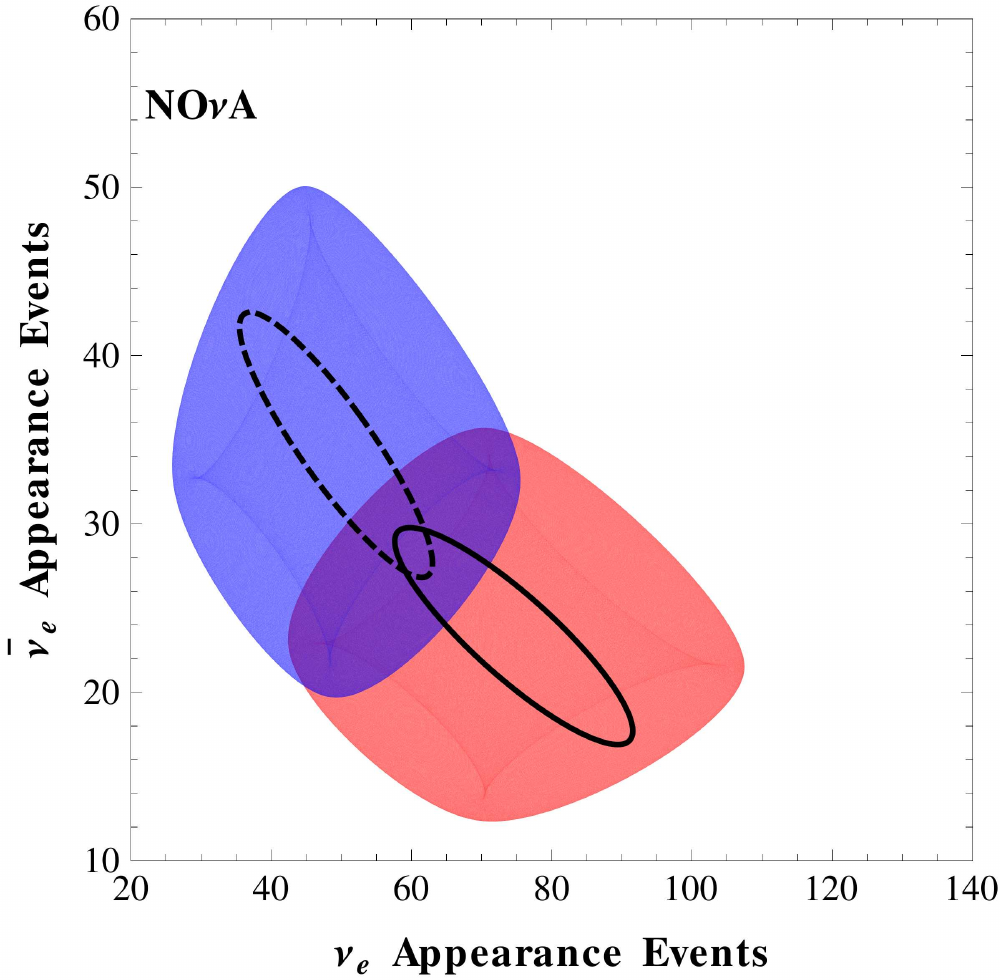}
}
\caption{The colored shaded blobs represent the convolution of the bi-events graphs 
for T2K (left panel) and NO$\nu$A (right panel) in the 3+1 scheme. The blobs are
obtained by superimposing several ellipses, each corresponding to 
a different value of the CP-phase $\delta_{14}$ taken in its range of variability $[-\pi,\pi]$.
The black curves show the 3-flavor ellipses as 
a benchmark. In both the 3-flavor and 4-flavor cases, the running parameter 
on the ellipses is the CP-phase $\delta_{13}$ in the range $[-\pi,\pi]$.  
For both T2K and NO$\nu$A we have assumed that half
of the full exposure will be used in the neutrino mode and the other half in the antineutrino mode.}
\label{fig:convoluted-bi-event}
\end{figure}

Figure~\ref{fig:NOvA-appearance-bi-events} shows the bi-events plots for NO$\nu$A.
The comparison with the corresponding bi-probability plot in Fig.~\ref{fig:NOvA-appearance-bi-probability}
shows that the geometrical properties of the ellipses are quite similar to those obtained 
for the probabilities. This is due to the fact the the energy spectrum
of NO$\nu$A is more sharp than the T2K one (see Fig.~\ref{fig:event-spectrum}).
 As a consequence the peak energy is more important in determining the global behavior of the 
total rate. Finally, in Fig.~\ref{fig:convoluted-bi-event} we show the convolution plot 
in the bi-event space, which gives a visual information on the degree of separation
of the two neutrino mass hierarchies.  
 
\section{Details of the Statistical Method}
\label{simulation-details}

This section deals with the numerical technique and analysis procedure
which we follow to compute our main results.
We use the GLoBES software~\cite{Huber:2004ka,Huber:2007ji} 
along with its new physics tools to obtain our results. We include the 4-flavor
effects both in the $\nu_\mu \to \nu_e$ appearance channel and in the 
$\nu_\mu \to \nu_\mu$ disappearance channel. We have found
that, for the $\nu_\mu \to \nu_\mu$ disappearance channel, the survival probability is very close
to the 3-flavor case, in agreement with the analytical considerations made in~\cite{Klop:2014ima}.
We consider
the true value of $\stch$ to be 0.085 (see table~\ref{tab:benchmark-parameters})
to generate the data and keep it fixed in the fit expecting that the Daya
Bay experiment would be able to measure $\tet$ with a very high-precision
($\sim$ 3\% relative precision at 1$\sigma$ C.L.) by the end 
of 2017~\cite{Zhan:2015aha}. As far as the atmospheric mass-squared splitting
is concerned, we take the true value of $\Delta m^2_{32}$ = $2.4 \times 10^{-3}$
eV$^2$ ($ - 2.4 \times 10^{-3}$ eV$^2$) for NH (IH). Accordingly, we take
 $\Delta m^2_{31}$ = $2.475 \times 10^{-3}$ eV$^2$ ($ - 2.4 \times 10^{-3}$ eV$^2$) for NH (IH). 
We also do not marginalize over this parameter in the fit since the present precision on this parameter is 
already quite good~\cite{Gonzalez-Garcia:2014bfa}, and the future data from
the running T2K and NO$\nu$A experiments would certainly improve this
further~\cite{Agarwalla:2013qfa,Abe:2014tzr}. This should remain true also in the presence
of sterile neutrino oscillations because
the value of $\Delta m_{31}^2$ is extracted from the $\nu_\mu \to \nu_\mu$ searches
which, for the small values of the mixing angles $\theta_{14}$ and $\theta_{24}$
considered in the present analysis (see table~\ref{tab:benchmark-parameters}), are almost unaffected 
by the 4-flavor effects. For $\tmt$, we consider the maximal mixing ($\pi/4$) as the true choice, and 
in the fit, we marginalize over the range given in table~\ref{tab:benchmark-parameters}.
We marginalize over both the 
choices of hierarchy in the fit for all the analyses, except for the mass hierarchy 
discovery studies where our aim is to exclude the wrong hierarchy in the fit.
We vary the true value of $\delta_{13}$ in its allowed range of $[-\pi, \pi]$,
and it has been marginalized over its full range in the fit if the performance 
indicator demands so. We take the line-averaged constant Earth matter 
density\footnote{The line-averaged constant Earth matter density has been 
computed using the Preliminary Reference Earth Model (PREM)~\cite{PREM:1981}.} 
of 2.8 g/cm$^{3}$ for both the baselines. 

We take the mass-squared splitting $\Delta m^2_{41} = 1\,$eV$^2$, which
is the value currently suggested by the SBL anomalies. However, we stress that our
results would remain unaltered for different choices of such parameter, provided 
that  $\Delta m^2_{41} \gtrsim 0.1\,$eV$^2$. For such values, the fast
oscillations induced by the new large frequency get completely averaged
because of the finite resolution of the detector. For the same reason, 
the LBL setups are insensitive to the sign of $\Delta m^2_{41}$ and
we can safely assume positive sign for it. Concerning the active-sterile 
mixing angles, we take the true value of 0.025 for both the $\sin^2\theta_{14}$ and 
$\sin^2\theta_{24}$ and keep them fixed in the fit.  
These values are close to the best fit obtained by the global 3+ fits~\cite{Giunti:2013aea}.
We vary the true value of $\delta_{14}$ in its allowed 
range of [$-\pi,\pi$], and it has been marginalized over its full range in the fit 
as needed. We assume $\sin^2\theta_{34}$ = 0 and $\delta_{34}$ = 0 in our
simulation.%
\footnote{We recall that the vacuum $\nu_\mu \to \nu_e$ transition probability is independent
of $\theta_{34}$ (and $\delta_{34}$). In matter, a tiny dependence appears which 
is more appreciable in NO$\nu$A than in T2K (see the appendix of~\cite{Klop:2014ima} for a detailed discussion).} 
In our analysis, we do not explicitly consider the near detectors 
of T2K and NO$\nu$A which may shed some light on $\theta_{14}$ and $\theta_{24}$, 
but certainly, the near detector data are not sensitive to the CP-phases which is the main thrust 
of this work. In our simulation, we have performed a full spectral analysis using
the binned events spectra for both experiments. In the statistical analysis, 
the Poissonian $\Delta\chi^{2}$ is marginalized over the 
uncorrelated systematic uncertainties (as mentioned in section~\ref{experimental-details})
using the method of pulls as discussed in Refs.~\cite{Huber:2002mx,Fogli:2002pt}. 
When showing the results, we display the $1,2,3\sigma$ confidence levels for 1 d.o.f.
using the relation $\textrm{n}\sigma \equiv \sqrt{\Delta\chi^2}$.
In~\cite{Blennow:2013oma}, it was shown that the above relation is valid in the
frequentist method of hypothesis testing.

\section{Results of the Sensitivity Study}
\label{results}

\subsection{CP-violation Searches in the Presence of Sterile Neutrinos}
\label{CPV}

In this section we explore the impact of sterile neutrinos in the CPV searches of T2K and NO$\nu$A.
As a first step we consider the discovery potential of the CPV induced by the standard 3-flavor CP-phase
$\delta_{13}$, which is proportional to $\sin \delta_{13}$. The discovery potential is defined as the
confidence level at which one can reject the test hypothesis of no CP-violation, i.e. the cases
$\delta_{13}=0$ and $\delta_{13}=\pi$. We have taken the best fit values of all the parameters at
the values specified in the second column of table~\ref{tab:benchmark-parameters}. In the 3-flavor scheme,
we marginalize over $\theta_{23}$ and over the hierarchy. In the 3+1 scheme,
in addition, we marginalize over the unknown value of $\delta_{14}$.

\begin{figure}[H]
\hskip0.5cm
\centerline{
\includegraphics[height=19 cm,width= 17cm]{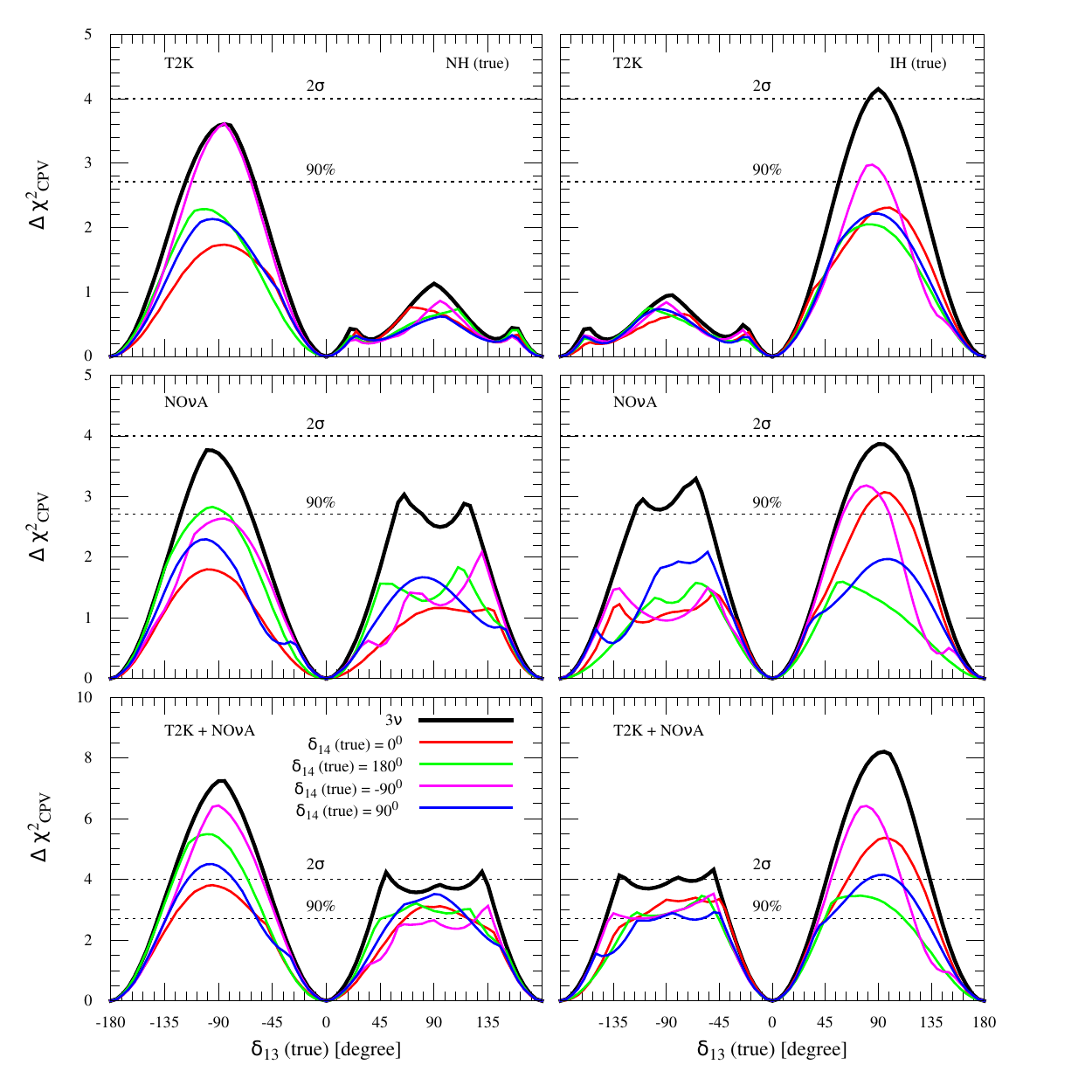}
}
\caption{Discovery potential of CP-violation induced by $\sin \delta_{13}$. 
Upper panels refer to T2K. Middle panels to NO$\nu$A. Lower panels 
to T2K and NO$\nu$A combined. In the left (right) panels, we consider
NH (IH) as the true hierarchy choice. In each panel, the black curve
corresponds to the 3-flavor case. The colored curves
are obtained in the 3+1 scheme for four different true values of $\delta_{14}$. 
We marginalize over $\theta_{23}$ and $\delta_{14}$ over their allowed ranges in the fit, 
and also over the hierarchy.} 
\label{fig:cpv}
\end{figure}

In Figure~\ref{fig:cpv} we display the results of the numerical analysis. The upper panels refer to T2K, 
the middle ones to NO$\nu$A, and the lower ones
to their combination. In the left (right) panels, we consider NH (IH) as the true hierarchy choice. In each panel, 
we present the results obtained for the 3-flavor case (black solid curve) and for the 3+1 scheme, in which case
we select four different values of the true value of $\delta_{14}$ (while its test value is left free to vary and is marginalized away). 
The values of the phase $\delta_{14}$ and the colors of the corresponding curves are the same of the previous plots. 
The 3-flavor sensitivities (black curves) are in agreement with
those shown in the official analyses~\cite{Abe:2014tzr}. We see that for all values of the new CP-phase $\delta_{14}$ the discovery
potential of the two experiments decreases with respect to that of the 3-flavor case. The loss of sensitivity
is imputable to the degeneracy between the two CP-phases $\delta_{13}$ and $\delta_{14}$.  Similar to the 3-flavor case, 
the discovery potential has a maximum for $\delta_{13} = -90^0$  ($\delta_{13} = 90^0$) for NH (IH). Abrupt changes
in the sensitivity are evident in the range $[45^0, 135^0]$ for the NH case and in the range $[-135^0, -45^0]$ 
for the IH case. This behavior can be traced to the degeneracy among the two CP-phases and the mass hierarchy. In fact,
in these ranges the best fit is obtained for the false hierarchy. In the bi-events plots these ranges correspond to
points where the ellipses of the two hierarchies tend to overlap. 

\begin{figure}[H]
\centerline{
\includegraphics[width=0.49\textwidth]{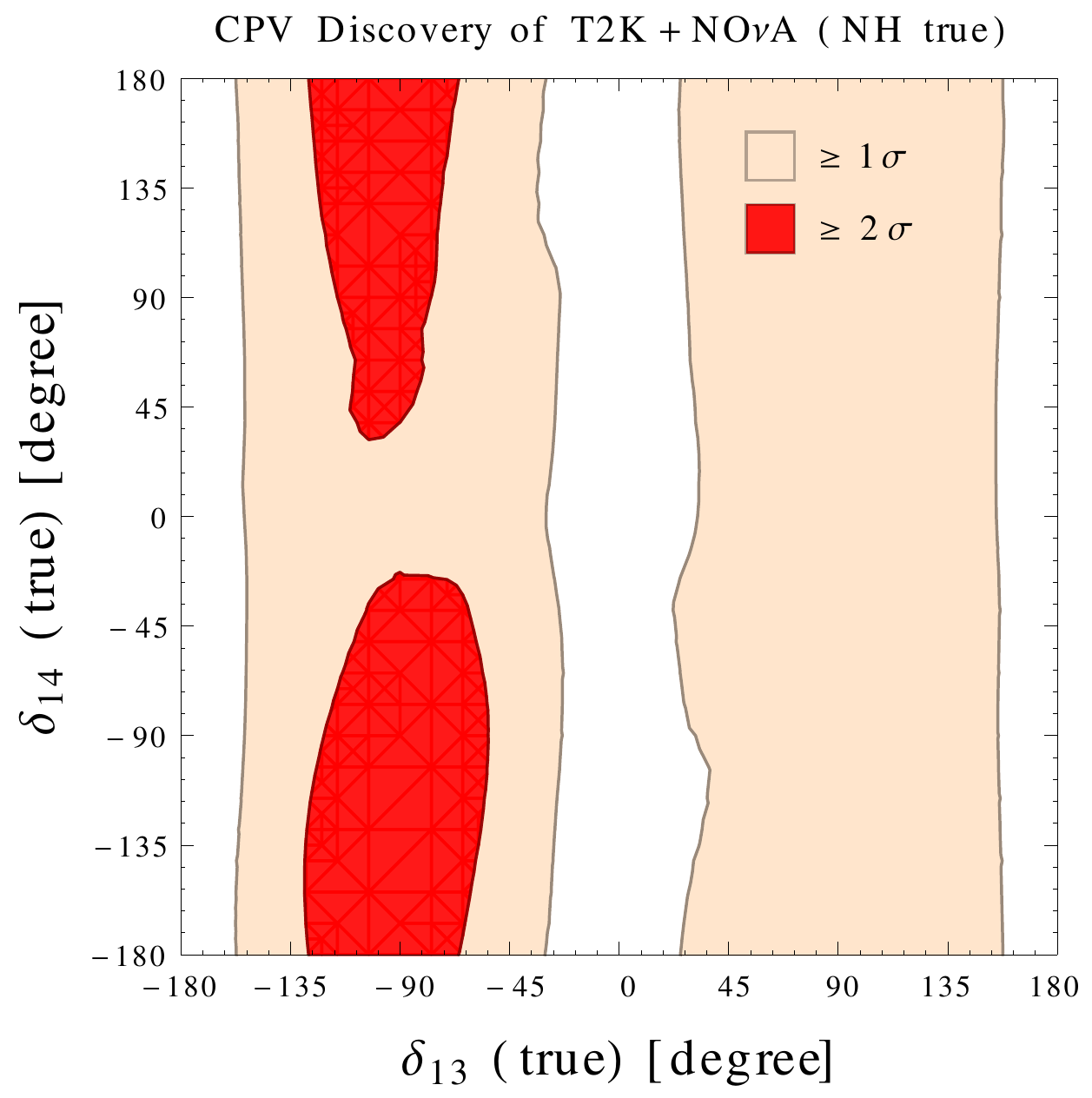}
\includegraphics[width=0.49\textwidth]{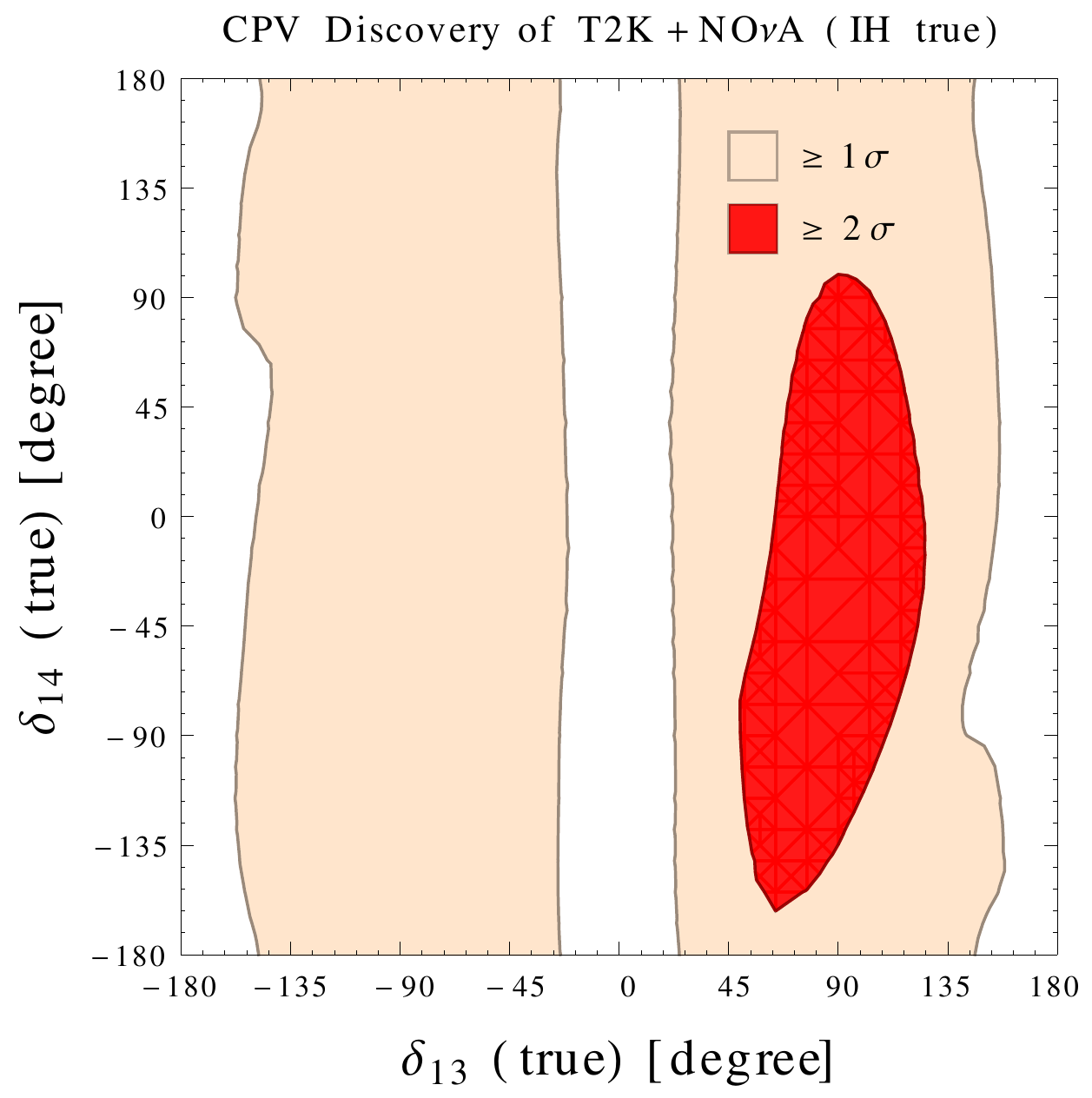}
}
\caption{Discovery potential of CP-violation induced by $\sin \delta_{13}$
in the 3+1 scheme for the T2K + NO$\nu$A combined setup.
 Inside the red regions the discovery potential is $\ge 2 \sigma$. 
Inside the beige regions it is $ \ge 1 \sigma$.}
\label{fig:cpv-true-delta13-delta14}
\end{figure}

Until now we have considered only four selected values of the CP-phase $\delta_{14}$. It is interesting
to see what happens for a generic choice of such a parameter. To this purpose we have generalized the 
analysis by treating $\delta_{14}$ as a free parameter. In Fig.~\ref{fig:cpv-true-delta13-delta14} 
we show the iso-contour lines of the discovery potential of the CP-violation induced by $\delta_{13}$
as a function of the true values of the two phases $\delta_{13}$ and $\delta_{14}$. Inside the red regions
the discovery potential is larger than $2 \sigma$. Inside the beige regions it is larger than $1 \sigma$.
The plots refer to the combination T2K + NO$\nu$A for the two cases of NH (left) and IH (right). One 
can easily check that horizontal cuts of the contour plots made in correspondence of the four particular
values of the phase $\delta_{14}$ considered in Fig.~\ref{fig:cpv} return the 1$\sigma$ and 2$\sigma$
intervals derivable from the last two panels of Fig.~\ref{fig:cpv}. 
   
\begin{figure}[H]
\centerline{
\includegraphics[width=0.49\textwidth]{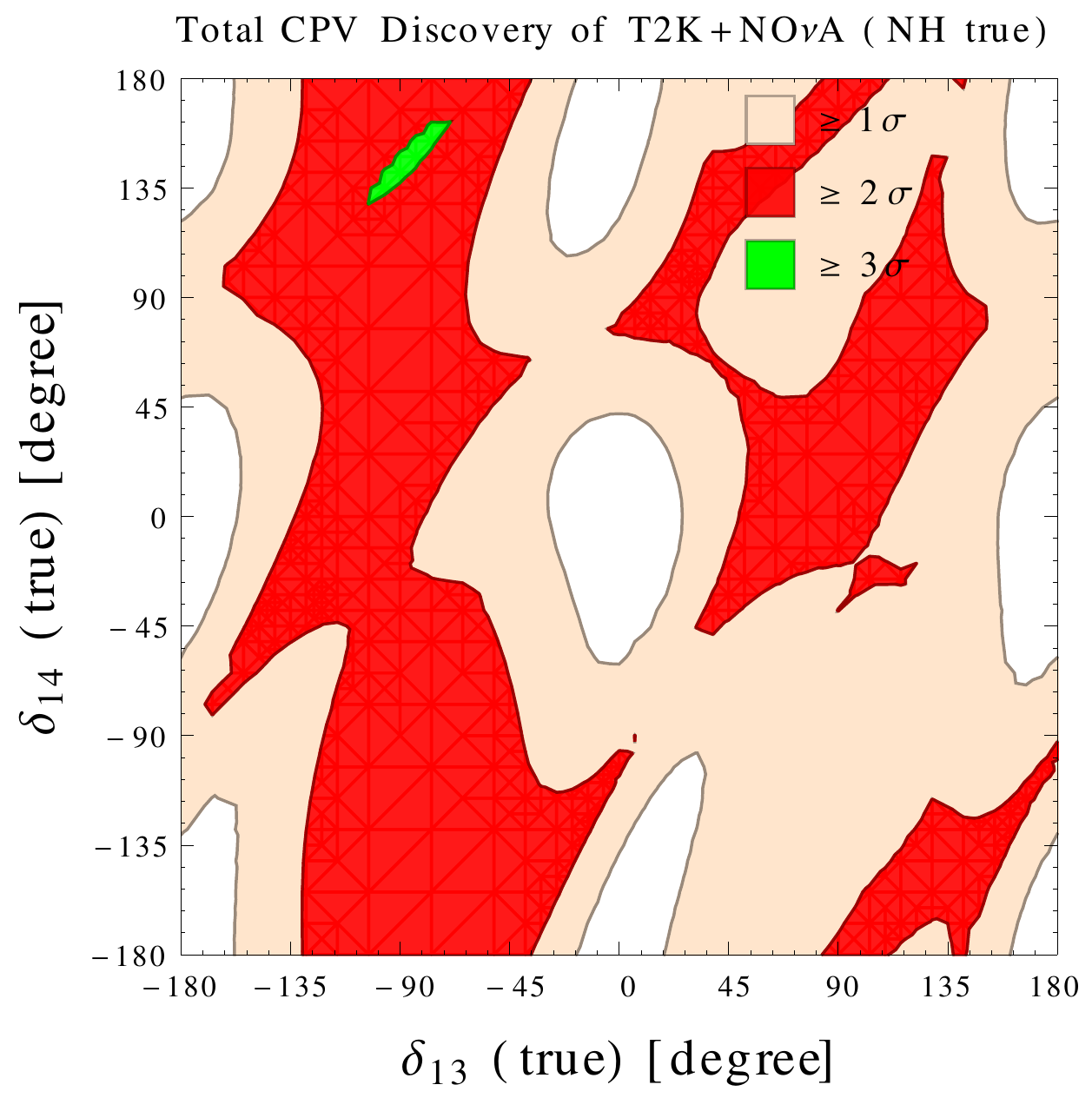}
\includegraphics[width=0.49\textwidth]{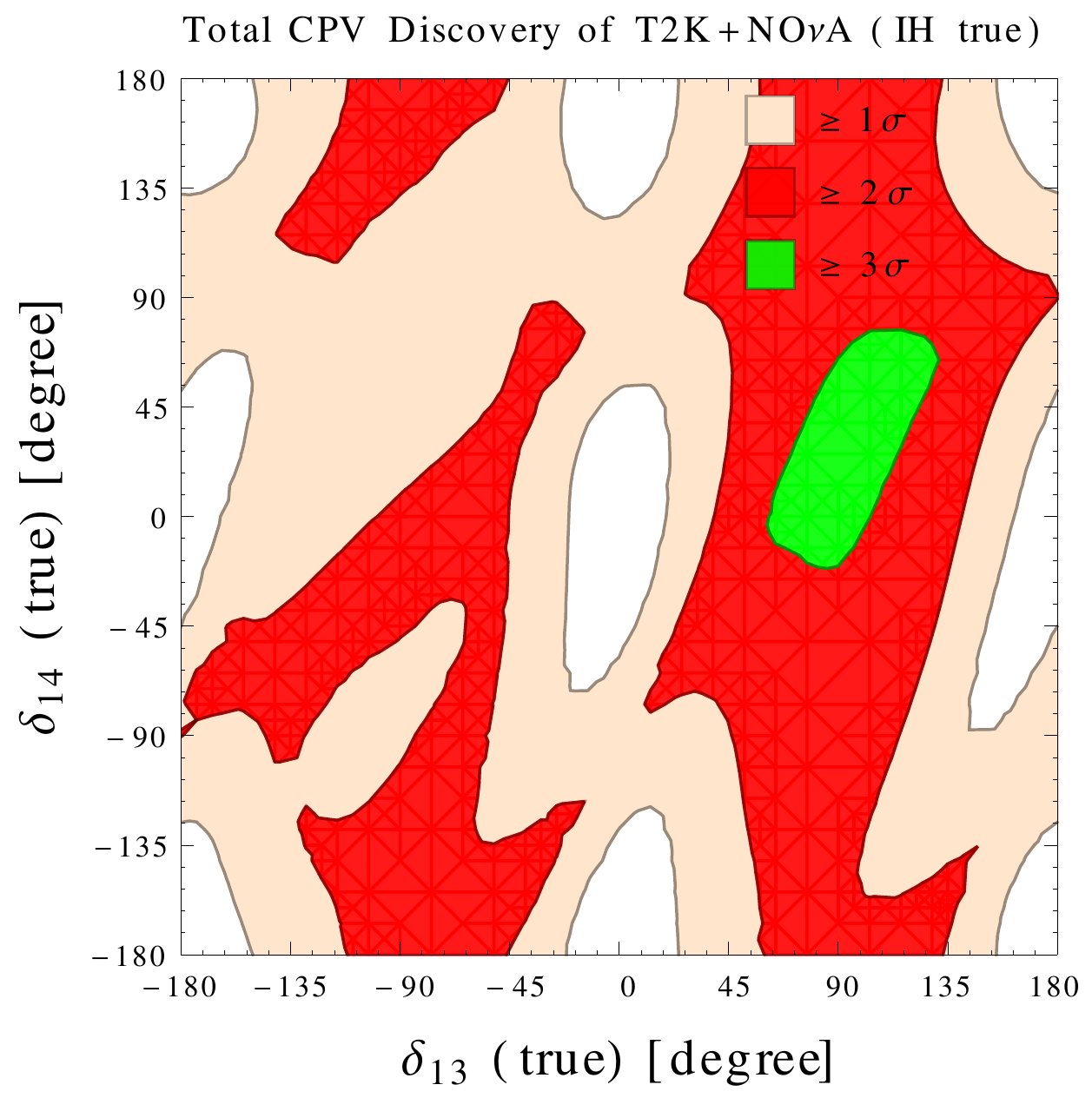}
}
\caption{Discovery potential of total CP-violation (induced simultaneously by
$\sin \delta_{13}$ and $\sin \delta_{14}$) in the 3+1 scheme for the T2K + NO$\nu$A
combined setup. Inside the green regions
the discovery potential is $\ge 3 \sigma$. Inside the red regions it is $\ge 2 \sigma$. 
Inside the beige regions it is $\ge 1 \sigma$.}\label{fig:total-cpv}
\end{figure}

In the 3+1 scheme also the new CP-phase $\delta_{14}$ can be a source of CP-violation.
Hence it is interesting to determine the discovery potential of the CPV induced
by $\sin \delta_{14}$ alone and the {\em total} CP-violation induced simultaneously by $\sin \delta_{13}$ and $\sin \delta_{14}$.
Our numerical analysis shows that the discovery potential of non-zero $\sin \delta_{14}$
is always below the 2$\sigma$ level so we do not show the corresponding plot.
Instead in Fig.~\ref{fig:total-cpv} we show the results for the {\em total} CPV discovery
since it is appreciably different (larger) from that induced by $\sin \delta_{13}$ alone. 
The plots refer to the combination T2K + NO$\nu$A for the two cases of NH (left) and IH (right).
Inside the small green regions the discovery potential is larger than $3 \sigma$. Inside the red regions
it is larger than $2 \sigma$. Inside the beige regions it is larger than $1 \sigma$.
As expected the regions in Fig.~\ref{fig:total-cpv} contain as sub-regions those of Fig.~\ref{fig:cpv-true-delta13-delta14} 
where the sole CPV induced by $\sin \delta_{13}$ is considered. In Fig.~\ref{fig:total-cpv}  we can recognize nine white regions
where the discovery potential is below 1$\sigma$. We stress that these nine regions correspond to four physical regions because
both CP-phases are cyclic variables. Geometrically one can view the square represented in Fig.~\ref{fig:total-cpv} 
as an unwrapped torus, provided one takes into account that the upper edge is connected with the lower edge, and 
that left edge with the right one\footnote{Formally one can see that the square is homeomorphic to the torus. 
Topologically the torus can be seen as the quotient space of the square.}.
These four regions contain the points where the total CPV is zero 
i.e. $\sin \delta_{13} = \sin \delta_{14} =0$. This condition is verified for the four (inequivalent) CP-conserving cases
 $[\delta_{13}, \delta_{14}] = [0,0]$, $[\pi,0]$, $[0,\pi]$, $[\pi, \pi]$ and all the other five combinations
obtainable by a change of sign of one of (or both) the phases equal to $\pi$. 

\subsection{Reconstruction of the CP phases}
\label{CP-phases}

\begin{figure}[H]
\centerline{
\includegraphics[width=0.49\textwidth]{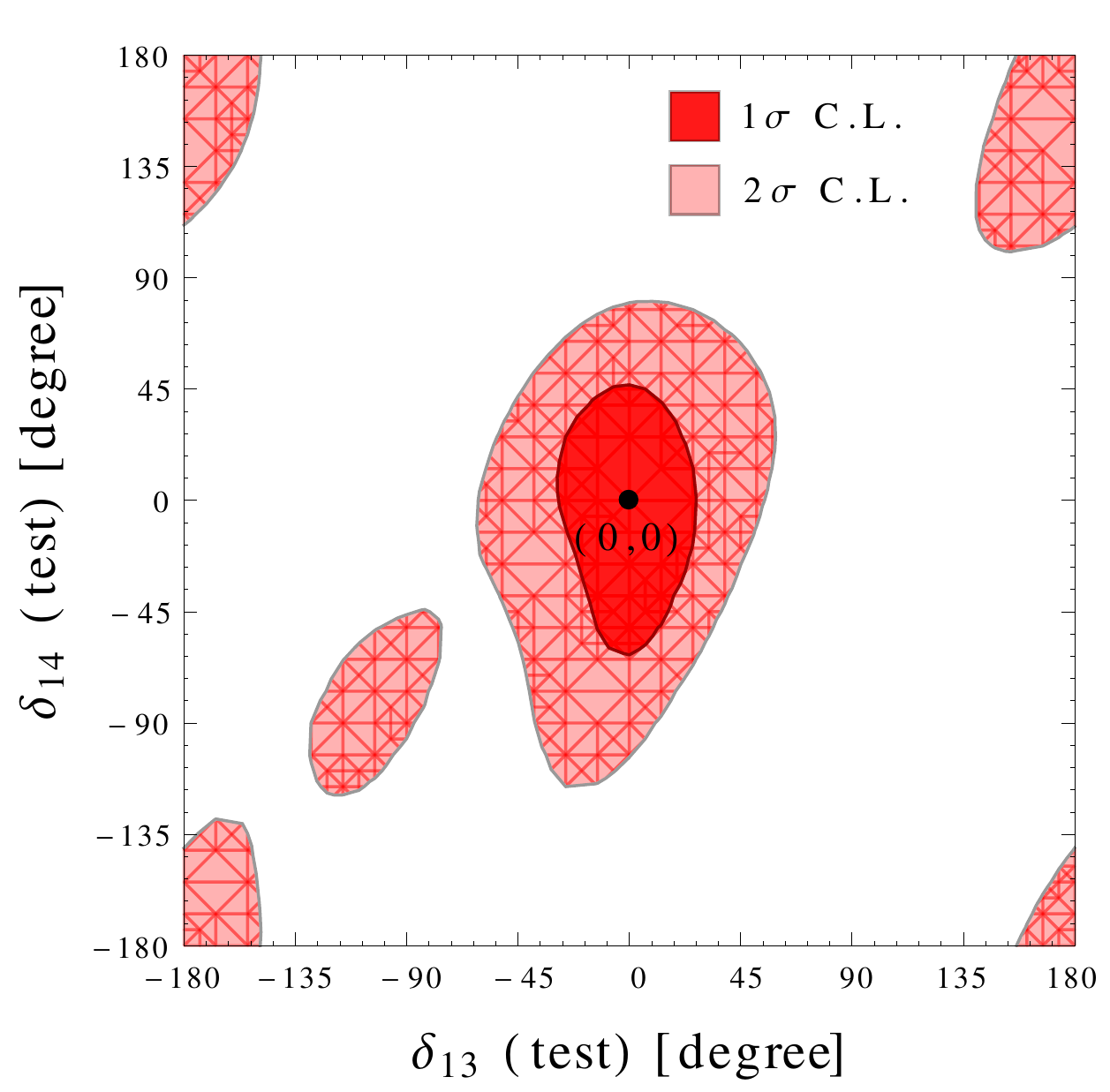}
\includegraphics[width=0.49\textwidth]{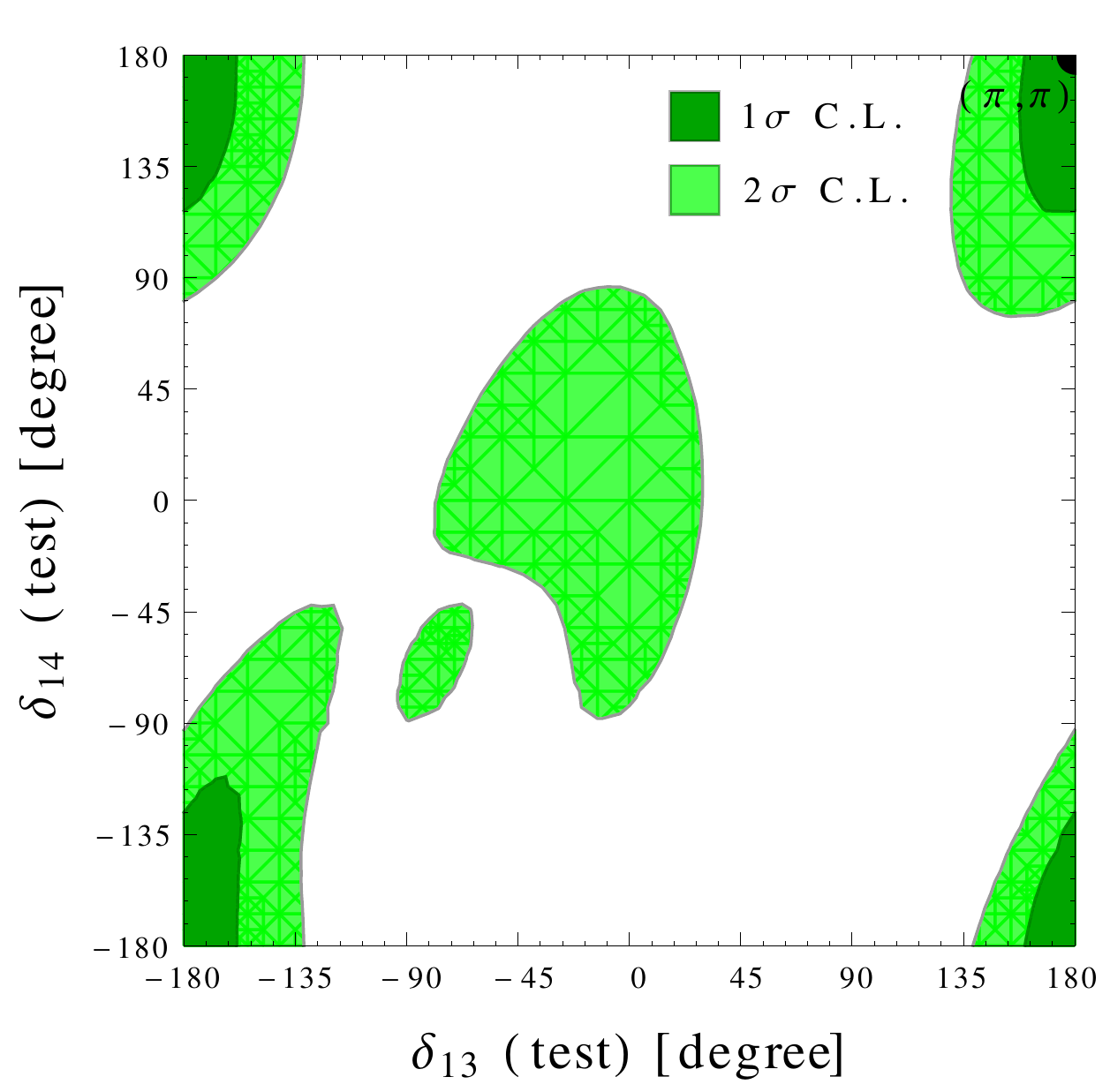}
}
\centerline{
\includegraphics[width=0.49\textwidth]{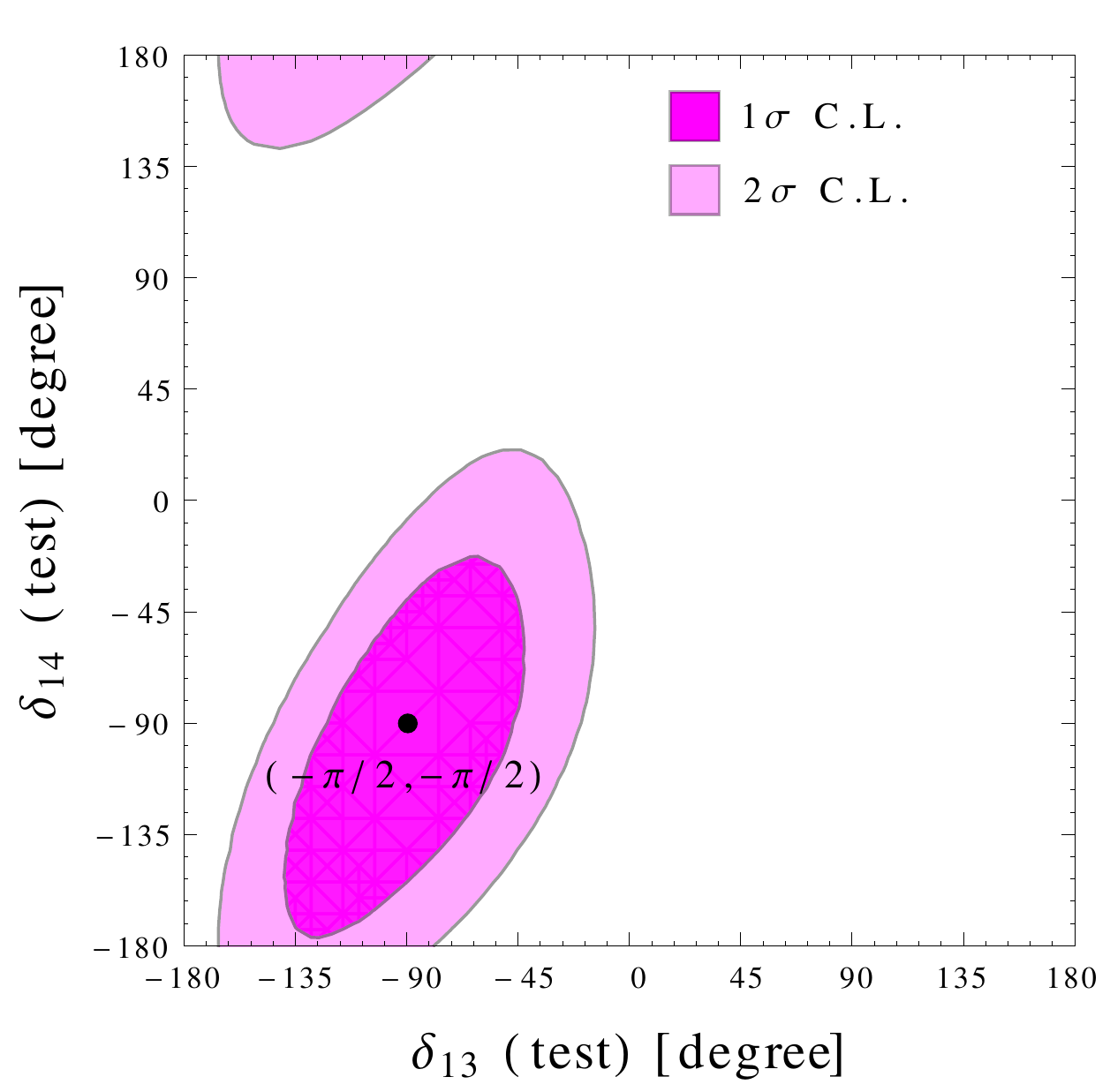}
\includegraphics[width=0.49\textwidth]{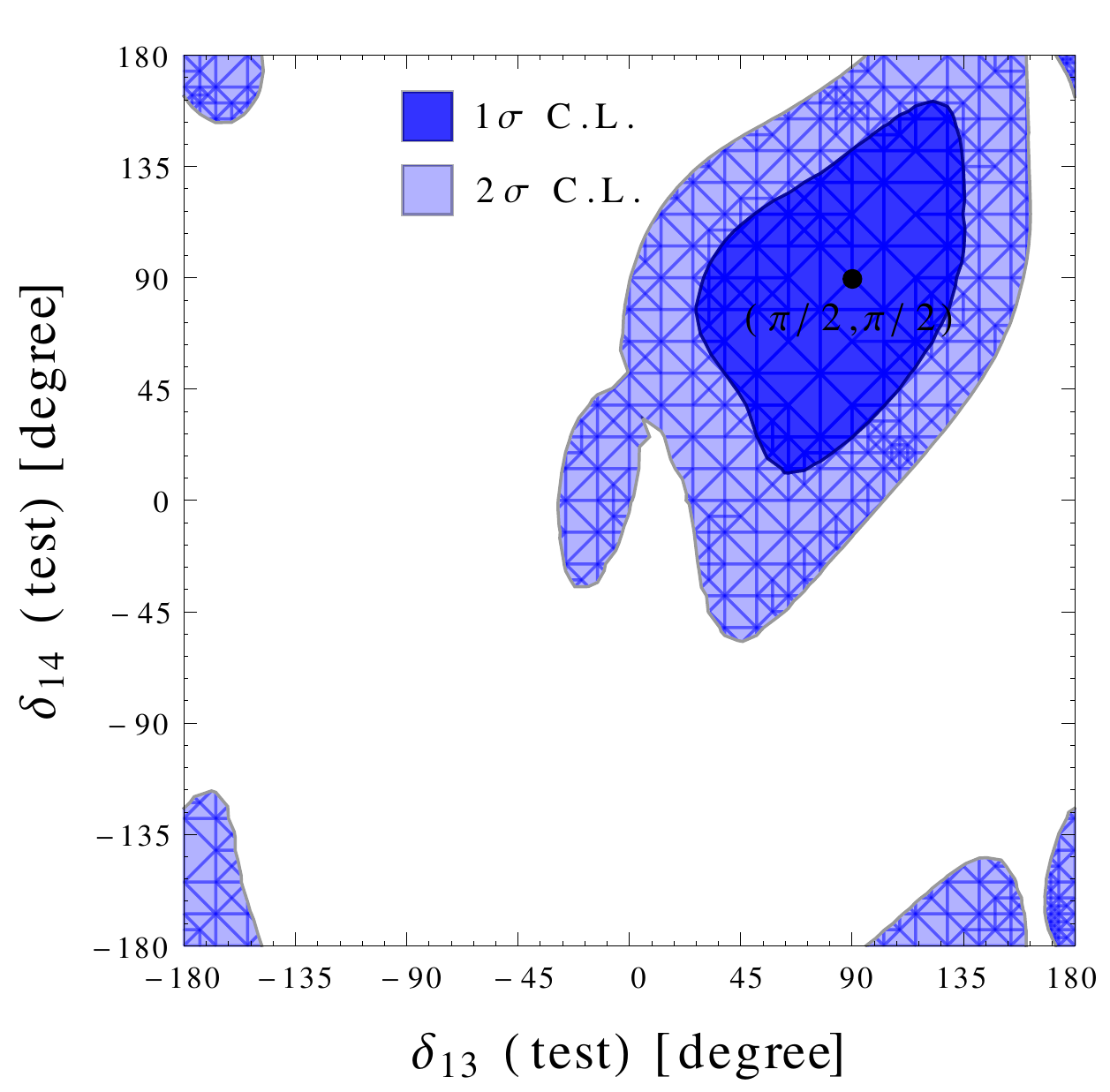}
}
\caption{Reconstructed regions for the two CP-phases $\delta_{13}$ and $\delta_{14}$ for
the T2K + NO$\nu$A combined setup, for the four choices of their true values indicated in each panel.
The NH is taken as the true hierarchy, while we have marginalized over the two possible hierarchies
in the test model. The contours refer to 1$\sigma$ and 2$\sigma$ levels.}
\label{fig:cp-reconstruction}
\end{figure}

The CP-violation discovery potential tells us how much one experiment will be able to rule out 
the case of CP conservation given a positive observation of CPV corresponding to a true
value of the phases $\delta_{13}$ and $\delta_{14}$. While this is certainly a very important feature,
a complementary information is provided by the capability of reconstructing the values
of the two CP-phases, independent of the amount of CP-violation (if any).  
Figure \ref{fig:cp-reconstruction} gives a quantitative answer to such a different kind of question.
The four plots represent the regions reconstructed around four representative
points in the plane [$\delta_{13}, \delta_{14}$]. In all cases we have taken the NH as the true hierarchy 
in the data and then marginalized over NH and IH in theory.
Similar results (not shown) were obtained for the IH case.
The two upper panels refer to the CP-conserving cases $[0,0]$ and $[\pi,\pi]$ respectively.
The third and fourth panels refer to the two (maximally) CP-violating cases $[-\pi/2, -\pi/2]$
and $[\pi/2, \pi/2]$. The two confidence levels refer to 1$\sigma$ and 2$\sigma$ (1 d.o.f.).
We see that in all cases we obtain a unique reconstructed region at the 1$\sigma$ level.
Note that this is true also in the second panel, because the four corners of the square
form a connected region due to the cyclic properties of the two CP-phases.
At the 2$\sigma$ level we obtain a unique region only in the case $[-\pi/2, -\pi/2]$ (bottom left).
Small spurious islands start to appear in the other cases. We have checked that these islands disappear if one
assumes the prior knowledge of the correct mass hierarchy. In all cases the typical
1$\sigma$ uncertainty is about $40^0$ ($50^0$) for $\delta_{13}$ ($\delta_{14}$).
As recently shown in the 4-flavor analysis performed in~\cite{Palazzo:2015gja}, the present data seem
to indicate a slight preference for the combination $[\delta_{13}, \delta_{14}] = [-\pi/2, -\pi/2]$.
If this trend gets confirmed in a few years (assuming the existence of
a sterile neutrino), the picture should resemble that of the left bottom panel of Fig.~\ref{fig:cp-reconstruction}.  

\subsection{Impact of Sterile Neutrinos on Mass Hierarchy Measurements}
\label{MH-Impact}

\begin{figure}[H]
\hskip0.5cm
\centerline{
\includegraphics[height=19 cm,width= 17cm]{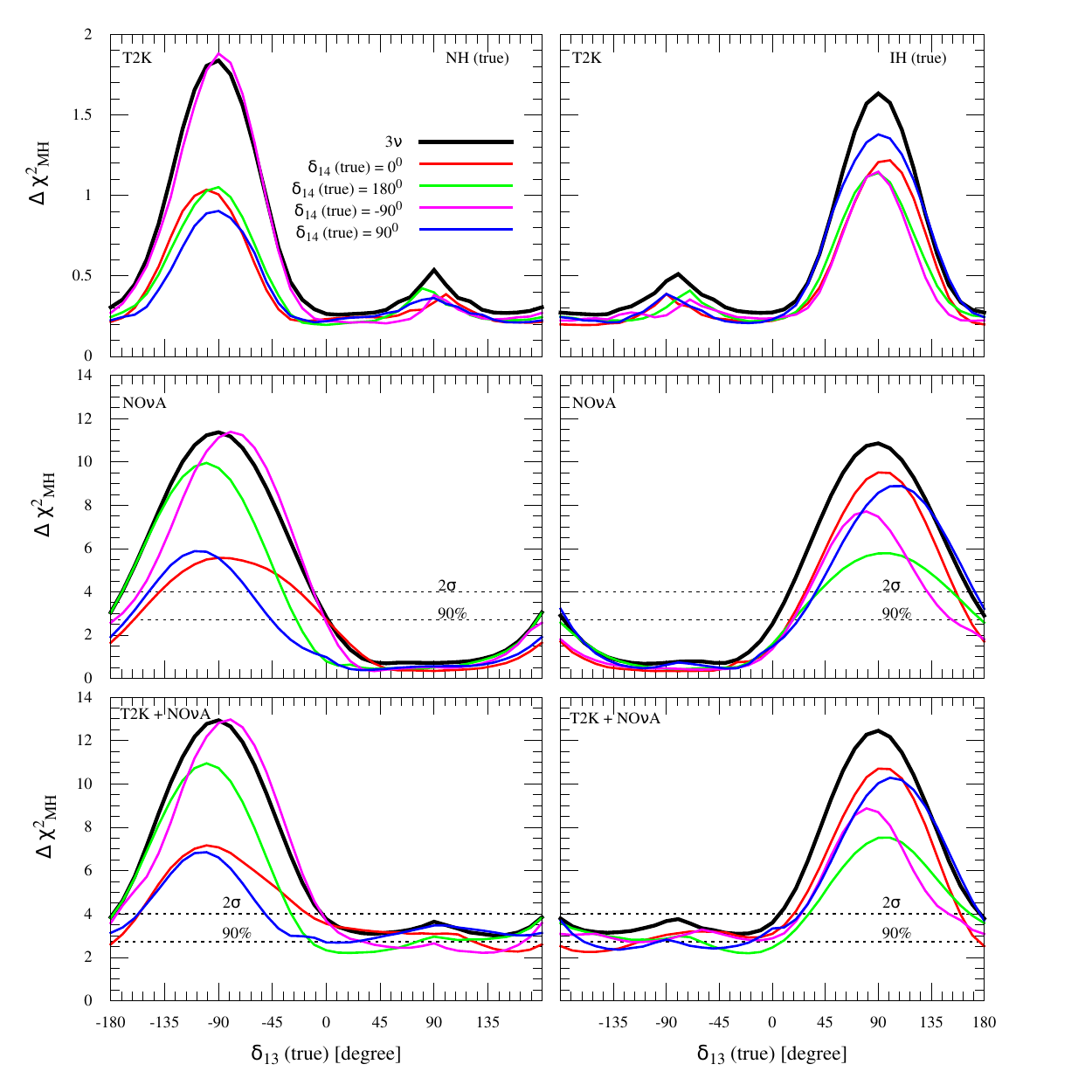}
}
\caption{Discovery potential for excluding the wrong hierarchy as a function 
of true $\delta_{13}$. Upper panels are for T2K. Middle panels are for NO$\nu$A. 
Lower panels are for T2K and NO$\nu$A combined. In the left (right) panels, 
we consider NH (IH) as true hierarchy choice. In each panel, we give the 
results for the 3-flavor case (black line) and for the 3+1 scheme for 
four different values of true $\delta_{14}$.} 
\label{fig:mh}
\end{figure}

In this section we assess the discovery potential of the neutrino mass hierarchy. 
This is defined as the confidence level at which one can exclude the false (or wrong)
test hierarchy given a data set generated with the true hierarchy. We have taken the best
fit values of all the parameters as given in table~\ref{tab:benchmark-parameters}.
In the test model, we have marginalized over 
$\delta_{13}$ and $\theta_{23}$  within their full $3\sigma$ range. In the 3+1 scheme
we have also marginalized over the CP-phase $\delta_{14}$. In Fig.~\ref {fig:mh} we
show the discovery potential of excluding the wrong hierarchy as a function 
of the true value of $\delta_{13}$. The upper panels refer to T2K, the middle ones to NO$\nu$A,
and the lower ones to their combination. In the left (right) panels, 
we consider NH (IH) as the true hierarchy choice. In each panel, we give the 
results for the 3-flavor case (thick black curve) and for the 3+1 scheme (colored curves) 
for four different values of the true $\delta_{14}$ (that is $-90^0$, $90^0$, $0^0$ and $180^0$).
The color convention is the same adopted in the rest of the paper. 

We observe that in T2K (upper panels) the discovery potential is quite limited both in the 3-flavor framework and in the 3+1
scheme. This is due to the fact that the matter effects are small in T2K. Comparing the results
of the 3+1 scheme (colored curves) with those of the 3-flavor one (black curve) we 
observe that, apart for the case $\delta_{14} = -90^0$ for NH ($\delta_{14} = 90^0$ for IH),
in all the other cases the discovery potential is smaller than the 3-flavor one. 
The overall behavior of the 4-flavor curves is similar to that of the 3-flavor one. 
In particular, the sensitivity presents a maximum at $\delta_{13} = -90^0$ for NH. This similar behavior can be 
understood by observing that in the bi-events plots the point $\delta_{13} = -90^0$
(the squares in the four panels of Fig.~\ref{fig:T2K-appearance-bi-events})
always provides the maximal separation from the cloud 
generated by the convolution of all the possible IH ellipses (see Fig.~\ref{fig:convoluted-bi-event}).
A similar observation can be done for the specular case of $\delta_{13} =+90^0$ and IH.

Concerning NO$\nu$A (middle panels) we can make the following observations.
Similarly to T2K the maximal discovery potential is obtained for $\delta_{13} = -90^0$ for NH 
and $\delta_{13} = 90^0$ for IH. For such two values the representative points
on the ellipse (the squares for NH and the circles for IH in Fig.~\ref{fig:NOvA-appearance-bi-events}) always provide the
maximal separation from the convolution of all the ellipses of the opposite MH (see again Fig.~\ref{fig:convoluted-bi-event}).
However, there are also important differences with respect to T2K.
First of all, we observe that the maximal discovery potential is much larger than that of T2K.
This is imputable to the fact that the matter effects are much bigger in NO$\nu$A (see the discussion 
in section~\ref{sec:probability}). Second, we can see that in the NH case (left middle panel) there is a good sensitivity 
not only for $\delta_{14} = -90^0$ (magenta curve) but also for $\delta_{14} = 180^0$ (green curve).
In the IH case (right middle panel) there is a good sensitivity for $\delta_{14} = 90^0$ (blue curve) 
and $\delta_{14} = 0^0$ (red curve). This different behavior with respect to T2K
can be traced to the fact that in NO$\nu$A the peak energy is not centered exactly at the first
oscillation maximum but at $\Delta = 0.4\pi$.  Finally, we notice that the combination of the two experiments
(lower panels) is dominated by NO$\nu$A.

\begin{figure}[H]
\centerline{
\includegraphics[width=0.49\textwidth]{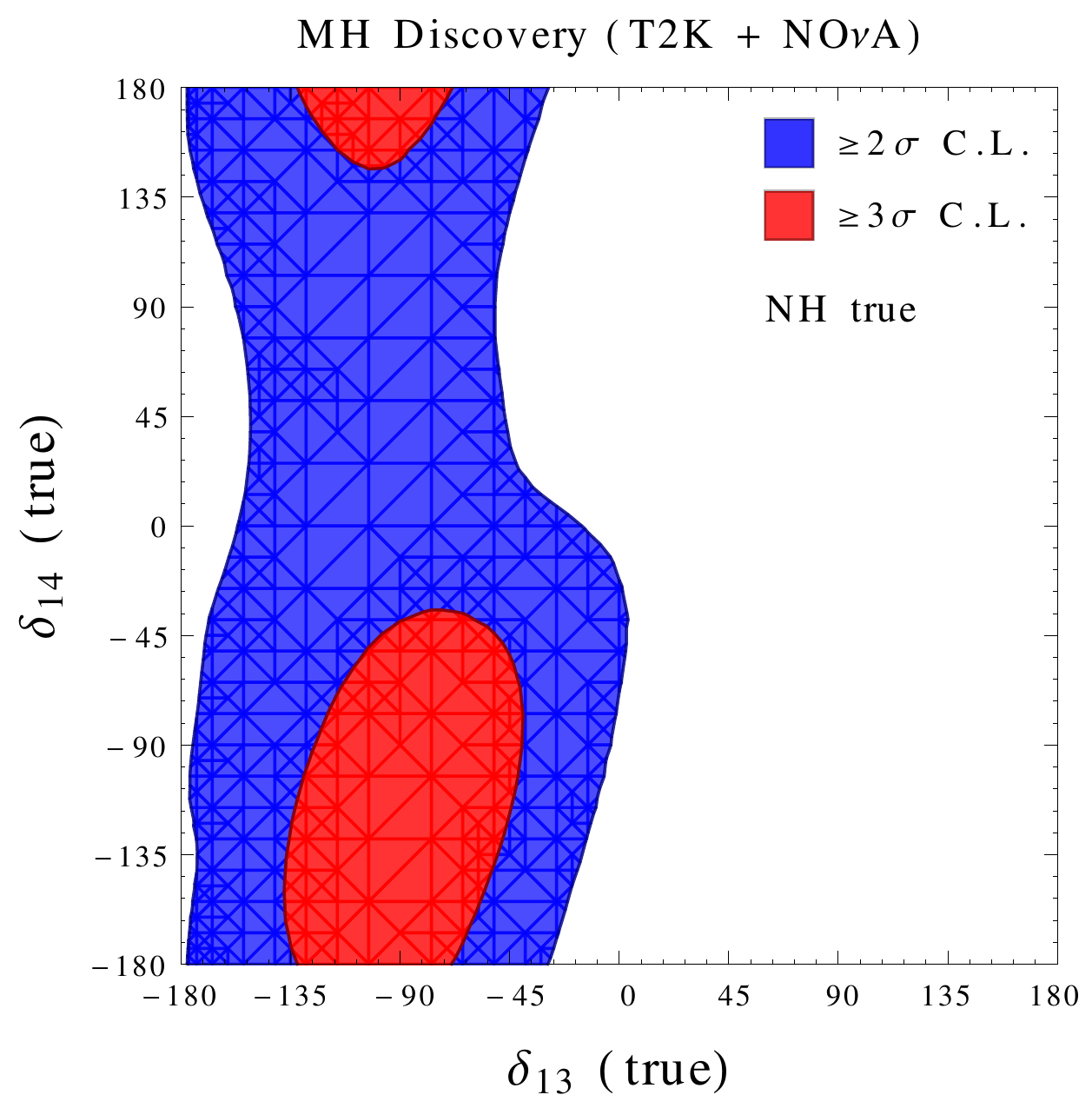}
\includegraphics[width=0.49\textwidth]{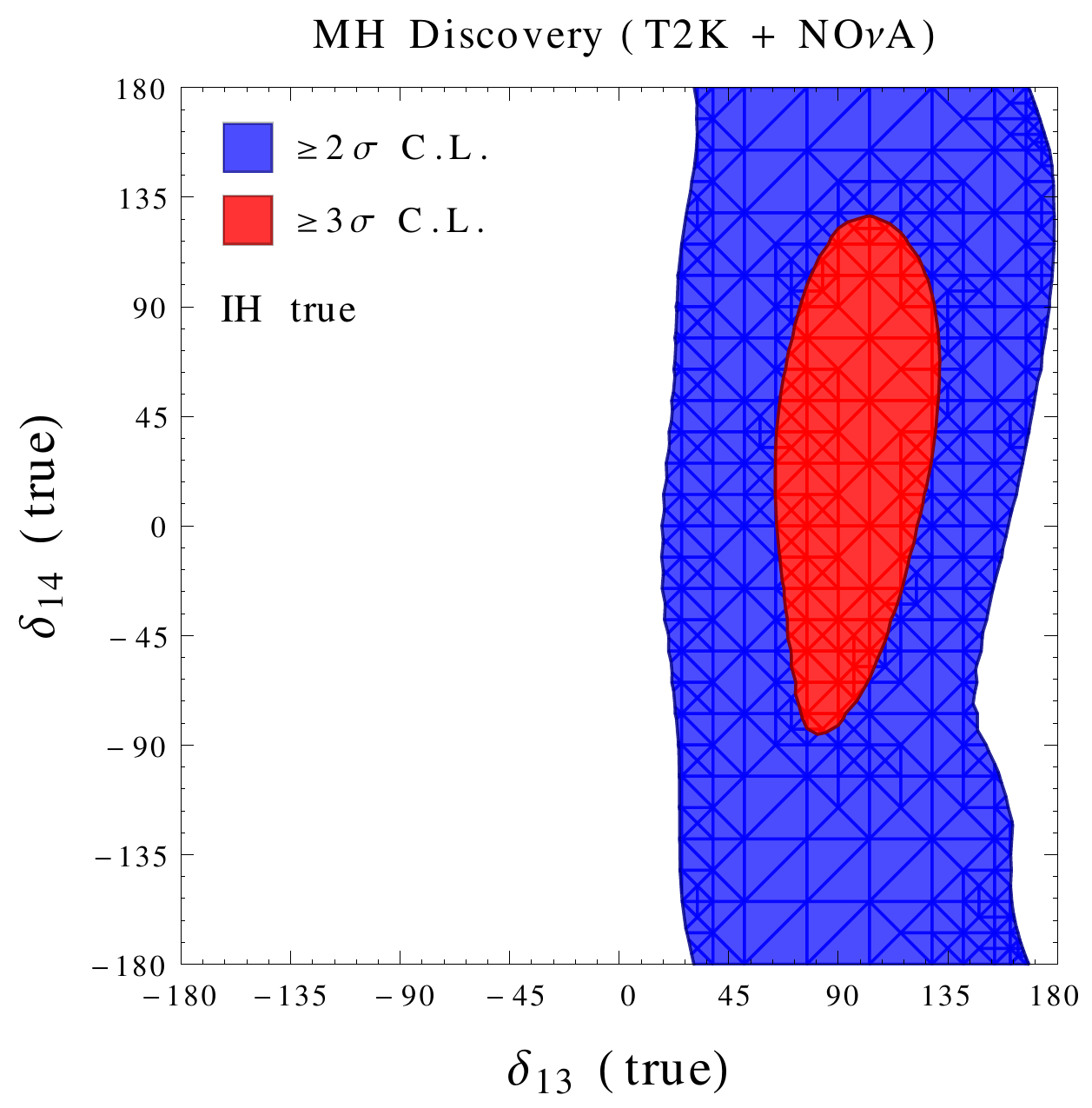}
}
\caption{Discovery potential for excluding the wrong hierarchy for the combination of T2K and NO$\nu$A
as a a function of the two CP-phases $\delta_{13}$ and $\delta_{14}$. Inside the red regions
the discovery potential is $\ge 3 \sigma$. Inside the blue regions it is $\ge 2 \sigma$.}

\label{fig:mh-true-delta13-delta14}
\end{figure}

The study of the discovery potential of the neutrino mass hierarchy can be generalized
to the case in which the CP-phase $\delta_{14}$ can assume any value in its variability range.
Figure~\ref{fig:mh-true-delta13-delta14} shows the results of such more general analysis,
where we have treated $\delta_{14}$ as a free parameter. We display the iso-contour lines of the 
discovery potential as a function of the true values of the two phases $\delta_{13}$ and $\delta_{14}$. 
Inside the red regions the discovery potential is larger than $3 \sigma$. Inside the blue regions it is larger than $2 \sigma$.
The plots refer to the combination T2K + NO$\nu$A for the two cases of NH (left) and IH (right). One 
can easily check that horizontal cuts of the contour plots made in correspondence of the four particular
values of the phase $\delta_{14}$ considered in Fig.~\ref{fig:mh} return the 1$\sigma$ and 2$\sigma$
intervals derivable from the last two panels of Fig.~\ref{fig:mh}.

\section{Conclusions and Outlook}
\label{Conclusions}

We have considered the impact of light sterile neutrinos on the prospective data expected
to come from the two long-baseline experiments T2K and NO$\nu$A when
the planned full exposure will be reached. We have presented a detailed discussion of the behavior
of the 4-flavor $\nu_\mu \to \nu_e$ and $\bar\nu_\mu \to \bar\nu_e$ transition probabilities, 
extending for the first time the bi-probability representation, commonly used in the 3-flavor framework, 
to the 3+1 scheme. We have also performed a comprehensive sensitivity study of the two experiments
(taken alone and in combination) in order to assess their discovery potential in the presence 
of a sterile neutrino species. We have considered realistic benchmark values of the 
3+1 mass-mixing parameters as inferred from the existing global short-baseline fits.
We found that the performance of both the experiments
in claiming the discovery of the CP-violation induced by the standard CP-phase
$\delta_{13} \equiv \delta$, and the neutrino mass hierarchy 
get substantially deteriorated.
The degree of loss of sensitivity depends on the value of the unknown CP-phase $\delta_{14}$.
We have also assessed the discovery potential of total CP-violation (i.e., induced simultaneously by 
the two CP-phases $\delta_{13}$ and $\delta_{14}$) and the capability
of the two experiments of reconstructing the true values of such CP-phases.
The typical (1$\sigma$ level) uncertainty on the reconstructed phases   
is approximately $40^0$ for $\delta_{13}$ and $50^0$ for $\delta_{14}$.

In the eventuality of a discovery of a sterile neutrino at the new short-baseline experiments,
we will face two challenges. First, we will have to reassess the status of the
3-flavor parameters whose best fit values will change in the 3+1 scheme.
Among the 3-flavor properties, the most sensitive to the perturbations
induced by the sterile neutrino oscillations are the CP-phase $\delta$, and 
the neutrino mass hierarchy. In both cases, their determination is based on 
the observation of very tiny effects which, in the LBL setups, can be appreciably
perturbed by new interference phenomena induced by the sterile neutrinos.
Our study gives the first quantitative assessment of the discovery potential of 
the 3-flavor CPV and of the MH for the two LBL experiments T2K and NO$\nu$A.
The second, perhaps more stimulating challenge, will be that of determining 
all the new parameters that govern the enlarged 3+1 scheme.  According to our
study, T2K and NO$\nu$A may be able to give the first indications on one
of the new CP-phases involved in the 3+1 scheme. The future
LBL experiments (DUNE, LBNO and T2HK) will be needed to extract 
more robust information on the enlarged CP-violation sector. We hope that
the comprehensive analysis presented in this paper may play an important role 
in exploring light sterile neutrinos at the long baseline facilities.

\subsubsection*{Acknowledgments}

S.K.A. is supported by the DST/INSPIRE Research Grant [IFA-PH-12],
Department of Science \& Technology, India. S.K.A. would like to thank 
Joachim Kopp for useful communications. A.P. is supported by the 
Grant ``Future In Research'' {\it Beyond three neutrino families},
contract no. YVI3ST4, of Regione Puglia, Italy. A.P. also acknowledges support from the 
Max-Planck-Institut f\"ur Physik (Werner Heisenberg Institut), M\"unchen, Germany,
where this work was begun.

\bibliographystyle{JHEP}
\bibliography{Sterile-References}

\providecommand{\href}[2]{#2}\begingroup\raggedright\begin{thebibliography}{10}

\bibitem{Nobel}
T.~Kajita and A.~B. McDonald, ``For the discovery of neutrino oscillations,
  which shows that neutrinos have mass.''
\newblock The Nobel Prize in Physics 2015,
  \url{http://www.nobelprize.org/nobel_prizes/physics/laureates/2015/}.

\bibitem{An:2015rpe}
{\bf Daya Bay} Collaboration, F.~P. An et~al., {\it {New Measurement of
  Antineutrino Oscillation with the Full Detector Configuration at Daya Bay}},
  {\em Phys. Rev. Lett.} {\bf 115} (2015), no.~11 111802,
  [\href{http://arxiv.org/abs/1505.03456}{{\tt arXiv:1505.03456}}].

\bibitem{RENO:2015ksa}
{\bf RENO} Collaboration, J.~H. Choi et~al., {\it {Observation of Energy and
  Baseline Dependent Reactor Antineutrino Disappearance in the RENO
  Experiment}},  \href{http://arxiv.org/abs/1511.05849}{{\tt
  arXiv:1511.05849}}.

\bibitem{Abe:2014bwa}
{\bf Double Chooz} Collaboration, Y.~Abe et~al., {\it {Improved measurements of
  the neutrino mixing angle $\theta_{13}$ with the Double Chooz detector}},
  {\em JHEP} {\bf 10} (2014) 086, [\href{http://arxiv.org/abs/1406.7763}{{\tt
  arXiv:1406.7763}}]. [Erratum: JHEP02,074(2015)].

\bibitem{Stanco:2015ejj}
L.~Stanco, {\it {Next Generation of Neutrino Studies and Facilities}},
  \href{http://arxiv.org/abs/1511.09409}{{\tt arXiv:1511.09409}}.

\bibitem{Abazajian:2012ys}
K.~N. Abazajian et~al., {\it {Light Sterile Neutrinos: A White Paper}},
  \href{http://arxiv.org/abs/1204.5379}{{\tt arXiv:1204.5379}}.

\bibitem{Palazzo:2013me}
A.~Palazzo, {\it {Phenomenology of light sterile neutrinos: a brief review}},
  {\em Mod. Phys. Lett.} {\bf A28} (2013) 1330004,
  [\href{http://arxiv.org/abs/1302.1102}{{\tt arXiv:1302.1102}}].

\bibitem{Gariazzo:2015rra}
S.~Gariazzo, C.~Giunti, M.~Laveder, Y.~F. Li, and E.~M. Zavanin, {\it {Light
  sterile neutrinos}},  \href{http://arxiv.org/abs/1507.08204}{{\tt
  arXiv:1507.08204}}.

\bibitem{Lasserre:2014ita}
T.~Lasserre, {\it {Light Sterile Neutrinos in Particle Physics: Experimental
  Status}},  {\em Phys. Dark Univ.} {\bf 4} (2014) 81--85,
  [\href{http://arxiv.org/abs/1404.7352}{{\tt arXiv:1404.7352}}].

\bibitem{Palazzo:2011rj}
A.~Palazzo, {\it {Testing the very-short-baseline neutrino anomalies at the
  solar sector}},  {\em Phys. Rev.} {\bf D83} (2011) 113013,
  [\href{http://arxiv.org/abs/1105.1705}{{\tt arXiv:1105.1705}}].

\bibitem{Palazzo:2012yf}
A.~Palazzo, {\it {An estimate of $\theta_{14}$ independent of the reactor
  antineutrino flux determinations}},  {\em Phys. Rev.} {\bf D85} (2012)
  077301, [\href{http://arxiv.org/abs/1201.4280}{{\tt arXiv:1201.4280}}].

\bibitem{Giunti:2009xz}
C.~Giunti and Y.~F. Li, {\it {Matter Effects in Active-Sterile Solar Neutrino
  Oscillations}},  {\em Phys. Rev.} {\bf D80} (2009) 113007,
  [\href{http://arxiv.org/abs/0910.5856}{{\tt arXiv:0910.5856}}].

\bibitem{Nunokawa:2003ep}
H.~Nunokawa, O.~L.~G. Peres, and R.~Zukanovich~Funchal, {\it {Probing the LSND
  mass scale and four neutrino scenarios with a neutrino telescope}},  {\em
  Phys. Lett.} {\bf B562} (2003) 279--290,
  [\href{http://arxiv.org/abs/hep-ph/0302039}{{\tt hep-ph/0302039}}].

\bibitem{IceCube-Sterile-CPAN-2015}
{\bf IceCube} Collaboration, J.~Salvad{\'o}~Serra, {\it {Sterile Neutrino
  Search in IceCube Neutrino Observatory}},  2015.
\newblock Talk given at the VII CPAN Days Conference, December 1-3, 2015,
  Segovia, Spain,
  \url{https://indico.ific.uv.es/indico/sessionDisplay.py?sessionId=3&confId=2541#20151201}.

\bibitem{Abe:2014gda}
{\bf Super-Kamiokande} Collaboration, K.~Abe et~al., {\it {Limits on sterile
  neutrino mixing using atmospheric neutrinos in Super-Kamiokande}},  {\em
  Phys. Rev.} {\bf D91} (2015) 052019,
  [\href{http://arxiv.org/abs/1410.2008}{{\tt arXiv:1410.2008}}].

\bibitem{Timmons:2015lga}
A.~Timmons, {\it {Searching for Sterile Neutrinos at MINOS}},  in {\em {Topical
  Research Meeting on Prospects in Neutrino Physics (NuPhys2014) London, UK,
  United Kingdom, December 15-17, 2014}}, 2015.
\newblock \href{http://arxiv.org/abs/1504.04046}{{\tt arXiv:1504.04046}}.

\bibitem{Adamson:2011ku}
{\bf MINOS} Collaboration, P.~Adamson et~al., {\it {Active to sterile neutrino
  mixing limits from neutral-current interactions in MINOS}},  {\em Phys. Rev.
  Lett.} {\bf 107} (2011) 011802, [\href{http://arxiv.org/abs/1104.3922}{{\tt
  arXiv:1104.3922}}].

\bibitem{Agafonova:2015neo}
{\bf OPERA} Collaboration, N.~Agafonova et~al., {\it {Limits on muon-neutrino
  to tau-neutrino oscillations induced by a sterile neutrino state obtained by
  OPERA at the CNGS beam}},  {\em JHEP} {\bf 06} (2015) 069,
  [\href{http://arxiv.org/abs/1503.01876}{{\tt arXiv:1503.01876}}].

\bibitem{Klop:2014ima}
N.~Klop and A.~Palazzo, {\it {Imprints of CP violation induced by sterile
  neutrinos in T2K data}},  {\em Phys. Rev.} {\bf D91} (2015), no.~7 073017,
  [\href{http://arxiv.org/abs/1412.7524}{{\tt arXiv:1412.7524}}].

\bibitem{Donini:2001xy}
A.~Donini and D.~Meloni, {\it {The 2+2 and 3+1 four family neutrino mixing at
  the neutrino factory}},  {\em Eur. Phys. J.} {\bf C22} (2001) 179--186,
  [\href{http://arxiv.org/abs/hep-ph/0105089}{{\tt hep-ph/0105089}}].

\bibitem{Donini:2001xp}
A.~Donini, M.~Lusignoli, and D.~Meloni, {\it {Telling three neutrinos from four
  neutrinos at the neutrino factory}},  {\em Nucl. Phys.} {\bf B624} (2002)
  405--422, [\href{http://arxiv.org/abs/hep-ph/0107231}{{\tt hep-ph/0107231}}].

\bibitem{Donini:2007yf}
A.~Donini, M.~Maltoni, D.~Meloni, P.~Migliozzi, and F.~Terranova, {\it {3+1
  sterile neutrinos at the CNGS}},  {\em JHEP} {\bf 12} (2007) 013,
  [\href{http://arxiv.org/abs/0704.0388}{{\tt arXiv:0704.0388}}].

\bibitem{Dighe:2007uf}
A.~Dighe and S.~Ray, {\it {Signatures of heavy sterile neutrinos at long
  baseline experiments}},  {\em Phys. Rev.} {\bf D76} (2007) 113001,
  [\href{http://arxiv.org/abs/0709.0383}{{\tt arXiv:0709.0383}}].

\bibitem{Donini:2008wz}
A.~Donini, K.-i. Fuki, J.~Lopez-Pavon, D.~Meloni, and O.~Yasuda, {\it {The
  Discovery channel at the Neutrino Factory: nu(mu) ---> nu(tau) pointing to
  sterile neutrinos}},  {\em JHEP} {\bf 08} (2009) 041,
  [\href{http://arxiv.org/abs/0812.3703}{{\tt arXiv:0812.3703}}].

\bibitem{Yasuda:2010rj}
O.~Yasuda, {\it {Sensitivity to sterile neutrino mixings and the discovery
  channel at a neutrino factory}},  in {\em {Physics beyond the standard models
  of particles, cosmology and astrophysics. Proceedings, 5th International
  Conference, Beyond 2010, Cape Town, South Africa, February 1-6, 2010}},
  pp.~300--313, 2011.
\newblock \href{http://arxiv.org/abs/1004.2388}{{\tt arXiv:1004.2388}}.

\bibitem{Meloni:2010zr}
D.~Meloni, J.~Tang, and W.~Winter, {\it {Sterile neutrinos beyond LSND at the
  Neutrino Factory}},  {\em Phys. Rev.} {\bf D82} (2010) 093008,
  [\href{http://arxiv.org/abs/1007.2419}{{\tt arXiv:1007.2419}}].

\bibitem{Bhattacharya:2011ee}
B.~Bhattacharya, A.~M. Thalapillil, and C.~E.~M. Wagner, {\it {Implications of
  sterile neutrinos for medium/long-baseline neutrino experiments and the
  determination of $\theta_{13}$}},  {\em Phys. Rev.} {\bf D85} (2012) 073004,
  [\href{http://arxiv.org/abs/1111.4225}{{\tt arXiv:1111.4225}}].

\bibitem{Donini:2012tt}
A.~Donini, P.~Hernandez, J.~Lopez-Pavon, M.~Maltoni, and T.~Schwetz, {\it {The
  minimal 3+2 neutrino model versus oscillation anomalies}},  {\em JHEP} {\bf
  07} (2012) 161, [\href{http://arxiv.org/abs/1205.5230}{{\tt
  arXiv:1205.5230}}].

\bibitem{Acciarri:2015uup}
{\bf DUNE} Collaboration, R.~Acciarri et~al., {\it {Long-Baseline Neutrino
  Facility (LBNF) and Deep Underground Neutrino Experiment (DUNE) Conceptual
  Design Report Volume 2: The Physics Program for DUNE at LBNF}},
  \href{http://arxiv.org/abs/1512.06148}{{\tt arXiv:1512.06148}}.

\bibitem{Hollander:2014iha}
D.~Hollander and I.~Mocioiu, {\it {Minimal 3+2 sterile neutrino model at
  LBNE}},  {\em Phys. Rev.} {\bf D91} (2015), no.~1 013002,
  [\href{http://arxiv.org/abs/1408.1749}{{\tt arXiv:1408.1749}}].

\bibitem{Berryman:2015nua}
J.~M. Berryman, A.~de~Gouv{\^e}a, K.~J. Kelly, and A.~Kobach, {\it {Sterile
  neutrino at the Deep Underground Neutrino Experiment}},  {\em Phys. Rev.}
  {\bf D92} (2015), no.~7 073012, [\href{http://arxiv.org/abs/1507.03986}{{\tt
  arXiv:1507.03986}}].

\bibitem{Gandhi:2015xza}
R.~Gandhi, B.~Kayser, M.~Masud, and S.~Prakash, {\it {The impact of sterile
  neutrinos on CP measurements at long baselines}},  {\em JHEP} {\bf 11} (2015)
  039, [\href{http://arxiv.org/abs/1508.06275}{{\tt arXiv:1508.06275}}].

\bibitem{Agarwalla:2011hh}
S.~K. Agarwalla, T.~Li, and A.~Rubbia, {\it {An Incremental approach to unravel
  the neutrino mass hierarchy and CP violation with a long-baseline Superbeam
  for large $\theta_{13}$}},  {\em JHEP} {\bf 1205} (2012) 154,
  [\href{http://arxiv.org/abs/1109.6526}{{\tt arXiv:1109.6526}}].

\bibitem{Agarwalla:2013kaa}
{\bf LAGUNA-LBNO} Collaboration, S.~Agarwalla et~al., {\it {The mass-hierarchy
  and CP-violation discovery reach of the LBNO long-baseline neutrino
  experiment}},  \href{http://arxiv.org/abs/1312.6520}{{\tt arXiv:1312.6520}}.

\bibitem{Abe:2014oxa}
{\bf Hyper-Kamiokande Working Group} Collaboration, K.~Abe et~al., {\it {A Long
  Baseline Neutrino Oscillation Experiment Using J-PARC Neutrino Beam and
  Hyper-Kamiokande}},  \href{http://arxiv.org/abs/1412.4673}{{\tt
  arXiv:1412.4673}}.

\bibitem{Abe:2015zbg}
{\bf Hyper-Kamiokande Proto-Collaboration} Collaboration, K.~Abe et~al., {\it
  {Physics potential of a long-baseline neutrino oscillation experiment using a
  J-PARC neutrino beam and Hyper-Kamiokande}},  {\em PTEP} {\bf 2015} (2015)
  053C02, [\href{http://arxiv.org/abs/1502.05199}{{\tt arXiv:1502.05199}}].

\bibitem{Palazzo:2015gja}
A.~Palazzo, {\it {3-flavor and 4-flavor implications of the latest T2K and
  NO$\nu$A electron (anti-)neutrino appearance results}},
  \href{http://arxiv.org/abs/1509.03148}{{\tt arXiv:1509.03148}}.

\bibitem{Palazzo:2015wea}
A.~Palazzo, {\it {Consistent analysis of the numu to nue sterile neutrinos
  searches of ICARUS and OPERA}},  {\em Phys. Rev.} {\bf D91} (2015), no.~9
  091301, [\href{http://arxiv.org/abs/1503.03966}{{\tt arXiv:1503.03966}}].

\bibitem{Antonello:2012pq}
M.~Antonello et~al., {\it {Experimental search for the ``LSND anomaly'' with
  the ICARUS detector in the CNGS neutrino beam}},  {\em Eur. Phys. J.} {\bf
  C73} (2013), no.~3 2345, [\href{http://arxiv.org/abs/1209.0122}{{\tt
  arXiv:1209.0122}}].

\bibitem{Antonello:2015jxa}
M.~Antonello et~al., {\it {Some conclusive considerations on the comparison of
  the ICARUS numu to nue oscillation search with the MiniBooNE low-energy event
  excess}},  \href{http://arxiv.org/abs/1502.04833}{{\tt arXiv:1502.04833}}.

\bibitem{Agafonova:2013xsk}
{\bf OPERA} Collaboration, N.~Agafonova et~al., {\it {Search for $\nu_\mu
  \rightarrow \nu_e$ oscillations with the OPERA experiment in the CNGS beam}},
   {\em JHEP} {\bf 07} (2013) 004, [\href{http://arxiv.org/abs/1303.3953}{{\tt
  arXiv:1303.3953}}]. [Addendum: JHEP07,085(2013)].

\bibitem{Giunti:2013aea}
C.~Giunti, M.~Laveder, Y.~F. Li, and H.~W. Long, {\it {Pragmatic View of
  Short-Baseline Neutrino Oscillations}},  {\em Phys. Rev.} {\bf D88} (2013)
  073008, [\href{http://arxiv.org/abs/1308.5288}{{\tt arXiv:1308.5288}}].

\bibitem{Kopp:2013vaa}
J.~Kopp, P.~A.~N. Machado, M.~Maltoni, and T.~Schwetz, {\it {Sterile Neutrino
  Oscillations: The Global Picture}},  {\em JHEP} {\bf 05} (2013) 050,
  [\href{http://arxiv.org/abs/1303.3011}{{\tt arXiv:1303.3011}}].

\bibitem{Capozzi:2013csa}
F.~Capozzi, G.~L. Fogli, E.~Lisi, A.~Marrone, D.~Montanino, and A.~Palazzo,
  {\it {Status of three-neutrino oscillation parameters, circa 2013}},  {\em
  Phys. Rev.} {\bf D89} (2014) 093018,
  [\href{http://arxiv.org/abs/1312.2878}{{\tt arXiv:1312.2878}}].

\bibitem{Forero:2014bxa}
D.~V. Forero, M.~Tortola, and J.~W.~F. Valle, {\it {Neutrino oscillations
  refitted}},  {\em Phys. Rev.} {\bf D90} (2014), no.~9 093006,
  [\href{http://arxiv.org/abs/1405.7540}{{\tt arXiv:1405.7540}}].

\bibitem{Gonzalez-Garcia:2014bfa}
M.~C. Gonzalez-Garcia, M.~Maltoni, and T.~Schwetz, {\it {Updated fit to three
  neutrino mixing: status of leptonic CP violation}},  {\em JHEP} {\bf 11}
  (2014) 052, [\href{http://arxiv.org/abs/1409.5439}{{\tt arXiv:1409.5439}}].

\bibitem{Cervera:2000kp}
A.~Cervera et~al., {\it {Golden measurements at a neutrino factory}},  {\em
  Nucl. Phys.} {\bf B579} (2000) 17--55,
  [\href{http://arxiv.org/abs/hep-ph/0002108}{{\tt hep-ph/0002108}}].
  [Erratum-ibid.B593:731-732,2001].

\bibitem{Asano:2011nj}
K.~Asano and H.~Minakata, {\it {Large-Theta(13) Perturbation Theory of Neutrino
  Oscillation for Long-Baseline Experiments}},  {\em JHEP} {\bf 06} (2011) 022,
  [\href{http://arxiv.org/abs/1103.4387}{{\tt arXiv:1103.4387}}].

\bibitem{Agarwalla:2013tza}
S.~K. Agarwalla, Y.~Kao, and T.~Takeuchi, {\it {Analytical Approximation of the
  Neutrino Oscillation Probabilities at large $\theta_{13}$}},
  \href{http://arxiv.org/abs/1302.6773}{{\tt arXiv:1302.6773}}.

\bibitem{Minakata:2001qm}
H.~Minakata and H.~Nunokawa, {\it {Exploring neutrino mixing with low energy
  superbeams}},  {\em JHEP} {\bf 10} (2001) 001,
  [\href{http://arxiv.org/abs/hep-ph/0108085}{{\tt hep-ph/0108085}}].

\bibitem{Friedland:2012tq}
A.~Friedland and I.~M. Shoemaker, {\it {Searching for Novel Neutrino
  Interactions at NOvA and Beyond in Light of Large $\theta_{13}$}},
  \href{http://arxiv.org/abs/1207.6642}{{\tt arXiv:1207.6642}}.

\bibitem{Itow:2001ee}
{\bf T2K} Collaboration, Y.~Itow et~al., {\it {The JHF-Kamioka neutrino
  project}},  \href{http://arxiv.org/abs/hep-ex/0106019}{{\tt hep-ex/0106019}}.

\bibitem{Abe:2011ks}
{\bf T2K} Collaboration, K.~Abe et~al., {\it {The T2K Experiment}},  {\em
  Nucl.Instrum.Meth.} {\bf A659} (2011) 106--135,
  [\href{http://arxiv.org/abs/1106.1238}{{\tt arXiv:1106.1238}}].

\bibitem{Ayres:2002ws}
D.~Ayres, G.~Drake, M.~Goodman, V.~Guarino, T.~Joffe-Minor, et~al., {\it
  {Letter of Intent to build an Off-axis Detector to study numu to nue
  oscillations with the NuMI Neutrino Beam}},
  \href{http://arxiv.org/abs/hep-ex/0210005}{{\tt hep-ex/0210005}}.

\bibitem{Ayres:2004js}
{\bf NOvA} Collaboration, D.~Ayres et~al., {\it {NOvA: Proposal to build a 30
  kiloton off-axis detector to study nu(mu) to nu(e) oscillations in the NuMI
  beamline}},  \href{http://arxiv.org/abs/hep-ex/0503053}{{\tt
  hep-ex/0503053}}.

\bibitem{Ayres:2007tu}
{\bf NOvA} Collaboration, D.~S. Ayres et~al., ``{The NOvA Technical Design
  Report}.'' FERMILAB-DESIGN-2007-01, 2007.

\bibitem{Patterson:2012zs}
{\bf NOvA} Collaboration, R.~Patterson, {\it {The NOvA Experiment: Status and
  Outlook}},  {\em Nucl.Phys.Proc.Suppl.} {\bf 235-236} (2013) 151--157,
  [\href{http://arxiv.org/abs/1209.0716}{{\tt arXiv:1209.0716}}].

\bibitem{Para:2001cu}
A.~Para and M.~Szleper, {\it {Neutrino oscillations experiments using off-axis
  NuMI beam}},  \href{http://arxiv.org/abs/hep-ex/0110032}{{\tt
  hep-ex/0110032}}.

\bibitem{Abe:2013hdq}
{\bf T2K} Collaboration, K.~Abe et~al., {\it {Observation of Electron Neutrino
  Appearance in a Muon Neutrino Beam}},  {\em Phys.Rev.Lett.} {\bf 112} (2014)
  061802, [\href{http://arxiv.org/abs/1311.4750}{{\tt arXiv:1311.4750}}].

\bibitem{T2K_antineutrino_EPS_HEP_2015}
{\bf T2K} Collaboration, M.~Ravonel, {\it {Antineutrino oscillations with
  T2K}},  2015.
\newblock Talk given at the {EPS-HEP} 2015 Conference, July 22-29, 2015,
  Vienna, Austria,
  \url{https://indico.cern.ch/event/356420/session/10/contribution/322}.

\bibitem{Salzgeber:2015gua}
{\bf T2K} Collaboration, M.~R. Salzgeber, {\it {Anti-neutrino oscillations with
  T2K}},  \href{http://arxiv.org/abs/1508.06153}{{\tt arXiv:1508.06153}}.

\bibitem{Abe:2014tzr}
{\bf T2K} Collaboration, K.~Abe et~al., {\it {Neutrino oscillation physics
  potential of the T2K experiment}},  {\em PTEP} {\bf 2015} (2015), no.~4
  043C01, [\href{http://arxiv.org/abs/1409.7469}{{\tt arXiv:1409.7469}}].

\bibitem{Childress:2013npa}
{\bf NuMI, NOvA, LBNE} Collaboration, S.~Childress and J.~Strait, {\it {Long
  baseline neutrino beams at Fermilab}},  {\em J. Phys. Conf. Ser.} {\bf 408}
  (2013) 012007, [\href{http://arxiv.org/abs/1304.4899}{{\tt
  arXiv:1304.4899}}].

\bibitem{NOvA_appearance_seminar_FNAL_2015}
{\bf NOvA} Collaboration, R.~Patterson, {\it {First oscillation results from
  NOvA}},  2015.
\newblock Talk given at the Joint Experimental-Theoretical Physics Seminar,
  Fermilab, 6th August, 2015,
  \url{http://nova-docdb.fnal.gov/cgi-bin/RetrieveFile?docid=13883&filename=20150806_nova_docdb.pdf&version=2}.

\bibitem{Bian:2015opa}
{\bf NOvA} Collaboration, J.~Bian, {\it {First Results of $\nu_e$ Appearance
  Analysis and Electron Neutrino Identification at NOvA}},  in {\em {Meeting of
  the APS Division of Particles and Fields (DPF 2015) Ann Arbor, Michigan, USA,
  August 4-8, 2015}}, 2015.
\newblock \href{http://arxiv.org/abs/1510.05708}{{\tt arXiv:1510.05708}}.

\bibitem{Adamson:2016tbq}
{\bf NOvA} Collaboration, P.~Adamson et~al., {\it {First measurement of
  electron neutrino appearance in NOvA}},
  \href{http://arxiv.org/abs/1601.05022}{{\tt arXiv:1601.05022}}.

\bibitem{Adamson:2016xxw}
{\bf NOvA} Collaboration, P.~Adamson et~al., {\it {First measurement of
  muon-neutrino disappearance in NOvA}},
  \href{http://arxiv.org/abs/1601.05037}{{\tt arXiv:1601.05037}}.

\bibitem{Agarwalla:2012bv}
S.~K. Agarwalla, S.~Prakash, S.~K. Raut, and S.~U. Sankar, {\it {Potential of
  optimized NOvA for large $\theta_(13)$ and combined performance with a LArTPC
  and T2K}},  {\em JHEP} {\bf 1212} (2012) 075,
  [\href{http://arxiv.org/abs/1208.3644}{{\tt arXiv:1208.3644}}].

\bibitem{Agarwalla:2013ju}
S.~K. Agarwalla, S.~Prakash, and S.~U. Sankar, {\it {Resolving the octant of
  theta23 with T2K and NOvA}},  {\em JHEP} {\bf 1307} (2013) 131,
  [\href{http://arxiv.org/abs/1301.2574}{{\tt arXiv:1301.2574}}].

\bibitem{Huber:2004ka}
P.~Huber, M.~Lindner, and W.~Winter, {\it {Simulation of long-baseline neutrino
  oscillation experiments with GLoBES (General Long Baseline Experiment
  Simulator)}},  {\em Comput.Phys.Commun.} {\bf 167} (2005) 195,
  [\href{http://arxiv.org/abs/hep-ph/0407333}{{\tt hep-ph/0407333}}].

\bibitem{Huber:2007ji}
P.~Huber, J.~Kopp, M.~Lindner, M.~Rolinec, and W.~Winter, {\it {New features in
  the simulation of neutrino oscillation experiments with GLoBES 3.0: General
  Long Baseline Experiment Simulator}},  {\em Comput.Phys.Commun.} {\bf 177}
  (2007) 432--438, [\href{http://arxiv.org/abs/hep-ph/0701187}{{\tt
  hep-ph/0701187}}].

\bibitem{Zhan:2015aha}
{\bf Daya Bay} Collaboration, L.~Zhan, {\it {Recent Results from Daya Bay}},
  \href{http://arxiv.org/abs/1506.01149}{{\tt arXiv:1506.01149}}.

\bibitem{Agarwalla:2013qfa}
S.~K. Agarwalla, S.~Prakash, and W.~Wang, {\it {High-precision measurement of
  atmospheric mass-squared splitting with T2K and NOvA}},
  \href{http://arxiv.org/abs/1312.1477}{{\tt arXiv:1312.1477}}.

\bibitem{PREM:1981}
A.~M. Dziewonski and D.~L. Anderson, {\it Preliminary reference earth model},
  {\em Physics of the Earth and Planetary Interiors} {\bf 25} (1981) 297--356.

\bibitem{Huber:2002mx}
P.~Huber, M.~Lindner, and W.~Winter, {\it {Superbeams versus neutrino
  factories}},  {\em Nucl. Phys.} {\bf B645} (2002) 3--48,
  [\href{http://arxiv.org/abs/hep-ph/0204352}{{\tt hep-ph/0204352}}].

\bibitem{Fogli:2002pt}
G.~L. Fogli, E.~Lisi, A.~Marrone, D.~Montanino, and A.~Palazzo, {\it {Getting
  the most from the statistical analysis of solar neutrino oscillations}},
  {\em Phys. Rev.} {\bf D66} (2002) 053010,
  [\href{http://arxiv.org/abs/hep-ph/0206162}{{\tt hep-ph/0206162}}].

\bibitem{Blennow:2013oma}
M.~Blennow, P.~Coloma, P.~Huber, and T.~Schwetz, {\it {Quantifying the
  sensitivity of oscillation experiments to the neutrino mass ordering}},  {\em
  JHEP} {\bf 1403} (2014) 028, [\href{http://arxiv.org/abs/1311.1822}{{\tt
  arXiv:1311.1822}}].

\end{thebibliography}\endgroup

\end{document}